\documentclass[aps, preprintnumbers, tightenlines, nofootinbib, 11pt]{revtex4-2}

\usepackage{amssymb,amsmath,amsfonts,amsbsy,epsfig,wrapfig,caption,cancel,graphicx, pstricks, slashed, mathtools}
\usepackage{comment}
\usepackage[colorlinks=true, linktocpage=true, allcolors=blue]{hyperref}
\usepackage{enumitem}
\usepackage{caption}
\usepackage{eqnarray}
\usepackage{tikz-feynman}
\usepackage{mathrsfs}

\usetikzlibrary{shapes.misc}

\tikzset{cross/.style={cross out, draw=black, minimum size=2*(#1-\pgflinewidth), inner sep=0pt, outer sep=0pt},
cross/.default={1pt}}

\newcommand{\eq}[2]{\begin{equation}\label{#1}#2 \end{equation}}

\newcommand{\wt}[1]{\widetilde{#1}}
\newcommand{\ovl}[1]{\overline{#1}}

\usepackage{xcolor}

\newcommand{\tnu}{\tilde{\nu}}

\newcommand{\N}{\noindent}

\newcommand{\calD}{{\cal D}}
\newcommand{\calE}{{\cal E}}
\newcommand{\calF}{{\cal F}}
\newcommand{\calL}{{\cal L}}

\newcommand{\calG}{{\cal G}}

\newcommand{\calO}{{\cal O}}

\newcommand{\bn}{{\mathbf{n}}}

\newcommand{\bJ}{{\mathbf{J}}}

\newcommand{\bk}{{\mathbf{k}}}

\newcommand{\bv}{{\mathbf{v}}}

\newcommand{\bx}{{\mathbf{x}}}
\newcommand{\by}{{\mathbf{y}}}

\newcommand{\bq}{{\mathbf{q}}}

\newcommand{\expect}[1]{\langle #1 \rangle}

\newcommand{\hbn}{\hat{\mathbf{n}}}
\newcommand{\field}{\psi}
\newcommand{\Field}{\Psi}
\newcommand{\diff}{\kappa}
\newcommand{\auxold}{\chi}
\newcommand{\auxnew}{\phi}
\newcommand{\noise}{\xi}
\newcommand{\scrA}{\mathscr{A}}
\newcommand{\scrB}{\mathscr{B}}
\newcommand{\scrI}{\mathscr{I}}

\newcommand{\dbar}{d\hspace*{-0.08em}\bar{}\hspace*{0.1em}}
\newcommand{\deltabar}{\delta\hspace*{-0.24em}\bar{}\hspace*{0.1em}}

\def\bea{\begin{eqnarray}}
	\def\eea{\end{eqnarray}}
\def\be{\begin{equation}}
	\def\ee{\end{equation}}
\def\ba{\begin{array}}
	\def\ea{\end{array}}

\def\nn{\nonumber}

\definecolor{Dgreen}{rgb}{0,0.7,0.0}

\makeatletter
\def\l@subsection#1#2{}
\def\l@subsubsection#1#2{}
\makeatother

\begin{document}
	\title{On the Renormalization Group Flow of Active Flocks}	
	\author{Kevin T. Grosvenor${}^1$}
	\email{ktgrosvenor@up.edu.ph}
	\author{Subodh P. Patil${}^2$}
	\email{patil@lorentz.leidenuniv.nl}	
	\affiliation{${}^1$National Institute of Physics, University of the Philippines, Diliman, Quezon City 1101, Philippines}
	\affiliation{${}^2$Instituut-Lorentz for Theoretical Physics, Leiden University, 2333 CA, Leiden, The Netherlands}

\begin{abstract}
We study the statistical field-theoretic renormalization of active flocks via the Martin–Siggia–Rose–De Dominicis–Janssen (MSRDJ) formulation of stochastic dynamics, focusing on the Toner–Tu theory of Malthusian flocks, or polar-ordered, momentum non-conserving active fluids in which density fluctuations relax sufficiently rapidly to be eliminated as hydrodynamic degrees of freedom. Working in the limit of isotropic diffusion in two spatial dimensions, we perform an all orders analysis of the perturbative renormalization of propagators and interaction vertices. We show that a nonlinearly realized diagonal rotation/shift symmetry gives rise to Ward identities that protect the trigonometric structure of the convective derivative, ensuring that the sine and cosine interaction potentials retain their functional form under renormalization up to a rescaling of their overall couplings. We demonstrate that the freedom to renormalize the dynamical critical exponent in this boost non-invariant theory prevents the determination of the precise values of the critical exponents and anomalous dimensions in the theory, highlighting parallels and crucial differences with sine-Gordon and XY models. We express all anomalous dimensions as functions of the dynamical critical exponent, which remains undetermined due to the absence of a relation ordinarily forced upon boost invariant systems. We revisit and contextualize several claims and counterclaims in the recent literature in light of these results.
\end{abstract}

\maketitle
	
\tableofcontents
	
	
\section{Preliminary remarks}

Recently, there has been renewed interest and debate over how hydrodynamic descriptions of two-dimensional polar flocks behave under renormalization, and in particular, the conditions under which long range order can persist in such systems \cite{CS1, Ikeda2024VicsekNG, inconvenienttruth, comment1, response1, comment2,Amoretti:2024obt}. These are all framed within the context of the Toner-Tu hydrodynamics of active flocks, within which it was first established that motile polar systems can evade the Coleman-Hohenberg-Mermin-Wagner constraints that forbid long-range order in equilibrium two-dimensional continuous-spin systems \cite{tonertu1, TonerTu1998, TonerTuRamaswamy2005, 	Vicsek1995, Marchetti2013, VicsekZafeiris2012}. That framework naturally lends itself to the language of critical exponents and canonical scaling predictions. The long-wavelength, long-time scaling of two-dimensional polar flocks is customarily characterized by three critical exponents: the dynamical critical exponent $z$ (so that time scales as $t \sim y^z$, where $y$ is a distance transverse to the flock propagation), the anisotropy exponent $\zeta$ ($x \sim y^{\zeta}$, where $x$ is a longitudinal distance), and the roughness exponent $\chi$ ($\delta \theta \sim y^{\chi}$, where $\delta \theta$ are fluctuations in the flocking direction).

In the case where one takes the \emph{Malthusian} limit -- where density fluctuations are on such short timescales that they can be eliminated as a hydrodynamic variable -- Chat\'e and Solon \cite{CS1, comment1,comment2} argue that a generalized Galilean invariance protects the nonlinear couplings of the theory, denoted $\lambda_x$ and $\lambda_y$ from receiving corrections\footnote{We define the Malthusian flock and its parameters in the next section. For the present discussion, $\lambda_x$ and $\lambda_y$ can be taken to parameterize the derivatively coupled operators $\lambda_x\partial_x\psi$ and $\lambda_y \partial_y\psi$, where $\psi$ is the flock angle, and where each operator can be dressed by powers or functions of $\psi$.}, leading to three relations among the three critical exponents, and therefore exact exponent values. On the other hand, although Chen et al. \cite{inconvenienttruth, response1} agree with two of the resulting exponent relations, they contest the claim that symmetries force $\lambda_x$ and $\lambda_y$ to have vanishing anomalous dimensions. Consequently, they obtain only two independent exponent relations and conclude that the long wavelength properties of the Malthusian-flock cannot be determined exactly from symmetry arguments alone. Both groups, however, agree that the variance of the noise that the flock is subject to, denoted $\Delta$, receives no explicit diagrammatic corrections, and therefore does not acquire an anomalous dimension.

In the context of immortal (i.e. number density conserving) flocks, the divergence is more substantial. Chen et al. argue that the known symmetries of the flocking equations are insufficient to derive any exact relations among the critical exponents, implying that no exact exponent values can presently be obtained. Chat\'e and Solon on the other hand, while not explicitly deriving exact exponent relations for immortal flocks, contend that Chen et al.'s renormalization-group analysis does not establish their no-go conclusion and therefore do not rule out the possibility that additional symmetry-based constraints remain to be discovered. In summary, while there is broad agreement that the Malthusian limit admits more constraints, the number-conserving case is evidently rather more unsettled, and there does not yet appear to be consensus on universal limitations as to how large, or how long-lived active flocks can become. More recent numerical and quantitative assessments of the predictions for hydrodynamic limits of flocks (continuum simulations and microscopic Vicsek-type models) have been crucial in testing analytic claims \cite{GinelliChate2010, SolonTailleur2013, SolonTailleur2015, QAssess, Solon2022, Karmakar2024}. These numerical studies are central to the current debate because finite-size effects, implementation differences between Malthusian and number-conserving systems, and the treatment of non-linearities all affect the interpretation of measured exponents.

At its core, the debate hinges upon consistent and complete renormalization group (RG) analyses of the theory. However, one is confronted with a number of difficulties in interpreting prior analyses as they have nominally been discussed in the literature. Firstly, while guided by essentially correct physical intuition, conclusions drawn solely from dynamical scaling arguments, or from phenomenological renormalization schemes such as dynamical RG should be interpreted with caution, and mindful of a number of caveats. Moreover, symmetry arguments by themselves do not establish the all-orders non-renormalization statements required to determine the critical exponents exactly. Such statements must instead follow from explicit Ward identities derived within a consistent renormalized field-theoretic framework. 

In contrast, a complete and correctly implemented functional or diagrammatic renormalization analysis would provide a controlled framework in which all symmetry-allowed operators generated under coarse-graining can be systematically identified and their contributions explicitly evaluated. Within such a framework, care must nevertheless be exercised in distinguishing constraints on operator structures from constraints on scaling dimensions. Indeed, while the operator structure of the theory of interest is preserved under renormalization, we nevertheless find that there is one more scaling dimension in the theory than there are renormalization constraints to determine them completely. Secondly, previous analyses have made the seemingly a priori reasonable assumption that the specific form of the equations of motion (involving derivatively coupled trigonometric functions of the flock angle $\psi$) will deform under renormalization, going so far as claiming that one is guaranteed to flow away from these special functional forms. If that is the case, then the RG analysis must be taken from the more general starting point with all possible higher-point interactions included in a consistent power counting expansion (as one does in any effective field theoretic treatment), rather than the specific ones that derive from the original Toner-Tu equation of motion. However, from this more general standpoint, the interpretation of the order parameter as a director angle for the flock becomes tenuous without additional qualifications. One additionally encounters analyses that invoke the assumption that any dimensionally regularized computation in $4 - \epsilon$ dimensions can be extrapolated down to 2 dimensions. These extrapolations are used to infer that corrections to the diffusion constants are all positive. While this may seem plausible given the validity of the extrapolation, this is far from proven, and as will be elaborated upon further, need not hold in general, and should therefore not be assumed without an explicit renormalization analysis.

Clearly, it behooves us to perform a consistent renormalization of any given hydrodynamic description of active flocks before attempting to resolve the questions currently under debate. In what follows, we adopt the Martin–Siggia–Rose–De Dominicis–Janssen (MSRDJ) formulation for stochastic systems [19–21] and perform an all-order renormalization of propagators and interaction vertices for the isotropic limit of the Malthusian flock. A central result of our analysis is that the nonlinear realization of a diagonal subgroup combining flock-angle shifts with spatial rotations gives rise to Ward identities that constrain the renormalization of the interaction vertices, ensuring that the trigonometric structure of the convective derivative is preserved under renormalization. We also derive exact relations among the anomalous dimensions of the couplings in terms of the dynamical critical exponent. Determining the latter, however, requires one additional independent relation that does not presently appear to follow from the known symmetries of the theory\footnote{\label{foot:mc} Previous versions of this preprint mistakenly identified an internal field space invariance, as well as a redundancy of the MSRDJ formalism as supplying this relation. We withdraw this claim and conclude that the fate of long range order is undetermined within the model confines of the Malthusian flock.}, and whose absence can be traced to subtleties specific to perturbative renormalization in the context of boost non-invariant systems.

We should also be clear about what our analysis will not attempt: The treatment that follows does not compute the RG flow of the full theory of Malthusian flocks, nor any extension to the theory of immortal flocks. We restrict our attention in what follows to the theory of Malthusian flocks in the limit of isotropic diffusion. That is, if $\diff_{||}$ and $\diff_{\perp}$ are the diffusion constants parallel and perpendicular to the direction of net motion of the flock, we only consider the limit $\diff_{||} = \diff_{\perp}$.\footnote{It turns out that the diagonal shift/rotation subgroup discussed in Section \ref{sec:symm} will still be preserved even in the anisotropic case (i.e. when $\diff_{||} \neq \diff_{\perp}$), but at the cost of introducing two additional non-linear interaction terms, each with its own coupling constant. We thank Sriram Ramaswamy and Ananyo Maitra for sharing this fact with us. We discuss extensions of our analysis to the anisotropic case in our concluding remarks.}

The main technical results that will follow in the subsequent sections are summarized as:  
\begin{itemize}
\item By recasting the stochastic differential equations defining the isotropic Malthusian flock into the MSRDJ formalism, we obtain a statistical field theory suitable for systematic perturbative renormalization.

\item We identify extensive diagrammatic cancellations and non-renormalization properties that follow from the kinematics and nonlinear realization of the underlying symmetries.

\item We prove that the Ward identities associated with the diagonal shift/rotation symmetry preserve the trigonometric form of the convective derivative to all orders, so that the renormalized interaction is completely characterized by the running of the single coupling $\lambda$.

\item We derive exact all-orders relations among the anomalous dimensions of the couplings in terms of the dynamical critical exponent $z$. These relations determine the renormalization-group flow up to one remaining independent scaling relation.
We discuss the consequences of these results for quasi-long-range order in isotropic Malthusian flocks and compare them with previous treatments in the literature.
\end{itemize}

As stated above, the latter point hinges on a subtlety regarding the renormalization procedure in boost non-invariant settings that we expand on in Appendix \ref{sec:wilsonianRG}. The properties of the stochastic fluctuations, in particular, its variance, also gets renormalized, as we argue must be the case in an interacting theory. The consequences of these results for quasi-long-range order are discussed in Section V. We conclude with a comparison to previous treatments of Malthusian flocks and discuss extensions to anisotropic diffusion and to theories in which density fluctuations are restored as hydrodynamic degrees of freedom.\footnote{It became incumbent upon us over the course of our study to identify where and how the treatment and conclusions in \cite{CS1, Ikeda2024VicsekNG, inconvenienttruth, comment1, response1, comment2,Amoretti:2024obt} diverged from each other, and to that of our own analysis. In the interest of clarity, we address certain key points in our concluding remarks and appendices, and welcome correspondence for additional clarification.}

\section{Setup}

In this section, we elaborate on the set up within which we renormalize the theory of Malthusian flocks, utilizing the MSRDJ formalism as our workhorse \cite{PhysRevA.8.423,Janssen:1976qag,PhysRevB.18.4913}. We begin with the hydrodynamic equation for the local direction angle $\field$ of a two dimensional flock, in terms of which we define the unit vectors longitudinal and transverse to the flocking direction:
\begin{align}
	\label{eq:nd}
\hbn &= \binom{\cos\field}{\sin\field}, &%
\hbn_{\perp} &= \binom{- \sin\field}{\cos \field}.
\end{align}
In order to proceed, we perform a background field expansion and take the net motion of the flock to be in the positive $x$-direction (cf. the discussion surrounding \eqref{eq:bg} in Appendix \ref{sec:feynrules}). This will be without loss of generality given the symmetries of the system, to be detailed in Section \ref{sec:symm}. Therefore, $\field$ denotes the local perturbation of the fluid flow away from net motion along the $x$-axis, so that $k_x$ and $k_y$ may always be thought of as $k_{||}$ and $k_{\perp}$, the momenta parallel and perpendicular to the flocking direction.

The equation of motion is given in eq. III.6 of \cite{inconvenienttruth}:
\begin{align} \label{eq:1}
\partial_t \field &= \lambda \nabla \cdot \hat{\bn}_{\perp} + \diff \nabla^2 \field + \noise,
\end{align}
where $\lambda$ and $\diff$ are constants and $\noise$ is taken to be Gaussian noise, with zero mean and two-point function 
\begin{equation}
\bigl\langle \noise ( \bx, t ) \, \noise ( \bx', t' ) \bigr\rangle = 2 \Delta \delta ( \bx - \bx' ) \delta ( t - t' ),
\end{equation}
where $\Delta$ is a constant (half the variance of the noise). The coefficient $\diff$ is the diffusion constant and can, more generally, be different along the two spatial directions. Here, we stick to the spatially isotropic case as a matter of conceptual clarity and simplicity. We discuss how to extend our calculations to the anisotropic case in our concluding remarks, along with extensions where density fluctuations are re-admitted as a hydrodynamic variable.

With the definition
\begin{equation}
\label{eq:calE}
\calE \equiv \partial_t \field - \lambda \nabla \cdot \hbn_{\perp} - \diff \nabla^2 \field,
\end{equation}
the equation of motion \eqref{eq:1} now reads $\calE - \noise = 0$. The partition function for this system is the path integral over all configurations of $\psi$ and the noise constrained by the equation of motion:
\begin{equation} \label{eq:Z}
Z = \int \calD \field \, \calD \noise \, e^{- \frac{1}{4 \Delta} \int dt \, d^d x \, \noise^2} \delta ( \calE - \noise ).
\end{equation}
%

The integral over the noise is trivial, of course, but the result will be more tractable if we integrate in an auxiliary field by Fourier transforming the delta function:
\begin{equation}
Z = \int \calD \field \, \calD \chi \, \calD \noise \, \exp \biggl[ - \int dt \, d^d x \biggl( \frac{1}{4 \Delta} \noise^2 - i \chi ( \calE - \noise ) \biggr) \biggr]
\end{equation}
Now, $\noise$ may be integrated out leaving
\begin{equation}
\label{eq:msrdj}
Z = \int \calD \field \, \calD \chi \, e^{- \int dt \, d^d x ( \Delta \chi^2 - i \chi \calE )} = \int \calD \field \, \calD \chi \, e^{-S},
\end{equation}
where the action $S$ is the spacetime integral of the Lagrangian density,\footnote{Although the MSRDJ functional integral closely resembles a Euclidean field theory, it represents a stochastic path integral whose precise form depends on the underlying stochastic calculus. In fact, the functional integral in the partition function \eqref{eq:Z} should contain the term $\det \frac{\delta \calE}{\delta \field}$. This is unity in the It\^{o} prescription employed in this work, but nontrivial in any other, such as the Stratonovich prescription. Furthermore, nonlinear field redefinitions generally require the corresponding It\^{o} transformation law rather than the naive chain rule, a point that becomes important when interpreting symmetries and redundancies of the stochastic theory.}
\begin{equation} \label{eq:origlag}
\calL = \Delta \chi^2 - i \chi \calE.
\end{equation}
The Euler-Lagrange equations of motion read
\begin{align} \label{eq:ABold}
\scrA &\equiv \calE + 2i \Delta \chi = 0, &%
\scrB &= \partial_t \chi + \lambda \partial_{||} \chi + \diff \nabla^2 \chi = 0,
\end{align}
where the longitudinal derivative is
\begin{equation}
\partial_{||} \chi = \hbn \cdot \nabla \chi = ( \cos \field ) \partial_x \chi + ( \sin \field ) \partial_y \chi.
\end{equation}
The equation of motion $\scrA = 0$ implies that $\chi$ must be pure imaginary if $\field$ is to be real. Therefore, we define the real field $\auxnew$ to be
\begin{equation}
    \auxnew = - i \chi,
\end{equation}
so that
\begin{equation}
    \calL = \auxnew \calE - \Delta \auxnew^2.
\end{equation}
This way, the equations of motion become
\begin{align} \label{eq:AB}
    \scrA &\equiv \calE - 2 \Delta \auxnew = 0, &%
    \scrB &\equiv \partial_t \auxnew + \lambda \partial_{||} \auxnew + \diff \nabla^2 \auxnew = 0.
\end{align}
Changing to this real field does not affect any observable quantities nor their running, as had better be the case since $\auxnew$ or $\chi$ is simply an auxiliary field. In fact, using the real response field $\auxnew$ is how MSRDJ is implemented from the start in \cite{heliasdahmen, Grosvenor:2021eol}, for example.

Proceeding in terms of $\phi$ henceforth, we note that $\nabla \cdot \hbn_{\perp}$ may also be written as $- \partial_{||} \field$, so that we can rewrite
\begin{equation}
\calE = \partial_t \field + \lambda \partial_{||} \field - \diff \nabla^2 \field,
\end{equation}
and
\begin{align}
\auxnew \calE &= \auxnew \bigl[ \partial_t \field + \lambda ( \cos \field ) \partial_x \field + \lambda ( \sin \field ) \partial_y \field - \diff \nabla^2 \field \bigr] \notag \\
&= \auxnew \bigl[ ( \partial_t + \lambda \partial_x - \diff \nabla^2 ) \field + \lambda ( \cos \field - 1 ) \partial_x \field + \lambda ( \sin \field ) \partial_y \field \bigr].
\end{align}
We can therefore decompose the Lagrangian into the following free and interaction parts:
\begin{subequations} \label{eq:L}
\begin{align}
\calL_{\rm free} &= \auxnew ( \partial_t + \lambda \partial_x - \diff \nabla^2 ) \field - \Delta \auxnew^2, \label{eq:Lfree} \\
\calL_{\rm int} &= \lambda \auxnew \bigl[ ( \cos \field - 1 ) \partial_x \field + ( \sin \field ) \partial_y \field \bigr]. \label{eq:Lint}
\end{align}
\end{subequations}
Operationally, the distinction between free and interacting terms is in whether they consist of field bilinears, or higher order terms, respectively. The RG flow of this system may be analyzed using the standard techniques of perturbative field theory. The details of this computation are relegated to the appendices. For this purpose, however, it is simpler and more convenient to integrate by parts in the interaction Lagrangian and instead write
\begin{align}
\calL_{\rm int} = - \lambda ( \sin \field - \field ) \partial_x \auxnew + \lambda ( \cos \field -1 ) \partial_y \auxnew.
\end{align}
The Feynman rules are derived in Appendix \ref{sec:feynrules}. The non-vanishing bare propagators are
\begin{subequations}
\begin{align}
\begin{tikzpicture}
\draw[thick,>=Latex,->] (-1,0) -- (0,0);
\draw[thick] (0,0) -- (1,0);
\end{tikzpicture} &= \expect{\field_{- \omega , \bk} \, \field_{\omega, \bk}}_0 = \frac{2 \Delta}{( \omega - \lambda k_x - i \diff k^2 ) ( \omega - \lambda k_x + i \diff k^2 )}, \label{eq:psipsitext} \\
\begin{tikzpicture}
\draw[thick,>=Latex,->] (-1,0) -- (0,0);
\draw[thick,dashed] (0,0) -- (1,0);
\end{tikzpicture} &= \expect{\auxnew_{- \omega, - \bk} \, \field_{\omega, \bk}}_0 = \frac{i}{\omega - \lambda k_x + i \diff k^2}, \label{eq:psichitext} \\
\begin{tikzpicture}
\draw[thick,dashed,>=Latex,->] (-1,0) -- (0,0);
\draw[thick] (0,0) -- (1,0);
\end{tikzpicture} &= \expect{\field_{- \omega, - \bk} \, \auxnew_{\omega, \bk}}_{0} = \frac{-i}{\omega - \lambda k_x - i \diff k^2}, \label{eq:chipsitext}
\end{align}
\end{subequations}
and the vertices are given as follows for $n=1, 2, 3, \ldots$,
\begin{equation} \label{eq:2nplus1vertex}
V_{2n+1} = \vcenter{
\hbox{
\begin{tikzpicture}
\draw[thick,dashed,>=Latex,>-] (-1,0) -- (0,0);
\node[left] at (-1,0) {$\omega, \bk$}; 
\draw[thick,>=Latex,-<] (0,0) -- (45:1cm);
\draw[thick,>=Latex,-<] (0,0) -- (30:1cm);
\draw[thick,>=Latex,-<] (0,0) -- (-30:1cm);
\draw[thick,>=Latex,-<] (0,0) -- (-45:1cm);
\draw[thick,dotted]  (-15:0.8) arc(-15:15:0.8) node[midway,right]{$\scriptstyle 2n$};
\end{tikzpicture}
}
}
= (-1)^{n} i \lambda k_y,
\end{equation}
and
\begin{equation} \label{eq:2nplus2vertextext}
V_{2n+2} = \vcenter{
\hbox{
\begin{tikzpicture}
\draw[thick,dashed,>=Latex,>-] (-1,0) -- (0,0);
\node[left] at (-1,0) {$\omega, \bk$}; 
\draw[thick,>=Latex,-<] (0,0) -- (45:1cm);
\draw[thick,>=Latex,-<] (0,0) -- (22.5:1cm);
\draw[thick,>=Latex,-<] (0,0) -- (-45:1cm);
\draw[thick,dotted]  (-30:0.8) arc(-30:10:0.8) node[midway,right]{$\scriptstyle 2n+1$};
\end{tikzpicture}
}
}
= - (-1)^{n} i \lambda k_x.
\end{equation}
Note that there is a special relationship between the vertices, which derives from the fact that they come from the cosine and sine of $\field$:
\begin{align} \label{eq:recursion}
V_{2n+1} + k_y \frac{\partial}{\partial k_x} V_{2n+2} &= 0, &%
V_{2n} - k_x \frac{\partial}{\partial k_y} V_{2n+1} &= 0.
\end{align}
It will turn out that these recursion relations are preserved under renormalization by the Ward identities associated with the nonlinearly realized diagonal rotation/shift symmetry as elaborated on in Section \ref{sec:symm}. Consequently, the trigonometric structure of the convective derivative is preserved to all orders, with renormalization affecting only the overall coupling parameter $\lambda$. 


Before detailing our findings, we ought to first understand the characteristic scales associated with the stochastic fluctuations and their dissipation. These are parameterized by $\Delta$ and $\kappa$, respectively, as opposed to an overall pre-factor in front of the action in \eqref{eq:msrdj} that would play a role analogous to that played by $\hbar$ or $\beta^{-1}$ in the context of quantum or thermal field theory. We could just as well have chosen to assign such an overall pre-factor with the dimensions of action, which would have consequently rescaled all couplings, and subsequently all anomalous dimensions generated by loop corrections. In analogous `natural' units, this pre-factor would be given by $\Delta/\kappa$, which in the limit of vanishing interactions corresponds to the characteristic amplitude of the equal time Fourier mode fluctuations: $k^2 \langle |\psi(t,\textbf{k})|^2\rangle \sim \Delta/\kappa$. Specifically, by taking the $\lambda \to 0$ limit of \eqref{eq:psipsitext} and integrating over the frequency domain, one finds
\eq{eq:fs}{k^2\langle | \psi(t,\mathbf{k})|^2\rangle= k^2\int \frac{d \omega}{2 \pi} \frac{2 \Delta}{\omega^2+\kappa^2 k^4}=\frac{\Delta}{\kappa}.}
We will show that $\kappa$ and $\Delta$ acquire anomalous dimensions, but that these are identical to all loop orders. Therefore, the exactly dimensionless ratio $\Delta/\kappa$ characterizes the relative strength of stochastic fluctuations and diffusion throughout our analysis. We will derive exact all-orders relations among the anomalous dimensions of the couplings in terms of the dynamical critical exponent. These relations imply that the renormalization-group flow is highly constrained by the Ward identities associated with the diagonal rotation/shift symmetry, although one additional independent relation is required to determine the scaling exponents completely. 


\section{All-Loop-Order Results}

A combination of a non-renormalization theorem and a series of power-counting arguments discussed in detail in Appendix \ref{sec:feynrules} prove that the only one-particle irreducible (1PI) diagrams that contribute either to propagator or vertex renormalizations are those with exactly one vertex, and any number of $\field$-$\field$ loops attached to the vertex. For example, for a propagator of type $a$ to $b$, the analog of the self-energy is
\begin{equation}
\Sigma_{ab} = 
\vcenter{
\hbox{
\begin{tikzpicture}[scale=0.5]
\draw[white] (0,-1.4) -- (0,1.4);
\draw[thick] (-2,0) -- (2,0);
\draw[thick] (0,0) to[out=170,in=180] (0,1.2);
\draw[thick] (0,0) to[out=10,in=0] (0,1.2);
\node[left] at (-2,0) {$a$};
\node[right] at (2,0) {$b$};
\end{tikzpicture}
}
}
+
\vcenter{
\hbox{
\begin{tikzpicture}[scale=0.5]
\draw[white] (0,-1.4) -- (0,1.4);
\draw[thick] (-2,0) -- (2,0);
\draw[thick] (0,0) to[out=170,in=180] (0,1.2);
\draw[thick] (0,0) to[out=10,in=0] (0,1.2);
\draw[thick] (0,0) to[out=-10,in=0] (0,-1.2);
\draw[thick] (0,0) to[out=190,in=180] (0,-1.2);
\node[left] at (-2,0) {$a$};
\node[right] at (2,0) {$b$};
\end{tikzpicture}
}
}
+ \cdots.
\end{equation}
One can either treat these as amputated, in which case they correct the \emph{inverse} propagator, or one can follow the usual random phase approximation (RPA) scheme. Both approaches are elaborated upon in the Appendices \ref{sec:onelooporder} and \ref{sec:twoloopandhigher}, and give identical results.

For the vertices, schematically, one has to resum  the following series of diagrams:
\begin{equation}
V^{\rm ren} = 
\vcenter{
\hbox{
\begin{tikzpicture}
\draw[thick,dashed,>=Latex,>-] (-1,0) -- (0,0);
\draw[thick,>=Latex,-<] (0,0) -- (45:1cm);
\draw[thick,>=Latex,-<] (0,0) -- (22.5:1cm);
\draw[thick,>=Latex,-<] (0,0) -- (-45:1cm);
\draw[thick,dotted]  (-30:0.8) arc(-30:10:0.8);
\end{tikzpicture}
}
}
\ + \
\vcenter{
\hbox{
\begin{tikzpicture}
\draw[thick,dashed,>=Latex,>-] (-1,0) -- (0,0);
\draw[thick,>=Latex,-<] (0,0) -- (45:1cm);
\draw[thick,>=Latex,-<] (0,0) -- (22.5:1cm);
\draw[thick,>=Latex,-<] (0,0) -- (-45:1cm);
\draw[thick,dotted]  (-30:0.8) arc(-30:10:0.8);
\draw[thick,->] (0,0) to[out=170,in=190] (-0.2,0.8);
\draw[thick] (0,0) to[out=55,in=10] (-0.2,0.8);
\end{tikzpicture}
}
}
\ + \
\vcenter{
\hbox{
\begin{tikzpicture}
\draw[thick,dashed,>=Latex,>-] (-1,0) -- (0,0);
\draw[thick,>=Latex,-<] (0,0) -- (45:1cm);
\draw[thick,>=Latex,-<] (0,0) -- (22.5:1cm);
\draw[thick,>=Latex,-<] (0,0) -- (-45:1cm);
\draw[thick,dotted]  (-30:0.8) arc(-30:10:0.8);
\draw[thick,->] (0,0) to[out=170,in=190] (-0.2,0.8);
\draw[thick] (0,0) to[out=55,in=10] (-0.2,0.8);
\draw[thick,->] (0,0) to[out=-170,in=-190] (-0.2,-0.8);
\draw[thick] (0,0) to[out=-55,in=-10] (-0.2,-0.8);
\end{tikzpicture}
}
} + \cdots
\end{equation}
A single loop integral has a leading logarithmic divergence of the form (see app. \ref{sec:onelooporder})
\begin{equation}
\scrI = \frac{\Delta}{4 \pi \diff} \ln \frac{\Lambda}{\mu},
\end{equation}
where $\Lambda$ and $\mu$ are UV and IR cut-offs, respectively, although the same conclusions would have been arrived at in dimensional regularization, as also elaborated upon in the appendix. The sum over all loops generates the factor
\begin{equation}
1 - \scrI + \frac{1}{2} \scrI^2 - \frac{1}{6} \scrI^3 + \cdots = e^{- \scrI} = \biggl( \frac{\Lambda}{\mu} \biggr)^{- \frac{\Delta}{4 \pi \diff}}.
\end{equation}
This contributes multiplicatively only to the convective derivative term. No other term in the Lagrangian gets diagrammatic corrections. Therefore, it would appear that $\diff$ and $\Delta$ do not get renormalized. In fact, this is precisely the argument given in the literature \cite{CS1,inconvenienttruth,Amoretti:2024obt} for the non-renormalization of $\Delta$ to derive the relationship $z - 2 \chi -1 - \zeta = 0$ among the critical exponents. Instead, we show that the possibility of assigning an anomalous dimension to the spatial coordinates relative to the temporal coordinate, or to spatial momenta relative to energy, which is equivalent to renormalizing the dynamical critical exponent $z$, leads to non-trivial anomalous dimensions for $\diff$ and $\Delta$ in spite of the fact that those terms do not get diagrammatic corrections. Furthermore, it is precisely this ambiguity that reduces the number of renormalization constraint equations by one and prevents the explicit determination of the exact values of the anomalous dimensions. In the end, we can only write the anomalous dimensions in terms of one undetermined parameter, which we take to be the dynamical critical exponent:
\begin{align}
\gamma_{\lambda} &= 1 - \frac{2}{z} - \frac{\Delta}{2 \pi \diff}, &
\gamma_{\diff} &= \gamma_{\Delta} = - \gamma_{\phi} = 2 \gamma_{\bx} = 2 - \frac{4}{z}.
\end{align}
Upon fixing renormalization conditions at some scale $\bar\mu$, the running of $\lambda$ is given in \eqref{eq:RGE} as
\begin{equation}
	\label{eq:RGE0}
	\lambda ( \mu ) = \biggl( \frac{\bar\mu}{\mu} \biggr)^{2 - \frac{2}{z} - \frac{\Delta}{2 \pi \diff}} \lambda ( \bar \mu ).
\end{equation}
This manifests the fact that the functional form of the cosine and sine derivative interactions are preserved to all orders under loop corrections, and therefore, the renormalized interactions are parameterized by a running $\lambda$. The former can be traced to the non-linear realization of a symmetry that mixes global shifts of $\psi$ compensated by a spatial rotation, along with its associated Ward-Takahashi identities as detailed in the next section. 

The net result is that $\lambda$ has a bare scaling dimension of 1 that is corrected to
\begin{equation} \label{eq:lambdaandim}
[ \lambda ] = 2 - \frac{2}{z} - \frac{\Delta}{2 \pi \diff}.
\end{equation}
Similarly, $\diff$ and $\Delta$ have vanishing classical scaling dimensions that are corrected to
\begin{equation}
	\label{eq:ad}
[ \diff ] = [ \Delta ] = 2 - \frac{4}{z}.
\end{equation}
Of course, this result implies that the ratio $\Delta / \diff$ must remain dimensionless even as $\Delta$ and $\diff$ gain anomalous dimensions. As we detail in the following section, the universality of the renormalization of $\lambda$ is a direct corollary of Ward-Takahashi identities. 

Given that our results are exact to all-orders, there is no restriction on the value of $\frac{\Delta}{\diff}$ other than positivity. Therefore, there is an entire line of fixed points parametrized by this ratio even at fixed $z$. Indeed, only the ratio $\frac{\Delta}{\diff}$ matters. This remains exactly marginal throughout renormalization and sets the scale of stochastic fluctuations via \eqref{eq:fs}. Compare this, for example, with the line of fixed points with a continuously varying dynamical critical exponent found in the eight-vertex model \cite{Isakov:2010he}. The line of fixed points corresponding to the $U(1)$-symmetric quantum Lifshitz models has fixed $z=2$, whereas $z$ varies continuously along the line of fixed points with only discrete Ising $\mathbb{Z}_2$ symmetry. The former falls under Model B (conserved order parameter) and the latter under Model A (no conservation laws) of the Halperin-Hohenberg classification of dynamic critical phenomena \cite{Hohenberg:1977ym}. 


\section{Symmetries and Ward-Takahashi Identities}
\label{sec:symm}

That the recursion relations \eqref{eq:recursion} for the vertex factors are preserved under renormalization turns out to be a consequence of a particular global symmetry and its associated Ward-Takahashi identities\footnote{Henceforth referred to simply as Ward identities, even though the truncated designation is typically used to refer to application of the Ward-Takahashi identities to transition matrix elements in the context of quantum field theory.}. The symmetry in question is a subset of a larger set of invariances (whose elaboration will immediately follow), and whose transformations are given by
\begin{align} \label{eq:transfs}
    \delta_\theta \auxnew &= - i \theta L \auxnew, &%
    \delta_\theta \field &= \theta ( 1- i L \field ),
\end{align}
where $L$ is the angular momentum operator, which is the generator of rotations,
\begin{equation}
L = - i ( x \partial_y - y \partial_x ).
\end{equation}
When computed not with respect to the Lagrangian in the form \eqref{eq:origlag}, but the one given by integration by parts:
\begin{equation}
\calL = - \field \scrB - \Delta \auxnew^2,
\end{equation}
where $\scrB$ is the $\field$ equation of motion given in \eqref{eq:AB}, the Noether current associated with the above symmetry is
\begin{subequations}
\begin{align}
J^t &= i \field L \auxnew + \auxnew, \\
\bJ &= ( \lambda \hat{\bn} \auxnew + D \nabla \auxnew ) ( 1 - i L \field ) - D \auxnew \nabla ( -i L \field ) + \Delta \auxnew^2 \binom{-y}{x} + D \nabla \auxnew.
\end{align}
\end{subequations}
The divergence of this current is
\begin{equation}
\partial_{\mu} J^{\mu} = i \scrA L \auxnew + \scrB ( 1 - i L \field ),
\end{equation}
which vanishes on-shell since $\scrA = 0$ and $\scrB = 0$ are the equations of motion.

For concreteness, let us consider the Ward identity applied to the four-point vertex operator:
\begin{equation}
\calF = \int_{t, \bx} \auxnew (t, \bx ) \, \field (t, \bx )^3.
\end{equation}
for which the corresponding Ward-Takahashi identity reads
\begin{equation}
\expect{\delta_\theta \calF} = \theta \int_{t' , \bx'} \expect{\calF \, \partial_{\mu} J^{\mu} (t', \bx' )}.
\end{equation}
As detailed in Appendix \ref{sec:feynrules}, the expectation value of any operator with two external $\auxnew$ fields vanishes. Since $\calF$ contains one factor of $\auxnew$ and every term in $\partial_{\mu} J^{\mu}$ contains one factor of $\auxnew$, it follows that the right hand side of the Ward identity vanishes. Hence,
\begin{equation}
\expect{\delta_\theta \calF} = 0.
\end{equation}
Now, in momentum space, the transformations \eqref{eq:transfs} read\footnote{Where in keeping with the conventions of Appendix \ref{sec:feynrules}, $\deltabar(\omega) := \delta(\omega)/(2\pi)$, and $\deltabar^{\,(d)}(\textbf{k}) := \delta(\textbf{k})/(2\pi)^d$.}
\begin{align}
\delta_\theta \auxnew_{\omega, \bk} &= -i \theta \wt{L} \auxnew_{\omega, \bk}, &%
\delta_\theta \field_{\omega, \bk} &= -i \theta \wt{L} \field_{\omega, \bk} + \theta \, \deltabar ( \omega ) \deltabar^{\,(d)} ( \bk ),
\end{align}
where
\begin{equation}
\wt{L} =  -i \biggl( k_y \frac{\partial}{\partial k_x} - k_x \frac{\partial}{\partial k_y} \biggr).
\end{equation}
Meanwhile, $\calF$ reads
\begin{equation}
\calF = \biggl( \prod_{i=1}^{3} \int_{\omega_i, \bk_i} \field_{\omega_i , \bk_i} \biggr) \int_{\omega_4, \bk_4} \auxnew_{\omega_4, \bk_4} \, \deltabar \biggl( \sum_{i=1}^{4} \omega_i \biggr) \deltabar^{\,(d)} \biggl( \sum_{i=1}^{4} \bk_i \biggr).
\end{equation}
Therefore,
\begin{align}
\label{eq:wir}
\delta_\theta \calF &= \theta \biggl( \prod_{i=1}^{4} \int_{\omega_i , \bk_i} \biggr) \biggl( \sum_{i=1}^{4} (-i \wt{L}_i ) \biggr) \biggl( \prod_{i=1}^{3} \field_{\omega_i , \bk_i} \biggr) \auxnew_{\omega_4, \bk_4} \, \deltabar \biggl( \sum_{i=1}^{4} \omega_i \biggr) \deltabar^{\,(d)} \biggl( \sum_{i=1}^{4} \bk_i \biggr) \notag \\
&\quad + \theta \sum_{i=1}^{3} \biggl( \prod_{j \neq i}^{3} \int_{\omega_j , \bk_j} \field_{\omega_j, \bk_j} \biggr) \int_{\omega_4, \bk_4} \auxnew_{\omega_4, \bk_4} \, \deltabar \biggl( \sum_{j \neq i}^{4} \omega_j \biggr) \deltabar^{\,(d)} \biggl( \sum_{j \neq i}^{4} \bk_j \biggr).
\end{align}
Once the expectation value is taken, the first line turns into the operator $-i \wt{L}$, where $\wt{L}$ is the total sum of the four $\wt{L}_i$'s, acting on the vertex $V_4$, while the second line turns into three copies of $V_3$. Since $V_4$ only depends on the momentum of the external $\field$ field being contracted with the internal $\auxnew$ field, the factor $\wt{L}_4$ does not contribute at all. Meanwhile, since $V_4$ depends only on $k_x$, the term $- k_x \frac{\partial}{\partial k_y}$ in $\wt{L}$ also does not contribute, only the term $k_y \frac{\partial}{\partial k_x}$ does. Thus, the Ward identity simplifies to
\begin{equation}
	\label{eq:vr}
V_3 + k_y \frac{\partial}{\partial k_x} V_4 = 0.
\end{equation}
This relation holds at all orders, and therefore for the fully dressed vertex factors. In other words, the relation \eqref{eq:recursion} for the bare vertices is satisfied at all loop orders. If we let $\calF$ be an odd-point function, then we will derive the other corresponding relation in \eqref{eq:recursion}. The Ward identity ensures that these recursion relations are preserved under renormalization.

The symmetry \eqref{eq:transfs} corresponds to a constant shift in the flock angle field $\psi$, along with a compensating rigid rotation of the underlying coordinates by the same angle. It seems obvious that this would have to be a symmetry of the action given that it relates indistinguishable physical configurations, provided the functional measure is also invariant under this transformation, as can be straightforwardly verified. 

The above transformation is supplemented with spacetime translation symmetries, whose associated Ward identities simply force correlation functions to depend on differences of space and time coordinates. We note for completeness that the Noether charge associated with time translation invariance does not correspond to a conserved energy within the MSRDJ formalism per se, rather, the conserved generator of the Hamiltonian flow underlying the saddle-point equations. On the other hand, spatial translation invariance does indeed imply the conservation of linear momentum in inter-flock interactions (even if momentum is not conserved in flock-reservoir interactions -- a generic feature of systems where boost invariance is kinematically absent \cite{deBoer:2017ing, Grosvenor:2024vcn}). Operationally, it is the Ward identities associated with the diagonal subgroup parameterized by $\theta$ that results in the renormalizability of the isotropic Malthusian flock model: The sine and cosine interaction potentials preserve their functional form under renormalization group improvement, up to an overall scale dependent rescaling that gets encoded into a running $\lambda$. This is a highly non-trivial result, which need not have followed directly from the known power counting renormalizability of all $z=2$ statistical field theories in $2+1$-dimensions with fewer than four derivatives, but is rather a consequence of a non-linearly realized symmetry that forces relations between correlation functions of different orders\footnote{That is, given that \eqref{eq:transfs} corresponds to transformations of the form $\delta_\theta\psi = \theta +$ (rotation), it immediately follows that $\delta_\theta \langle \psi(t_1,\textbf{x}_1)... \psi(t_n,\textbf{x}_n) \rangle$ returns a sum of $(n-1)$-pt correlation functions plus rotated $n$-point functions, which is the content of \eqref{eq:wir}, with the eventual corollary \eqref{eq:vr}.}.


\section{\label{sec:qlro}The (Undetermined) Fate of Quasi-Long-Range Order}

In this section, we address the nature and phase structure of quasi-long-range order in isotropic Mathusian flocks. We do so by directly examining the relevant limits of the renormalized propagators of the theory, after which we speculate on the physical principles behind our results given the symmetries and invariances discussed in the previous section.  

We begin by noting that at weak coupling, the existence and nature of long-range order can be inferred from the static, low-momentum limit of the renormalized $\field$-$\field$ propagator:
\begin{equation}
\calG_{2,0}^{\rm ren} := \calG_{\field\field}^{\rm ren} = \frac{2 \Delta}{( \omega - \lambda k_x - i \diff k^2 )( \omega - \lambda k_x + i \diff k^2 )},
\end{equation}
where all the parameters above now stand for their renormalized versions. Although two-point correlation functions no longer suffice in completely determining the nature of the long-range correlations between the collective excitations that will have condensed in any strongly coupled regime, they can still relay valuable information about the nature of this phase. Consequently, we consider the static limit of the above for all values of $\lambda$, by sending $\omega \to 0$:
\begin{equation} \label{eq:Gpsipsiren}
\calG_{\field\field}^{\rm ren} \xrightarrow{\omega = 0} \frac{2 \Delta}{\lambda^2 ( \bk \cdot \widehat{\bk}_b )^2 + \diff^2 k^4},
\end{equation}
where we have explicitly restored the implicit background field dependence in the propagator, with $\widehat{\bk}_b$ denoting the unit vector along the net flock direction, which we have assumed without loss of generality taken to point along the $x$-axis. Consider now the behavior of the above static \emph{structure factor} in the two cases: (1) as a function of $k_x$ when $k_y = 0$; and (2) as a function of $k_y$ when $k_x = 0$. In case (1), if the function has the same scaling dimension as $\frac{1}{k_{x}^{2}}$ in the limit that $k_x \rightarrow 0$, we can infer the existence of quasi-long-range order in the $x$-direction. This describes a two-point function in real space that decays exponentially in the $x$ direction with some characteristic correlation length $\xi_x$ in the limit in which $\xi_x \rightarrow \infty$. This follows from the fact that the Fourier transform of $e^{-x/ \xi_x}$ along $x$ is given by $\frac{2}{( 1/ \xi_x )^2 + k_{x}^{2}}$, which goes to $\frac{2}{k_{x}^{2}}$ in the limit $\xi_x \rightarrow \infty$. The same statement holds for case (2) with $k_x$ replaced with $k_y$.

More precisely, we are interested in the asymptotic behavior of the static real space correlation function either in the limit that
\eq{}{\lim_{|x|\to \infty}\langle \psi(x,y)\psi(0,0)\rangle \sim |x|^{-\eta_x} \mathcal{F}\left(\frac{y}{|x|}\right),}
where the ratio $y/|x|$ is kept fixed, or in the limit
\eq{}{\lim_{|y|\to \infty}\langle \psi(x,y)\psi(0,0)\rangle \sim |y|^{-\eta_y} \mathcal{K}\left(\frac{x}{|y|}\right)}
where the ratio $x/|y|$ is kept fixed. Quasi-long-range order along the $x$-axis requires $\eta_x > 0$, and similarly, $\eta_y > 0$ if it is to exist along the $y$-axis, which is the direction perpendicular to the flock in terms of our background field expansion. The translation of these statements into Fourier space imply that by taking the limit of the renormalized propagator 
\eq{}{\calG_{\field\field}^{\rm ren}(\omega,k_x,k_y)\big|_{\omega, k_y = 0} \sim \frac{1}{|k_x|^{2 - \eta_x}} \equiv \frac{1}{|k_x|^{Z_x}}, \qquad {\rm case~}(1)}
one has quasi-long-range order along the $x$-direction if the scaling exponent $Z_x < 2$. Similarly, one has quasi-long-range order along the $y$-direction if in the corresponding limit
\eq{}{\calG_{\field\field}^{\rm ren}(\omega,k_x,k_y)\big|_{\omega, k_x = 0} \sim \frac{1}{|k_y|^{2 - \eta_y}} \equiv \frac{1}{|k_x|^{Z_y}}, \qquad {\rm case~}(2)}
one also finds $Z_y < 2$. For case (1), one finds that along the $x$-direction that
\begin{equation}
\calG_{\field\field}^{\rm ren} \xrightarrow{\omega = k_y = 0} \frac{2 \Delta}{\lambda^2 k_{x}^{2}}.
\end{equation}
Stripping off the bare dimensions of the coefficient $\frac{2 \Delta}{\lambda^2}$ of $\frac{1}{k_{x}^{2}}$ just gives the anomalous dimension of this coefficient, namely $\gamma_{\Delta} - 2 \gamma_{\lambda} = \frac{\Delta}{2 \pi \kappa}$, where the anomalous dimensions are given in terms of $z$ in \eqref{eq:anomalousdims}. Factoring in the dimension of $\frac{1}{k_{x}^{2}}$ simply gives $- 2 [ k_x ] + \frac{\Delta}{2 \pi \kappa}$. Measuring this in units of $[k_x]$ gives
\begin{equation}
\frac{-2 [k_x] + \frac{\Delta}{2 \pi \kappa}}{[k_x]} = -2 + \frac{\frac{\Delta}{2 \pi \kappa}}{2/z} = -2 + \frac{z \Delta}{4 \pi \diff},
\end{equation}
so that,
\eq{}{Z_x = 2 - \frac{z \Delta}{4 \pi \diff}.}
Since $\Delta, \diff > 0$, the only condition needed for quasi-long-range order along the flocking direction is $z>0$, which is certainly met by any causal system, Malthusian flocks included. However, this is not a sufficient condition, since we furthermore observe that the dimension of $\lambda$ is $2 - \frac{\Delta}{4 \pi \diff} - \frac{2}{z}$, as given in Eq.~\eqref{eq:lambdadim}, and so there is a critical $z$ value
\begin{equation} \label{eq:zc}
z_c = \frac{1}{1 - \frac{\Delta}{8 \pi \diff}},
\end{equation}
where $\lambda$ is irrelevant if $z < z_c$. If $z$ is below this critical value, at long wavelengths, the system effectively has no drift term, only diffusion and noise. That system will therefore have no quasi-long-range order.

On the other hand, along the $y$-axis, one finds instead that
\begin{equation}
\calG_{\field\field}^{\rm ren} \xrightarrow{\omega = k_x = 0} \frac{2 \Delta}{\diff^2 k_{y}^{4}},
\end{equation}
which has scaling dimension $Z_y = 6 - \frac{4}{z}$, measured in units of $k_y$. There is quasi-long-range order perpendicular to the direction of the flock if $z<1$ and not if $z>1$. However, note that $z_c > 1$, and so if $z<1$, then quasi-long-range order is already lost along the direction of the flock, which invalidates the starting point altogether of a well-defined and stable flocking direction around which we can perturb. We conclude, therefore, that there is no quasi-long-range order in the direction perpendicular to the flock.

We can understand these results intuitively as follows: \textit{If the value of $z$ happens to be above the critical value}, quasi-long-range order should tend to be destroyed by increasing the noise $\Delta$ or decreasing diffusion $\diff$. Increasing noise will tend to randomize $\field$ from point to point while decreasing diffusion inhibits information flow or communication through the system. Both effects should tend to destroy quasi-long-range order. In the linearized theory, increasing $\Delta/\kappa$ enhances fluctuations but does not change their logarithmic scaling, so quasi–long-range order persists for any finite ratio. However, in the full nonlinear theory, large $\Delta/\kappa$ drives the system into a strong-coupling regime where nonlinear effects enhance fluctuations beyond logarithmic growth, thereby destroying the possibility of quasi–long-range order. When $\Delta/\kappa$ is below the critical value, RG flow to the IR implies finite values for $\lambda$ at finite separation, with quasi-long-range order along the flocking direction indicated though the IR asymptotics of the fully renormalized propagator.

\subsection{Soft limits and symmetry-protected gapless excitations}

Perhaps the most interesting aspect about the isotropic Malthusian flock, is that many of its properties follow directly from the fact that its fields non-linearly realize an exact diagonal rotation/shift symmetry, and inherits a large set of field space redundancies from the special form of its MSRDJ action. These properties include the fact that the advection terms corresponding to the sine and cosine derivative interactions preserve their functional form under RG improvement, and that the couplings $\lambda$, $\Delta$, and $\kappa$ are forced to have the same anomalous dimensions as a result of the Ward identities discussed in Section \ref{sec:symm}. Factoring certain additional subtleties that arise in the boost non-invariant context, one furthermore infers that the absence of diagrammatic corrections to a given coupling does not preclude them from acquiring an anomalous dimension, as discussed in Appendix \ref{sec:wilsonianRG}. It remains to elaborate on perhaps the most significant corollary of diagonal shift symmetry as far as the present study is concerned: The fact that the gapless excitations are protected by this symmetry. The gapless property is a necessary condition for correlation functions to have the requisite IR scaling for quasi-long-range order, since gapped excitations decay exponentially at long wavelengths.

We proceed by first establishing how the Ward identities satisfied by the effective action imply soft identities for its functional derivatives\footnote{Soft theorems relate $n$-point momentum space correlation functions where one of the external momenta is taken to be vanishing, with linear combinations of (possibly symmetry transformed) lower point correlators \cite{Cheung:2014dqa, Cheung:2016drk}.}, from which the existence of symmetry-protected gapless excitations follow as a corollary.  We first recall the fact that the MSRDJ action is invariant under the action of the symmetry defined in \eqref{eq:transfs}. Defining the generating functional
\begin{equation}
Z[J_\psi,J_\phi]
=
\int D\psi\,D\phi\,
\exp\left[
-S[\psi,\phi]
+\int d^3x\,
\left(J_\psi\psi+J_\phi\phi\right)
\right],
\end{equation}
where $J_\psi$ and $J_\phi$ are independent sources, and using the fact that the action and measure are invariant under the diagonal rotation/shift symmetry, one obtains the functional identity
\begin{equation}
\int d^3x\,
\left[
J_\psi\left(1-i L\frac{\delta W}{\delta J_\psi}\right)
-
i J_\phi L\frac{\delta W}{\delta J_\phi}
\right] = 0,
\label{eq:ward_W_diag}
\end{equation}
where $W=\ln Z$. Legenndre transforming to the effective action,
\begin{equation}
\Gamma[\bar\psi,\bar\phi]
=
W-\int d^3x\,
\left(J_\psi\bar\psi+J_\phi\bar\phi\right),
\end{equation}
where we've defined the background fields $\bar\psi=\delta W/\delta J_\psi$ and $\bar\phi= \delta W/\delta J_\phi$ and using the fact that $J_\psi=-\delta\Gamma/\delta\bar\psi$, and $J_\phi=-{\delta\Gamma}/{\delta\bar\phi}$, the Ward identity becomes
\begin{equation}
\int d^3x\,
\left[
\frac{\delta\Gamma}{\delta\bar\psi}
\left(1-iL\bar\psi\right)
-
\frac{\delta\Gamma}{\delta\bar\phi}
iL\bar\phi
\right]
=0 .
\label{eq:ward_Gamma_diag}
\end{equation}
This identity directly constrains the one-particle-irreducible vertices.
To extract the zero-momentum two-point vertex, differentiate
Eq.~\eqref{eq:ward_Gamma_diag} once with respect to $\bar\psi(y)$ and
then evaluate at a homogeneous background,
\begin{equation}
\bar\psi=\psi_0,\qquad \bar\phi=0 .
\end{equation}
Since $L_z\psi_0=0$, one obtains
\begin{equation}
\int d^3x\,
\Gamma^{(2)}_{\psi\psi}(x,y)=0 .
\end{equation}
By translation invariance,
\begin{equation}
\Gamma^{(2)}_{\psi\psi}(x,y)
=
\Gamma^{(2)}_{\psi\psi}(x-y),
\end{equation}
so that finally
\begin{equation}
\Gamma^{(2)}_{\psi\psi}(\omega=0,\mathbf q=\mathbf 0)=0 .
\label{eq:gapless_condition}
\end{equation}
The Ward identity therefore implies a soft theorem associated with the generator of the exact diagonal rotational symmetry, which is spontaneously broken in the ordered phase. In particular, the inverse propagator vanishes at zero frequency and momentum $\Gamma_{\psi \psi}^{(2)}(0,0)=0$, guaranteeing a gapless flock angle mode. Equivalently, any local potential or mass term would violate Eq.~\eqref{eq:ward_Gamma_diag}, because the constant part of the transformation shifts the homogeneous mode of $\psi$. The effective action may therefore depend on $\psi$ only through symmetry-compatible derivative and rotationally locked structures, as is implicit in the form of the action of the isotropic Malthusian flock. Consequently, $\psi$ remains gapless to all orders in perturbation theory\footnote{This protection ultimately relies only on the exact diagonal rotation/shift symmetry of the interacting action. It did not require an Adler-zero soft theorem associated with the incorrectly identified Galileon invariance in a previous version of our paper.}.


\subsection{\label{sec:BKT} Similarities and differences with sine-Gordon and dual XY models}

The renormalization group running described by \eqref{eq:RGE0} implies that the IR behavior of the isotropic Malthusian flock is determined by a marginal instability across a line of fixed points, rather than by an isolated interacting fixed point of the Wilson--Fisher type. That is, given that $\lambda$ runs as
\begin{equation}
	\nn
	\lambda ( \mu ) = \biggl( \frac{\bar\mu}{\mu} \biggr)^{2- \frac{2}{z} - \frac{\Delta}{4 \pi \diff}} \lambda ( \bar \mu ),
\end{equation}
we see that the infrared scaling dimension of the nonlinear coupling flips sign at $z=z_c$ given in Eq.~\eqref{eq:zc}. Since $\mu$ is an energy scale, the infrared limit corresponds to $\mu\rightarrow 0$, such that for $z > z_c$ the nonlinear coupling grows under coarse graining and is therefore infrared relevant for any finite separation, possibly crossing into a strong coupling regime. Conversely, for $z<z_c$, the coupling flows to zero and is thus irrelevant in the IR. The coupling is exactly marginal at the critical value \eqref{eq:zc}, and the resulting phase diagram is therefore characterized by a line of fixed points with a marginal instability separating two distinct regimes, with a disordered phase on one side, and a potentially strongly coupled gapless phase on the other, where moreover, quasi-long-range order persists. 

In order to better understand the nature of the transition, it is useful to rewrite the derivatively coupled advection potential \eqref{eq:Lint} in terms of so-called \textit{vertex operators}. That is, we rewrite
\begin{equation}\nn
	\mathcal{L}_{\rm int}
	=
	\lambda\phi
	\left[
	(\cos\psi-1)\partial_x\psi
	+
	(\sin\psi)\partial_y\psi
	\right],
\end{equation}
as
\eq{eq:vertex_form}{\mathcal{L}_{\rm int} = \frac{\lambda}{2}\phi \Big[\left(e^{i\psi}+e^{-i\psi}-2 	\right)\partial_x\psi -i \left(e^{i\psi} - e^{-i\psi} \right)\partial_y\psi \Big],}
so that the nonlinear interactions can be constructed as a combination of vertex operators
\begin{equation} 
	V_\pm = e^{\pm i\psi},
\end{equation}
with derivative self couplings\footnote{Or equivalently, derivatively coupled to the response field upon sufficient integrations by parts.}, and whose infrared scaling properties determine the fate of the theory. This structure bears a resemblance to the role played by vertex operators in the Coulomb-gas and sine-Gordon descriptions of the Berezinskii--Kosterlitz--Thouless (BKT) transition \cite{Berezinskii1971, Berezinskii1972, KosterlitzThouless1973, Kosterlitz1974, Kosterlitz1977, JoseKadanoffKirkpatrickNelson1977, AmitGoldschmidtGrinstein1980}, where the infrared relevance of the vertex operators determines whether vortex fugacity grows or decreases under coarse graining. Furthermore, the leading-order renormalization-group structure found in \cite{sezik2026} for the isotropic Malthusian flock near the special point $\Delta / \kappa = 4 \pi$ is structurally identical to that of sine-Gordon. This resemblance, however, should not be equated with the XY/sine-Gordon duality itself. In the conventional XY model, the BKT interpretation follows from an explicit weak–strong duality that relates the compact spin-wave theory to a gas of vortices and, in turn, to a sine-Gordon theory. The crucial question for the Malthusian flock is whether an analogous chain of dual descriptions exists.

In the XY-model case, the action is simply quadratic in the gradient of the XY angle field $\vartheta$. This allows for the change of variables from $\vartheta$ to $\mathbf{u} = \nabla \vartheta$. The latter is then expanded in terms of a smooth spin-wave piece $\mathbf{u}_{\rm sw}$ and a singular vortex piece $\mathbf{u}_{\rm v}$. The identification between the vortex density $\rho_{\rm v}$ and the vortex curl $\frac{1}{2 \pi} \epsilon^{ij} \partial_i \mathbf{u}_j$ is imposed by a Lagrange multiplier field $\psi$ and the original XY field $\mathbf{u}$ is integrated out. This latter step can be done explicitly because the XY action is quadratic in $\mathbf{u}$. At this point, one can integrate out the dual field $\psi$ to get the theory of a Coulomb gas of vortices or one can sum over vortex/anti-vortex ensembles. Keeping only the the lowest-order ensembles, which contain, for example, no vortices at all, one vortex, or one anti-vortex, generates the sine-Gordon theory for $\psi$. The crucial step is that of integrating out $\mathbf{u}$. The analog of this step cannot be completed in closed form for the Malthusian flock because the non-linear drift term proportional to $\lambda$ in the action is neither a function purely of the derivative of $\psi$ nor quadratic in $\psi$.

If we instead wish to interpret the Malthusian flock action as though it were the dual to a putative theory whose phase structure is supposed to be BKT-like, then we also encounter a roadblock. The sine-Gordon interaction term $\cos \psi$ simply arises from a sum of $e^{i q \psi}$ with $q = \pm 1$ and $q$ representing the vortex charge. In the Malthusian flock case, the vertices are $e^{\pm i \psi}$ dressed with $\pm \partial_{\pm} \phi$, where $\partial_{\pm} = \partial_x \pm i \partial_y$. Clearly, these coefficients are not simply a vortex fugacity. Instead, each vertex comes with an additional response insertion. It is not at all clear if or how this could arise from integrating out the fields of some putative dual field theory analogous to the XY-model. Since the BKT interpretation of the RG structure of sine-Gordon involves the phases of vortices in the dual XY-model, without an explicit duality, we cannot establish a similar phase interpretation of the RG structure of the isotropic Malthusian flock.

Another way to highlight the distinction between the isotropic Malthusian flock and sine-Gordon theory is that, unlike the conventional BKT transition, the instability in our case emerges from an anisotropic soft mode rather than a logarithmic Coulomb-gas sector. In both cases, the infrared behavior is controlled by a line of fixed points. However, the analogy should be understood at the level of universality classes rather than microphysics -- unlike the conventional BKT setup, isotropic Malthusian flocks are governed by an anisotropic soft mode, whose equal-time propagator takes the form \eqref{eq:Gpsipsiren}. Consequently, the infrared fluctuations are not generated by logarithmically scaling Green's functions that underlie the Coulomb-gas and sine-Gordon descriptions of the BKT transition. Rather, the transition emerges from the interplay between anisotropic soft fluctuations and symmetry-protected vertex nonlinearities in a genuinely $2+1$-dimensional stochastic field theory. 

A further distinction from the conventional BKT scenario arises from the nonlinearly realized symmetry of the MSRDJ action \eqref{eq:transfs}, which combines shifts of the flock angle field with time-dependent spatial rotations. This symmetry forbids the generation of non-derivatively coupled terms capable of pinning the soft mode. Consequently, although the vertex operators become infrared relevant, and drive the theory into a symmetry-protected and potentially strongly coupled regime for $z > z_c$, their interactions cannot generate a local potential for $\psi$ or destroy the gapless excitation. The strong-coupling phase therefore remains gapless despite the onset of nonperturbative infrared dynamics. Whether there exists a non-trivial Lifshitz scaling analog of the dual Coulomb-gas or sine-Gordon picture for the isotropic Malthusian flock presents itself as an important open question to us, and warrants further investigation.


\section{Concluding remarks and future directions}

In our attempt to make sense of the nature of quasi-long-range order in active flocks, we inadvertently did what anyone who naturally thinks along the lines of a particle physicist would have done --- proceeded to work in the simplest, most symmetric case to get our bearings, regardless of whether it represented a phenomenologically interesting corner of parameter space. By setting up the renormalization of the Malthusian flock within the MSRDJ formalism in the limit of isotropic diffusion, we chanced upon a theory with a remarkable set of invariances that facilitated its statistical field-theoretic renormalization to all orders. In doing so, we set up a theoretical laboratory within which we could meaningfully compare and contrast various claims and counterclaims in the literature (with a technical summary provided in Appendix F), and arrived at an understanding as to how certain assumptions made in the literature may not always withstand the scrutiny of a complete diagrammatic or functional analysis.

A central lesson of our analysis is that the RG calculation itself only determines the counterterms required to cancel the divergences generated by integrating out short-distance modes, together with the subsequent rescaling of dimensionful quantities needed to restore the theory to its original form. This procedure necessarily involves a convention for which quantities are held fixed under the RG transformation. In the present formulation, we have argued that the dimension of time should be kept fixed to properly implement an energy-based separation of scales. Thus, a counterterm multiplying the operator $\phi \, \partial_t \psi$ must be absorbed into the renormalization of $\phi$ rather than into that of $\psi$, since $\psi$ is a compact angular variable\footnote{A non-trivial renormalization of a compact angular field would result in the domain acquiring a conical singularity.}. Once the renormalization of $\phi$ is fixed in this way, a counterterm multiplying the spatial diffusion operator $-\kappa\phi\nabla^2\psi$ determines the scaling of the dispersion relation, $\omega \sim \kappa k^{2+\gamma}$ and hence the dynamical critical exponent $z=2+\gamma$. How the exponent $\gamma$ is distributed between anomalous dimensions assigned to $\kappa$ and to the spatial coordinates is ultimately a matter of convention; the physically relevant quantity for the dynamics is the resulting value of $z$. This is the origin of the remaining freedom in our determination of the anomalous dimensions: in the absence of an additional symmetry or scaling relation that fixes the dynamical exponent, our all-orders relations determine the RG flow only up to this one undetermined parameter.

The resulting RG structure is nevertheless highly constrained. In the isotropic Malthusian theory, the ratio $\Delta/\kappa$ remains exactly dimensionless, while the Ward identities associated with the diagonal rotation/shift symmetry preserve the trigonometric structure of the convective derivative to all orders. The nonlinear interaction is therefore parameterized by the running coupling $\lambda$, and the latter exhibits a marginal instability at
\begin{equation}
z_c = \frac{1}{1-\frac{\Delta}{8\pi\kappa}}.
\end{equation}
Away from the special renormalon point $\Delta/\kappa=4\pi$, the perturbative flow separates regions in which $\lambda$ is infrared irrelevant from regions in which it grows under coarse graining and eventually takes the theory beyond perturbative control. This establishes a precise statement about the perturbative RG structure of the isotropic Malthusian flock. It does not, by itself, determine the full infrared physics once the flow reaches strong coupling.

This distinction is particularly important when comparing the resulting RG flow of the Malthusian flock to the sine-Gordon and BKT descriptions. Near the special point $\Delta/\kappa=4\pi$, the beta functions exhibit a structure formally analogous to that of sine-Gordon theory near its corresponding special point. Moreover, the perturbative RG flow naturally separates regimes in which $\lambda$ flows to zero from those in which it grows. However, in the conventional sine-Gordon/XY correspondence, the identification of the sine-Gordon coupling with vortex fugacity, together with the weak--strong duality between the two descriptions, is what permits one to interpret the RG flow in terms of vortex-antivortex binding and unbinding and hence to draw conclusions about the strongly coupled phase. No analogous dual description is presently known for the Malthusian flock. Consequently, the RG calculation establishes the change in relevance of the nonlinear coupling, but does not by itself establish that this change corresponds to vortex proliferation, nor does it suffice to determine the nature of the strongly-coupled phase.

The same point can be appreciated by comparison with the KPZ problem. Both the KPZ model and the present Malthusian flock possess a Gaussian regime with $z=2$ scaling in the ultraviolet, and in the regime $\Delta/\kappa<4\pi$ the nonlinear interaction is relevant and drives the theory towards strong coupling. In KPZ, however, Galilean invariance\footnote{\label{fn:KPZ}In the context of boost-invariant systems, or in systems where boost-invariance is spontaneously (as opposed to kinematically) broken, this can also be viewed as a \textit{Galileon} symmetry for some field $\pi$, which corresponds to a spacetime-dependent field-space shift symmetry of the form $\pi \to \pi + c + b_\mu x^\mu$ \cite{Nicolis:2008in, Nicolis:2015sra}). When boosts are kinematically broken, as is the case in systems where momentum can be exchanged with the reservoir or bath \cite{deBoer:2017ing, Grosvenor:2024vcn}, one must separately categorize time and spatial translations. In this sense the KPZ model can be viewed as exhibiting a spatial generalized Galileon invariance, where the height field $h$ transforms as $h \to h + \bv\cdot \bx$ along with a compensating spatial boost $\bx \to \bx + \lambda \bv t$ \cite{kardar1986dynamic, tauber2014critical}. Many special properties of the KPZ model, which include enhanced soft limits follow from this symmetry.} supplies an additional exact relation, $[\lambda]=\chi+z-2$, between the dimension of the non-linear interaction coupling $\lambda$, the roughness exponent $\chi$ and the dynamical critical exponent $z$. Together with the known value $\chi = \frac{1}{2}$ of the roughness exponent in $(1+1)$ dimensions, this determines the dynamical critical exponent at the strongly-coupled fixed point to be $z= \frac{3}{2}$. We have found no analogous symmetry or Ward identity in the Malthusian flock. In particular, the diagonal rotation/shift symmetry that protects the trigonometric interaction structure does not supply the missing relation required to determine $z$. We therefore cannot determine the value of $z$ at the putative strongly-coupled infrared fixed point from the present analysis.

It is therefore useful to distinguish carefully between what is established within the controlled perturbative regime and what remains a conjecture about the infrared completion. The Ward identities guarantee that the flock-angle excitation remains gapless and forbid the generation of a local potential that would explicitly pin $\psi$. The perturbative RG likewise establishes a marginal vertex instability separating a regime in which the nonlinear coupling is irrelevant from one in which it becomes relevant. What happens after the latter flow enters strong coupling, however, requires information not contained in the perturbative calculation alone. In particular, while the structure of the flow is reminiscent of BKT physics, the absence of a known weak--strong duality or Coulomb-gas representation prevents us from identifying the strong-coupling regime with a conventional vortex-unbound phase, or from establishing a BKT transition in the usual sense. Whether a non-trivial dual description adapted to the anisotropic soft mode of the Malthusian flock exists remains an important open question.

There are several immediate directions in which the present analysis should be extended. Foremost, the isotropic limit studied here is highly restrictive. It behooves us to study the perturbative renormalization of the anisotropic deformation of the Malthusian flock, in which additional derivatively coupled operators appear. Although the diagonal shift/rotation symmetry survives in the anisotropic theory, its realization is accompanied by additional nonlinear interaction terms and therefore by additional running couplings. Determining whether the structural simplifications and Ward identities identified here continue to constrain the theory sufficiently to permit an all-orders treatment would provide an important test of the robustness of our conclusions.

A second natural extension is to restore density fluctuations as hydrodynamic degrees of freedom and thereby move beyond the Malthusian limit to the theory of immortal flocks. The additional hydrodynamic fields introduce new operators and couplings and substantially enlarge the space of possible RG flows. Our present analysis therefore makes no claim concerning the critical behavior of number-conserving flocks, and considerably more work is required before the methods developed here can be applied to that setting.

Finally, the relation between the present theory and more familiar descriptions of topological defects remains to be clarified. The appearance of vertex operators and a marginal instability is suggestive of the Coulomb-gas and sine-Gordon structures familiar from equilibrium statistical mechanics, but the anisotropic soft mode, the genuinely $(2+1)$-dimensional stochastic dynamics, and the derivative coupling of the vertex operators distinguish the present problem from the conventional BKT setting. Establishing whether an appropriate dual description exists --- perhaps one in which the relevant topological excitations and their fugacity can be identified directly --- would be especially valuable. Such a construction could determine whether the strong-coupling regime encountered here admits a phase interpretation analogous to that of the XY model, or whether the similarity to BKT flows is instead limited to the perturbative RG structure.

Thus, while the present work does not settle the broader question of the fate of long-range or quasi-long-range order in active flocks, it does provide an all-orders field-theoretic framework within which that question can be posed more sharply. The isotropic Malthusian flock supplies a particularly tractable setting in which symmetry, compactness, stochastic dynamics, and the freedom to renormalize spatial relative to temporal scales can be disentangled explicitly. The resulting Ward identities and renormalization structure constrain the theory considerably, while the remaining freedom in the dynamical critical exponent and the absence of a known weak--strong dual description identify precisely what additional input is needed to determine its ultimate infrared behavior. Much work remains to be done before these results can be extended to more general and phenomenologically relevant models of active flocks.


\section{Concluding remarks and future directions}

In our attempt to make sense of the nature of quasi-long-range order in active flocks, we inadvertently did what anyone who naturally thinks along the lines of a particle physicist would have done -- proceeded to work in the simplest, most symmetric case to get our bearings, regardless of whether it represented a phenomenologically interesting corner of parameter space. By setting up the renormalization of the Malthusian flock within the MSRDJ formalism in the limit of isotropic diffusion, we chanced upon a theory with a remarkable set of invariances that facilitated its statistical field theoretic renormalization to all orders. In doing so, we set up a theoretical laboratory within which we could meaningfully compare and contrast various claims and counterclaims in the literature (with a technical summary provided in Appendix \ref{sec:comp}), and arrived at an understanding as to how certain assumptions made in the literature may not always withstand the scrutiny of a complete diagrammatic or functional analysis.

We uncovered a rich phase structure, controlled by the dynamical critical exponent $z$ and the ratio of the variance of the noise $\Delta$ to the isotropic diffusion coefficient $\kappa$. The theory exhibits a continuous line of fixed points together with a marginal vertex instability at the critical value $z = z_c = \frac{1}{1 - \frac{\Delta}{8 \pi \diff}}$. This instability marks the boundary between a Gaussian phase and a strongly interacting symmetry-protected gapless phase, thereby realizing a form of nonequilibrium criticality distinct from conventional Wilson–Fisher universality. Gapless excitations persist on both sides of the transition,  implying the persistence of quasi-long-range order for $z > z_c$. 
  
We must of course, caution against extrapolating too far from the conclusions presented here. Foremost, the debate on the nature of long-range order in active flocks more broadly is far from settled by our work. It behooves us to study the perturbative renormalization of the anisotropic deformation of the Malthusian flock, in which numerous additional derivatively coupled operators feature, to say nothing of the extension to immortal flocks. Much work remains to be done if the framework presented in this paper is to extend to more interesting models of active flocks. 

\acknowledgements

We wish to thank Sriram Ramaswamy for suggesting the problem to us as well as valuable discussions, for which we also wish to thank Ananyo Maitra, Stefano Gagliani, Silke Henkes, Niels Obers, Wim van Saarloos, Koenraad Schalm, and Emir Sezik. KG wishes to thank the Delta-ITP for generous support, and the Lorentz Institute at Leiden University for hospitality over the period this work was initiated. KG and SP also wish to thank the organizers of the STATPHYS29 satellite meeting on Recent Advances in Active Matter and Statistical Physics held in Leiden in July of 2025 for the opportunity to present earlier work, and for the opportunity to interact and discuss with many members of the vibrant soft and active matter research community.
\appendix


\section{Loop Corrections in the MSRDJ Formalism}
\label{sec:feynrules}

In the various subsections of this appendix, we detail the computation of loop corrections to the parameters and fields of Malthusian flocking theory within the MSRDJ formulation. We start by elaborating on the Feynman rules by first identifying the propagators and vertices in the background field method. We then perform a power counting analysis of the perturbative expansion, and immediately infer various non-renormalization theorems that follow directly from the kinematics of the theory alone. We then proceed to compute vertex and propagator corrections, and elaborate on the one-loop corrections to the latter, where one immediately encounters diagrams that diverge. Upon summation over all diagrams contributing to the renormalization of a particular process, we find cancellations that reduce the superficial degree of divergence down to logarithmic divergences for the couplings and anomalous dimensions of the theory. We then extend our analysis to all loop orders, facilitated by the non-linear realization of the symmetries and field redundancies discussed in Section \ref{sec:symm} and their associated Ward identities. In the final section of this appendix, we re-frame our results within the framework of Wilsonian renormalization.

We begin by noting that the free Lagrangian \eqref{eq:Lfree} can be expressed as
\begin{equation}
\calL_{\rm free} = \frac{1}{2} \Field^{\intercal} M \Field,
\end{equation}
where the field vector $\Psi$ and operator matrix $M$ are given by
\begin{align}
\Field &= \binom{\field}{\auxnew}, &%
M &= \begin{pmatrix}
0 & \calO \\
\wt{\calO} & -2 \Delta
\end{pmatrix},
\end{align}
and where the differential operators $\calO$ and $\wt{\calO}$ are defined as
\begin{align}
\calO &\equiv - ( \partial_t + \lambda \partial_x + \diff \nabla^2 ), &%
\wt{\calO} &\equiv \partial_t + \lambda \partial_x - \diff \nabla^2.
\end{align}
We Fourier transform the fields as
\begin{equation}
\Field (t , \bx ) = \int_{\omega , \bk} e^{-i ( \omega t - \bk \cdot \bx )} \, \Field_{\omega , \bk},
\end{equation}
where $\int_{\omega, \bk} := \int \dbar \omega \, \dbar^d k$ and $\dbar \omega = \frac{d \omega}{2 \pi}$ and $\dbar^d k = \frac{d^d k}{(2 \pi )^d}$. The free action can thus be written as:
\begin{align}
S_{\rm free} = \frac{1}{2} \int_{t, \bx, \nu, \bq, \omega, \bk} e^{-i [ ( \omega + \nu ) t - ( \bk + \bq ) \cdot \bx ]} \Field_{\nu, \bq}^{\intercal} 
\begin{pmatrix}
0 & \calO_{\omega, \bk} \\
\wt{\calO}_{\omega, \bk} & 2 \Delta
\end{pmatrix}
\Field_{\omega, \bk},
\end{align}
where
\begin{align}
\calO_{\omega, \bk} &= i ( \omega - \lambda k_x - i \diff k^2 ), &%
\wt{\calO}_{\omega, \bk} &= - i ( \omega - \lambda k_x + i \diff k^2 ).
\end{align}
We note that
\begin{align}
\calO_{- \omega , - \bk} &= \wt{\calO}_{\omega, \bk}, &%
\wt{\calO}_{- \omega, - \bk} &= \calO_{\omega, \bk}.
\end{align}
Implicit in the above and in what follows is the fact that we have invoked the background field method:
\eq{eq:bg}{\Psi(\textbf{x},t) = \psi_0 + \psi(\textbf{x},t),~~\Phi(\textbf{x},t) = \phi_0 + \phi(\textbf{x},t),} 
where $\psi_0$ and $\phi_0$ correspond to stable saddle points of the MSRDJ, which we take to be $\psi_0 = \phi_0 \equiv 0$. The latter choice singles out the $x$-direction as the net direction of the flock, around which we calculate the scale-dependent effects of statistical fluctuations, having taken the flocking transition for granted\footnote{See e.g. \cite{Armas:2024iuy} for a critical discussion on the conditions under which this can be presumed.}.

The spacetime integral in $S_{\rm free}$ sets $\nu = - \omega$ and $\bq = - \bk$ as delta function constraints, so that:
\begin{equation}
S_{\rm free} = \frac{1}{2} \int_{\omega, \bk} \Psi_{- \omega , - \bk}^{\intercal} M_{\omega, \bk}^{\phantom{\intercal}} \Psi_{\omega, \bk}^{\phantom{\intercal}},
\end{equation}
where
\begin{equation}
M_{\omega , \bk} = 
\begin{pmatrix}
0 & \calO_{\omega, \bk} \\
\wt{\calO}_{\omega, \bk} & -2 \Delta
\end{pmatrix}.
\end{equation}
We note that
\begin{align} \label{eq:PT}
M_{- \omega, - \bk}^{\phantom{\intercal}} &= M_{\omega, \bk}^{\intercal}, &%
( M_{\omega, \bk} )^{\dagger} &= M_{\omega , \bk}.
\end{align}
The inverse of $M$ defines the matrix of propagators:
\begin{equation}
\calG_{\omega, \bk} = M_{\omega, \bk}^{-1} = \frac{1}{- \calO_{\omega, \bk} \wt{\calO}_{\omega, \bk}} 
\begin{pmatrix}
- 2 \Delta & - \calO_{\omega, \bk} \\
- \wt{\calO}_{\omega, \bk} & 0
\end{pmatrix}
=
\begin{pmatrix}
\frac{2 \Delta}{( \omega - \lambda k_x - i \diff k^2 ) ( \omega - \lambda k_x + i \diff k^2 )} & \frac{i}{\omega - \lambda k_x + i \diff k^2} \\
\frac{-i}{\omega - \lambda k_x - i \diff k^2} & 0
\end{pmatrix}.
\end{equation}
The matrix operator $\calG_{\omega, \bk}$ inherits the property \eqref{eq:PT},
\begin{align}
\calG_{- \omega, - \bk} &= \calG_{\omega, \bk}^{\intercal}, &%
( \calG_{\omega , \bk} )^{\dagger} &= \calG_{\omega , \bk},
\end{align}
from $M$, its inverse. Thus, the free propagator, denoted with a 0 subscript, is given by the transpose:
\begin{align}
\expect{\Field_{- \omega, - \bk}^{\phantom{\intercal}} \Field_{\omega, \bk}^{\intercal}}_0 &= \calG_{- \omega, -\bk} \notag \\
&= \calG_{\omega, \bk}^{\intercal} \notag \\
&= 
\begin{pmatrix}
\frac{2 \Delta}{( \omega - \lambda k_x - i \diff k^2 ) ( \omega - \lambda k_x + i \diff k^2 )} & \frac{-i}{\omega - \lambda k_x - i \diff k^2} \\
\frac{i}{\omega - \lambda k_x + i \diff k^2} & 0
\end{pmatrix}.
\end{align}
Diagrammatrically, we represent the individual propagators as
\begin{subequations}
\begin{align}
\begin{tikzpicture}
\draw[thick,>=Latex,->] (-1,0) -- (0,0);
\draw[thick] (0,0) -- (1,0);
\end{tikzpicture} &\equiv \expect{\field_{- \omega , \bk} \, \field_{\omega, \bk}}_0 = \frac{2 \Delta}{( \omega - \lambda k_x - i \diff k^2 ) ( \omega - \lambda k_x + i \diff k^2 )}, \label{eq:psipsi} \\
\begin{tikzpicture}
\draw[thick,>=Latex,->] (-1,0) -- (0,0);
\draw[thick,dashed] (0,0) -- (1,0);
\end{tikzpicture} &\equiv \expect{\auxnew_{- \omega, - \bk} \, \field_{\omega, \bk}}_0 = \frac{i}{\omega - \lambda k_x + i \diff k^2}, \label{eq:psichi} \\
\begin{tikzpicture}
\draw[thick,dashed,>=Latex,->] (-1,0) -- (0,0);
\draw[thick] (0,0) -- (1,0);
\end{tikzpicture} &\equiv \expect{\field_{- \omega, - \bk} \, \auxnew_{\omega, \bk}}_{0} = \frac{-i}{\omega - \lambda k_x - i \diff k^2}. \label{eq:chipsi}
\end{align}
\end{subequations}
In particular, we note the usual absence of a free $\auxnew$-$\auxnew$ response field propagator \cite{heliasdahmen, tauber2014critical}.



From \eqref{eq:Lint}, one can read off the $(2n+1)$- and $(2n+2)$-point interaction Lagrangians, where $n = 1, 2, 3, \ldots$, which read:
\begin{align} \label{eq:Lints}
\calL_{\rm int}^{(2n+1)} &= - \frac{(-1)^n}{(2n-1)!} \lambda \auxnew \field^{2n-1} \partial_y \field, &%
\calL_{\rm int}^{(2n+2)} &= \frac{(-1)^n}{(2n)!} \lambda \auxnew \field^{2n} \partial_x \field.
\end{align}
As usual, the factorials drop out of the final vertex factor due to the combinatorics of contracting external and internal fields, and with propagators attached at the vertex being amputated along with an overall momentum-conserving delta function. Thus, the vertex factors are
\begin{align} \label{eq:vertices}
\vcenter{
\hbox{
\begin{tikzpicture}
\draw[thick,dashed,>=Latex,>-] (-1,0) -- (0,0);
\node[left] at (-1,0) {$\omega, \bk$}; 
\draw[thick,>=Latex,-<] (0,0) -- (45:1cm);
\draw[thick,>=Latex,-<] (0,0) -- (30:1cm);
\draw[thick,>=Latex,-<] (0,0) -- (-30:1cm);
\draw[thick,>=Latex,-<] (0,0) -- (-45:1cm);
\draw[thick,dotted]  (-15:0.8) arc(-15:15:0.8) node[midway,right]{$\scriptstyle 2n$};
\end{tikzpicture}
}
}
&= (-1)^{n} i \lambda k_y, &%
\vcenter{
\hbox{
\begin{tikzpicture}
\draw[thick,dashed,>=Latex,>-] (-1,0) -- (0,0);
\node[left] at (-1,0) {$\omega, \bk$}; 
\draw[thick,>=Latex,-<] (0,0) -- (45:1cm);
\draw[thick,>=Latex,-<] (0,0) -- (22.5:1cm);
\draw[thick,>=Latex,-<] (0,0) -- (-45:1cm);
\draw[thick,dotted]  (-30:0.8) arc(-30:10:0.8) node[midway,right]{$\scriptstyle 2n+1$};
\end{tikzpicture}
}
}
&= - (-1)^{n} i \lambda k_x,
\end{align}
where in the above, we have taken $\omega , \bk$ to be the incoming frequency and momentum of the external $\field$ field that is contracted with the internal $\auxnew$ field. 

We note that despite the fact that the derivative does not act on the $\auxnew$ field in the interaction Lagrangian \eqref{eq:Lint}, the vertex factor does eventually depend only on the momentum entering the vertex via the internal $\auxnew$ field upon enforcing momentum conservation at the vertex, which is equivalent to integrating by parts at the level of the action\footnote{All variations are consistent with the boundary conditions $\psi, \phi \to 0$ as $|\textbf{x}| \to \infty$. Note that the symmetries and field redundancies discussed in Sec. \ref{sec:symm} admit a one-parameter asymptotic symmetry that relates all boundary configurations of the form $\phi = 0, \psi = c$ for some constant $c$, the significance of which will be followed up on in future work.}. That is, an equivalent way of proceeding would have been by integrating the interaction Lagrangian by parts from the outset, so that
\begin{align}
\calL_{\rm int} \equiv - \lambda ( \sin \field - \field ) \partial_x \auxnew - \lambda ( \cos \field ) \partial_y \auxnew.
\end{align}
Adding a factor of $\lambda\partial_y \auxnew$ to the above leaves the action invariant (given that it is a total derivative), so that equivalently, the interaction Lagrangian can be expressed as:
\begin{align}
\calL_{\rm int} = - \lambda ( \sin \field - \field ) \partial_x \auxnew - \lambda ( \cos \field -1 ) \partial_y \auxnew,
\end{align}
corresponding to the interaction terms
\begin{align}
\calL_{\rm int}^{(2n+1)} &= \frac{(-1)^n}{(2n)!} \lambda \field^{2n} \partial_y \auxnew, &%
\calL_{\rm int}^{(2n+2)} &= - \frac{(-1)^n}{(2n+1)!} \lambda \field^{2n+1} \partial_x \auxnew,
\end{align}
which reproduce the same vertex factors as \eqref{eq:vertices}.


\subsection{Non-Renormalization Theorems and Power Counting}
\label{sec:NRT}

\N There are a set of non-renormalization theorems that one can immediately infer directly from the kinematic structure of the propagators. The proof proceeds exactly as presented in \cite{Chapman:2020vtn}\footnote{We elaborate on additional relations between anomalous dimensions, along with the form invariance of the advection potential under renormalization group improvement enforced by various Ward identities in Sec. \ref{sec:symm}.}. Any diagram that has a subdiagram which is a loop of internal lines which are all of $\field$-$\auxnew$ or all of $\auxnew$-$\field$ type, with insertions in between, must vanish. 
\begin{equation}
\Gamma_N \equiv \vcenter{
\hbox{
\begin{tikzpicture}
\draw[thick,dashed,>=Latex,->] (180:1.5) arc(180:150:1.5);
\draw[thick] (150:1.5) arc(150:120:1.5);
\draw[thick,dashed,>=Latex,->] (120:1.5) arc(120:90:1.5);
\draw[thick] (90:1.5) arc(90:60:1.5);
\draw[thick,loosely dotted] (37:1.5) arc(37:23:1.5);
\draw[thick,dashed,>=Latex,->] (0:1.5) arc(0:-30:1.5);
\draw[thick] (-30:1.5) arc(-30:-60:1.5);
\draw[thick,dashed,>=Latex,->] (-60:1.5) arc(-60:-90:1.5);
\draw[thick] (-90:1.5) arc(-90:-120:1.5);
\draw[thick,loosely dotted] (-143:1.5) arc(-143:-157:1.5);
\node[red] at (-1.5,0) {x};
\node[red,right] at (-1.5,0) {$V_N$};
\node[red] at (-0.75,1.3) {x};
\node[red,below right] at (-0.75,1.3) {$V_1$};
\node[red] at (0.75,1.3) {x};
\node[red,below] at (0.75,1.3) {$V_2$};
\node[red] at (0.75,-1.3) {x};
\node[red,above left] at (0.75,-1.3) {$V_n$};
\begin{scope}[very thick,decoration={
    markings,
    mark=at position 0.6 with {\arrow{<}}}
    ] 
    \draw[thick,red,dotted,postaction={decorate}] (-1.5,0) -- +(180:1);
    \draw[thick,red,dotted,postaction={decorate}] (-0.75,1.3) -- +(120:1);
    \draw[thick,red,dotted,postaction={decorate}] (0.75,1.3) -- +(60:1);
    \draw[thick,red,dotted,postaction={decorate}] (0.75,-1.3) -- +(-60:1);
\end{scope}
\node[red] at (-3,0) {$k_N$};
\node[red] at (-1.5,2.6) {$k_1$};
\node[red] at (1.5,2.6) {$k_2$};
\node[red] at (1.5,-2.6) {$k_n$};
\node[left] at (-1.35,0.75) {$q$};
\node[above] at (0,1.55) {$q+k_1$};
\node[right] at (1.35,-0.75) {$q+k_1+ \cdots + k_{n-1}$};
\node[below] at (-1,-1.55) {$q+k_1+ \cdots + k_{n}$};
\end{tikzpicture}
}
}
\end{equation}
Each internal propagator has the form
\begin{align}
\mathcal{G}_{\auxnew\field} &= \frac{-i}{\omega - \Omega ( \bk )}, &%
\Omega ( \bk ) &= \lambda k_x + i \diff k^2.
\end{align}
The important thing to note is that the imaginary part of $\Omega ( \bk )$ is always positive (assuming $\diff$ is positive). Then,
\begin{align}
\Gamma_N &= \int \dbar \nu \, \dbar^d q \, \frac{-i V_1 (k_1, -q-k_1)}{\nu + \omega_1 - \Omega ( \bq + \bk_1 )} \cdot \frac{-i V_2 (k_2, -q - k_1 - k_2)}{\nu + \omega_1 + \omega_2 - \Omega ( \bq + \bk_1 + \bk_2)} \cdots \frac{-i V_N (k_N, -q)}{\nu - \Omega ( \bq )}.
\end{align}
Of course, for the theory at hand, the vertex only actually depends on the incoming momentum of the dashed line at the vertex: $V_N ( k_N, -q ) = V_N (-q)$, etc.

Note that all of the poles of the integrand are located in the upper half-plane in $\nu$. If there are no factors of frequency appearing in the vertices (i.e., the interaction vertices do not involve time derivatives), and if $N \geq 2$, then we can always complete the integration contour in the lower half-plane, thereby proving that $\Gamma_N = 0$. For $N \geq 2$, the assumption about there being no factors of the frequency in the vertices may be relaxed to the assumption that there be at most $N-2$ factors of frequency appearing in the vertices in total, for, in that case, the integrand still decays at least as fast as $| \nu |^{-2}$ at large $| \nu |$ and we may still close the contour in the lower half-plane, thereby proving that $\Gamma_N = 0$. This relaxed assumption is not necessary anyway because there are no interactions involving time derivatives in this theory. Note that $N = 1$ is not technically captured under this argument because the integrand goes like $\nu^{-1}$ in that case and that is not fast enough to justify the usual procedure of closing the contour (the contribution of the infinite semi-circle at infinity does not necessarily vanish), and will be dealt with as a separate case shortly.

We also note that a diagram that contains $\Gamma$ as a subdiagram will automatically be accompanied by the same diagram with $\Gamma$ replaced with the subdiagram $\Gamma'$, which is given by changing all of the internal propagators to be $\mathcal{G}_{\field \auxnew}$ rather than $\mathcal{G}_{\auxnew\field}$:
\begin{equation}
\Gamma_N' \equiv 
\vcenter{
\hbox{
\begin{tikzpicture}
\draw[thick,>=Latex,->] (180:1.5) arc(180:150:1.5);
\draw[thick,dashed] (150:1.5) arc(150:120:1.5);
\draw[thick,>=Latex,->] (120:1.5) arc(120:90:1.5);
\draw[thick,dashed] (90:1.5) arc(90:60:1.5);
\draw[thick,loosely dotted] (37:1.5) arc(37:23:1.5);
\draw[thick,>=Latex,->] (0:1.5) arc(0:-30:1.5);
\draw[thick,dashed] (-30:1.5) arc(-30:-60:1.5);
\draw[thick,>=Latex,->] (-60:1.5) arc(-60:-90:1.5);
\draw[thick,dashed] (-90:1.5) arc(-90:-120:1.5);
\draw[thick,loosely dotted] (-143:1.5) arc(-143:-157:1.5);
\node[red] at (-1.5,0) {x};
\node[red,right] at (-1.5,0) {$V_N'$};
\node[red] at (-0.75,1.3) {x};
\node[red,below right] at (-0.75,1.3) {$V_1'$};
\node[red] at (0.75,1.3) {x};
\node[red,below] at (0.75,1.3) {$V_2'$};
\node[red] at (0.75,-1.3) {x};
\node[red,above left] at (0.75,-1.3) {$V_n'$};
\begin{scope}[very thick,decoration={
    markings,
    mark=at position 0.6 with {\arrow{<}}}
    ] 
    \draw[thick,red,dotted,postaction={decorate}] (-1.5,0) -- +(180:1);
    \draw[thick,red,dotted,postaction={decorate}] (-0.75,1.3) -- +(120:1);
    \draw[thick,red,dotted,postaction={decorate}] (0.75,1.3) -- +(60:1);
    \draw[thick,red,dotted,postaction={decorate}] (0.75,-1.3) -- +(-60:1);
\end{scope}
\node[red] at (-3,0) {$k_N$};
\node[red] at (-1.5,2.6) {$k_1$};
\node[red] at (1.5,2.6) {$k_2$};
\node[red] at (1.5,-2.6) {$k_n$};
\node[left] at (-1.35,0.75) {$q$};
\node[above] at (0,1.55) {$q+k_1$};
\node[right] at (1.35,-0.75) {$q+k_1+ \cdots + k_{n-1}$};
\node[below] at (-1,-1.55) {$q+k_1+ \cdots + k_{n}$};
\end{tikzpicture}
}
}
\end{equation}
This is given by
\begin{align}
\Gamma_N' &= \int \dbar \nu \, \dbar^d q \, \frac{i V_1' (q, k_1)}{\nu + \omega_1 - \Omega ( \bq + \bk_1 )^*} \cdot \frac{i V_2' (q+k_1, k_2)}{\nu + \omega_1 + \omega_2 - \Omega ( \bq+ \bk_1 + \bk_2)^*} \cdots \frac{i V_N' (q- k_N, k_N)}{\nu - \Omega ( \bq )^*},
\end{align}
where we have used the fact that $\mathcal{G}_{\field \auxnew} = \mathcal{G}_{\auxnew \field}^{*}$. Now, all of the poles are in the lower half-plane. Again, if $N \geq 2$ and there are fewer than $N-2$ factors of frequency appearing in the vertices in total, then, we can always close the contour in the upper half-plane, thereby proving that $\Gamma_N' = 0$. 

Although $\Gamma_1$ and $\Gamma_1'$ do not necessarily vanish separately, it turns out that their sum, given by
\begin{equation}
\Gamma_1 + \Gamma_1' = \int \dbar \nu \, \dbar^d q \biggl( \frac{-i V(-q)}{\nu - \Omega ( \bq )} + \frac{i V(q)}{\nu - \Omega ( \bq )^*} \biggr).
\end{equation}
does indeed vanish. This follows from the fact that the vertices \eqref{eq:vertices} are linear in momentum and purely imaginary, $V(-q) = - V(q)$ and $[ -i V(-q) ]^* = i V(q)$, so that $\Gamma_1' = \Gamma_{1}^{*}$, and
\begin{equation}
\Gamma_1 + \Gamma_1' = 2i \int \dbar^d q \, V(q) \int \dbar \nu \, \frac{\nu - \text{Re}\, \Omega ( \bq )}{( \nu - \text{Re}\, \Omega ( \bq ) )^2 + ( \text{Im} \Omega ( \bq ))^2}.
\end{equation}
One can change the integration variable to the shifted frequency $\tnu = \nu - \text{Re}\, \Omega ( \bq )$. Provided $\text{Im}\, \Omega ( \bq ) \neq 0$, the numerator is odd in $\tnu$ while the denominator is even, and so the integral vanishes, so that
\begin{equation} \label{eq:NRT}
\Gamma_1 + \Gamma_1' = \Gamma_1 + \Gamma_{1}^{*} = 0.
\end{equation}
Therefore, the nonrenormalization theorem applies to the case $N=1$ as well as $N \geq 2$.


\label{sec:powercounting}

Let us momentarily relax the isotropic assumption so that $\diff$ splits into $\diff_x$ and $\diff_y$ in each direction, respectively\footnote{\label{fn:iso}We note that this is not to be understood in terms of the more general model for Malthusian flocks referenced in \cite{CS1, inconvenienttruth, comment1, response1, comment2}, where additional derivatively-coupled operators appear with coefficients that vanish in the limit $\kappa_x\to\kappa_y$. Rather, we consider the deformation of the isotropic diffusion term alone so that it can have differing $\kappa_x$ and $\kappa_y$.}. Assign the following \textit{bare} dimensions:
\begin{align}
[ t ] &= -z, &%
[ x] &= - \zeta, &%
[ y ] &= - 1, &%
[ \calL ] &= 1 + \zeta + z.
\end{align}
The form of the free Lagrangian \eqref{eq:Lfree} then implies that:
\begin{align}
\label{eq:dim}
1+ \zeta + z &= [ \auxnew ] + z + [ \field ] = [ \auxnew ] + [ \diff_{x} ] + 2 \zeta + [ \field ] = [ \auxnew ] + [ \diff_y ] + 2 + [ \field ] = [ \lambda ] + [ \field ] + [ \auxnew ] + \zeta \notag \\
&= [ \Delta ] + 2 [ \auxnew ] = [ \lambda ] + [ \auxnew ] + 1,
\end{align}
which are equivalent to the relations
\begin{subequations}
\begin{align}
[ \auxnew ] + [ \field ] &= 1+ \zeta,
\hspace{50pt}[ \diff_x ] = z - 2 \zeta, \\
[ \lambda ] &= z - \zeta,
\hspace{50pt}[ \diff_y ] = z -2,\\ 
[ \Delta ] + 2 [ \auxnew ] &= 1 + \zeta + z,
\hspace{32pt}[ \lambda ] + [ \auxnew ] = \zeta + z.
\end{align}
\end{subequations}
Of course, $\field$ must be dimensionless in order to retain its interpretation as an angle, for otherwise, we would not be able to take its cosine or sine. This is of course, a direct corollary of the power-counting renormalizability of all $z=2$ actions in 2+1 dimensions with up to four spatial derivatives \cite{Horava:2009uw, Visser:2009fg, Fujimori:2015wda, Fujimori:2015mea}. Therefore,
\begin{equation}
[ \field ] = 0,
\end{equation}
from which it follows that
\begin{subequations}
\begin{align}
[ \auxnew ] &= 1+ \zeta,\hspace{50pt}
[ \diff_x ] = z - 2 \zeta, \\
[ \diff_y ] &= z -2, \hspace{50pt}
[ \lambda ] = z - \zeta, \label{eq:l1} \\
[ \Delta ] &= z - \zeta - 1, \hspace{31pt}
[ \lambda ] = z -1. \label{eq:l2} 
\end{align}
\end{subequations}
The last two equations of \eqref{eq:l1} and \eqref{eq:l2} then imply that 
\begin{equation}
\zeta = 1,
\end{equation}
and so the dimensions $[x]$ and $[y]$ are identical, and
\begin{subequations}
\begin{align}
[ \auxnew ] &= 2, \hspace{50pt}
[ \diff ] = z - 2, \\
[ \lambda ] &= z - 1,\hspace{32pt}
[ \Delta ] = z - 2,
\end{align}
\end{subequations}
where $[ \diff_x ] = [ \diff_y ] = [ \kappa ]$ follows by corollary. At this point, it is argued within the context of a dynamical renormalization group analysis in \cite{inconvenienttruth} that in order to keep $\diff_{x}$, $\diff_{y}$ and $\Delta$ fixed, or, in other words, to make them dimensionless, we have to pick
\begin{equation} \label{eq:zis2}
z = 2.
\end{equation}
It follows immediately that $[ \lambda ] = 1$ and so $\lambda$ is a \emph{relevant} parameter at bare level.


\subsection{Organization of the Diagrammatic Expansion}
\label{sec:diagramorganization}

\N Define the following quantities:
\begin{subequations}
\begin{align}
    L &= \#\ \text{of loops}, \\
    I &= \#\ \text{of internal lines}, \\
    E &= \#\ \text{of external lines}, \\
    V_{2n+1} &= \#\ \text{of $(2n+1)$-vertices}, \\
    V_{2n+2} &= \#\ \text{of $(2n+2)$-vertices}.
\end{align}
\end{subequations}
There are two simple relations one can immediately derive:
\begin{subequations}
\begin{align}
    L &= I - \sum_{n=1}^{\infty} (V_{2n+1} + V_{2n+2}) + 1, \\
    2I + E &= \sum_{n=1}^{\infty} \bigl[ (2n+1) V_{2n+1} + (2n+2) V_{2n+2} \bigr].
\end{align}
\end{subequations}
The first equation comes from the fact that every internal line gets its own momentum to be integrated over, but each vertex imposes one momentum conservation relation among those momenta, except that those relations are themselves not all independent as there is one relation among them for overall momentum conservation. 

The second equation simply counts the total number of endpoints of lines that must be attached on all of the vertices. Every vertex of type $V_{2n+1}$ has $2n+1$ lines and every vertex of type $V_{2n+2}$ has $2n+2$ lines. These must be exhausted by all of the internal and external lines, except that internal lines contribute both of their endpoints while external lines only contribute one.


\subsubsection{Corrections to the Propagators}

\N For propagators, $E=2$ and we can write
\begin{equation} \label{eq:Iprop}
I = -1 + \sum_{n=1}^{\infty} \biggl( (2n+1) \frac{V_{2n+1}}{2} + (n+1) V_{2n+2} \biggr),
\end{equation}
and thus,
\begin{equation} \label{eq:Lprop}
L = \sum_{n=1}^{\infty} \biggl( (2n-1) \frac{V_{2n+1}}{2} + n V_{2n+2} \biggr).
\end{equation}
Clearly, the sum of all the $V_{2n+1}$ coefficients must be even. This is because each $2n+1$ vertex comes with an odd number of fields, and given that they must be pairwise contracted in order to form a diagram, there is no way this can be done if the total number of fields is odd. Therefore we need the total number of such $2n+1$ vertices, i.e. the sum of all the $V_{2n+1}$ to be an even number. 


\subsubsection{Corrections to the Vertices}

\N The corrections to the $(2n+2)$-vertex will have $E = 2n+2$, so that
\begin{equation} \label{eq:In}
I = - (n+1) + \sum_{m=1}^{\infty} \biggl( (2m+1) \frac{V_{2m+1}}{2} + (m+1) V_{2m+2} \biggr),
\end{equation}
and
\begin{equation} \label{eq:Ln}
    L = -n + \sum_{m=1}^{\infty} \biggl( (2m-1) \frac{V_{2m +1}}{2} + m V_{2m+2} \biggr).
\end{equation}
Note that \eqref{eq:Iprop} and \eqref{eq:Lprop} are simply the $n=0$ versions of \eqref{eq:In} and \eqref{eq:Ln}, respectively. As with the propagator corrections, the sum of the $V_{2m+1}$ coefficients must be even.

On the other hand, the corrections to $V_{2n+1}$ have $E=2n+1$, and so
\begin{equation} \label{eq:In2}
I = - \biggl( n+ \frac{1}{2} \biggr) + \sum_{m=1}^{\infty} \biggl( (2m+1) \frac{V_{2m+1}}{2} + (m+1) V_{2m+2} \biggr),
\end{equation}
and
\begin{equation} \label{eq:Ln2}
    L = -n + \frac{1}{2} + \sum_{m=1}^{\infty} \biggl( (2m-1) \frac{V_{2m +1}}{2} + m V_{2m+2} \biggr).
\end{equation}
In this case, the sum of the $V_{2m+1}$'s must be odd.


\subsection{One-Loop Corrections}
\label{sec:oneloopcorrections}

\N For the one-loop corrections to the propagators, $L=1$ in \eqref{eq:Lprop} implies that $V_{\geq 5} = 0$, and so
\begin{equation}
    1 = \frac{V_3}{2} + V_4.
\end{equation}
The possible values of $V_3$ and $V_4$ are tabulated in Table \ref{table:L1prop} along with the structure of the corresponding diagram. The diagram form merely shows the structure of the diagram; it does not show the dashed lines for $\auxnew$.
\begin{table}[h!]
\centering
\renewcommand{\arraystretch}{1.5}
\begin{tabular}{c c c c} 
 \hline
 $V_3$ & $V_4$ & $I$ & Diagram Form \\
 \hline\hline
 0 & 1 & 1 &
$\vcenter{
\hbox{
\begin{tikzpicture}[scale=0.5]
\draw[white] (0,-0.3) -- (0,2.1);
\draw[thick] (-2,0) -- (2,0);
\draw[thick] (0,0) to[out=170,in=180] (0,1.8);
\draw[thick] (0,0) to[out=10,in=0] (0,1.8);
\end{tikzpicture}
}
}$
\\
 \hline
\end{tabular}
%
\hspace{2cm}
\begin{tabular}{c c c c} 
 \hline
 $V_3$ & $V_4$ & $I$ & Diagram Form \\
 \hline\hline
 2 & 0 & 2 &
$\vcenter{
\hbox{
\begin{tikzpicture}[scale=0.5]
\draw[white] (0,-1.2) -- (0,1.2);
\draw[thick] (-2,0) -- (-1,0);
\draw[thick] (1,0) -- (2,0);
\draw[thick] (0,0) circle [radius=1];
\end{tikzpicture}
}
}$
\\
 \hline
\end{tabular}
\caption{Classification of 1-loop corrections to the propagator according to the numbers of 3-point and 4-point vertices ($V_3$ and $V_4$) or the number of internal lines ($I$).}
\label{table:L1prop}
\end{table}

For the one-loop corrections to the $(2n+2)$-vertex, we set $L=1$ in \eqref{eq:Ln} and find that the sum must terminate at $m=n$, so that the equation simplifies to
\begin{equation}
2n+2 = V_3 + 2V_4 + 3V_5 + 4 V_6 + \cdots + (2n+1) V_{2n+3} + (2n+2) V_{2n+4}.
\end{equation}
This is complicated to analyze for general $n$. However, the stucture of diagrams is fairly easy to see for specific values of $n$ and essentially follows the same pattern as the $n=1$ case. The case $n=1$, corresponding to corrections to the 4-point vertex, gives
\begin{equation}
4 = V_3 + 2V_4 + 3V_5 + 4V_6.
\end{equation}
There are five possible diagram structures given in Table \ref{table:L1V4}.
\begin{table}[h!]
\centering
\renewcommand{\arraystretch}{1.5}
\begin{tabular}{c c c c c c} 
 \hline
 $V_3$ & $V_4$ & $V_5$ & $V_6$ & $I$ & Diagram Form \\
 \hline\hline
 0 & 0 & 0 & 1 & 1 &
$\vcenter{
\hbox{
\begin{tikzpicture}[scale=0.5]
\draw[white] (0,-0.7) -- (0,1.4);
\draw[thick] (0,0) -- +(150:1);
\draw[thick] (0,0) -- +(-150:1);
\draw[thick] (0,0) -- +(30:1);
\draw[thick] (0,0) -- +(-30:1);
\draw[thick] (0,0) to[out=110,in=180] (0,1.2);
\draw[thick] (0,0) to[out=70,in=0] (0,1.2);
\end{tikzpicture}
}
}$
\\
 \hline
 1 & 0 & 1 & 0 & 2 &
$\vcenter{
\hbox{
\begin{tikzpicture}[scale=0.5]
\draw[white] (0,-1.2) -- (0,1.2);
\draw[thick] (-2,0) -- (-1,0);
\draw[thick] (1,0) -- (2,0);
\draw[thick] (1,0) -- +(30:1);
\draw[thick] (1,0) -- +(-30:1);
\draw[thick] (0,0) circle [radius=1];
\end{tikzpicture}
}
}$
\\
 \hline
\end{tabular}
%
\hspace{0.5cm}
\begin{tabular}{c c c c c c} 
 \hline
 $V_3$ & $V_4$ & $V_5$ & $V_6$ & $I$ & Diagram Form \\
 \hline\hline
 0 & 2 & 0 & 0 & 2 &
$\vcenter{
\hbox{
\begin{tikzpicture}[scale=0.5]
\draw[white] (0,-1.2) -- (0,1.2);
\draw[thick] (1,0) -- +(30:1);
\draw[thick] (1,0) -- +(-30:1);
\draw[thick] (-1,0) -- +(150:1);
\draw[thick] (-1,0) -- +(-150:1);
\draw[thick] (0,0) circle [radius=1];
\end{tikzpicture}
}
}$
\\
 \hline
 2 & 1 & 0 & 0 & 3 &
$\vcenter{
\hbox{
\begin{tikzpicture}[scale=0.5]
\draw[white] (0,-1.2) -- (0,1.2);
\draw[thick] (0.866,0.5) -- +(30:1);
\draw[thick] (0.866,-0.5) -- +(-30:1);
\draw[thick] (-1,0) -- +(150:1);
\draw[thick] (-1,0) -- +(-150:1);
\draw[thick] (0,0) circle [radius=1];
\end{tikzpicture}
}
}$
\\
 \hline
\end{tabular}
\hspace{0.5cm}
\begin{tabular}{c c c c c c} 
 \hline
 $V_3$ & $V_4$ & $V_5$ & $V_6$ & $I$ & Diagram Form \\
 \hline\hline
 4 & 0 & 0 & 0 & 4 &
$\vcenter{
\hbox{
\begin{tikzpicture}[scale=0.5]
\draw[white] (0,-1.2) -- (0,1.2);
\draw[thick] (0.866,0.5) -- +(30:1);
\draw[thick] (0.866,-0.5) -- +(-30:1);
\draw[thick] (-0.866,0.5) -- +(150:1);
\draw[thick] (-0.866,-0.5) -- +(-150:1);
\draw[thick] (0,0) circle [radius=1];
\end{tikzpicture}
}
}$
\\
 \hline
 \end{tabular}
\caption{Classification of 1-loop corrections to the 4-point vertex according to the numbers of 3-point up to 6-point vertices ($V_3$ to $V_6$) or the number of internal lines ($I$).}
\label{table:L1V4}
\end{table}

For the one-loop corrections to the $(2n+1)$-vertex, we set $L=1$ in \eqref{eq:Ln2}, we find that the sum must terminate at $m=n+1$ excluding $V_{2n+4}$, and the equation simplifies to
\begin{equation}
2n+1 = V_3 + 2V_4 + 3V_5 + \cdots + 2n V_{2n+2} + (2n+1) V_{2n+3}.
\end{equation}
For $n=1$, corresponding to the corrections to the 3-point vertex,
\begin{equation}
3 = V_3 + 2 V_4 + 3 V_5.
\end{equation}
There are three possible structures, shown in Table \ref{table:L1V3}.
\begin{table}[h!]
\centering
\renewcommand{\arraystretch}{1.5}
\begin{tabular}{c c c c c} 
 \hline
 $V_3$ & $V_4$ & $V_5$ & $I$ & Diagram Form \\
 \hline\hline
 0 & 0 & 1 & 1 &
$\vcenter{
\hbox{
\begin{tikzpicture}[scale=0.5]
\draw[white] (0,-1.2) -- (0,1.2);
\draw[thick] (1,0) -- (2,0);
\draw[thick] (1,0) -- +(30:1);
\draw[thick] (1,0) -- +(-30:1);
\draw[thick] (0,0) circle [radius=1];
\end{tikzpicture}
}
}$
\\
 \hline
\end{tabular}
%
\hspace{0.5cm}
\begin{tabular}{c c c c c} 
 \hline
 $V_3$ & $V_4$ & $V_5$ & $I$ & Diagram Form \\
 \hline\hline
 1 & 1 & 0 & 2 &
$\vcenter{
\hbox{
\begin{tikzpicture}[scale=0.5]
\draw[white] (0,-1.2) -- (0,1.2);
\draw[thick] (-2,0) -- (-1,0);
\draw[thick] (1,0) -- +(30:1);
\draw[thick] (1,0) -- +(-30:1);
\draw[thick] (0,0) circle [radius=1];
\end{tikzpicture}
}
}$
\\
 \hline
\end{tabular}
\hspace{0.5cm}
\begin{tabular}{c c c c c} 
 \hline
 $V_3$ & $V_4$ & $V_5$ & $I$ & Diagram Form \\
 \hline\hline
 3 & 0 & 0 & 3 &
$\vcenter{
\hbox{
\begin{tikzpicture}[scale=0.5]
\draw[white] (0,-1.2) -- (0,1.2);
\draw[thick] (-2,0) -- (-1,0);
\draw[thick] (0.866,0.5) -- +(30:1);
\draw[thick] (0.866,-0.5) -- +(-30:1);
\draw[thick] (0,0) circle [radius=1];
\end{tikzpicture}
}
}$
\\
 \hline
 \end{tabular}
\caption{Classification of 1-loop corrections to the 3-point vertex according to the numbers of 3-, 4-, and 5-point vertices ($V_3$, $V_4$, and $V_5$) or the number of internal lines ($I$).}
\label{table:L1V3}
\end{table}
%


\subsection{Superficial Degree of Divergence}
\label{sec:supdegdiv}

The power-counting classification presented in sec. \ref{sec:powercounting} has important consequences for the superficial degree of divergence of the various Feynman diagrams. A diagram with a $\field$-$\field$ propagator in an internal line will always have a lower degree of divergence compared to a diagram with the same structure, but where the $\field$-$\field$ propagator is replaced with either a $\auxnew$-$\field$ or a $\field$-$\auxnew$ propagator. In fact, the degree of divergence is reduced by 2 for each internal $\auxnew$-$\field$ or $\field$-$\auxnew$ propagator replaced with a $\field$-$\field$ propagator, and further reduced by 1 because one vertex factor will now contain an external rather than an internal momentum factor. 

Let us compute the superficial degree of divergence for arbitrary $z$ and spatial dimension $d$ before plugging in $z=d=2$. A diagram with $L$ loops will have $L$ internal frequency-momentum integrals for a total dimension of $(z+d)L$. Each internal line comes with a factor of a propagator. Let us split $I$ up into the number of $\auxnew$-$\field$ or $\field$-$\auxnew$ internal lines and the number of $\field$-$\field$ internal lines:
\begin{equation}
I = I_{\auxnew\field} + I_{\field\field}.
\end{equation}
Each internal line comes with a corresponding propagator factor, which has dimension $-z$ for a $\auxnew$-$\field$ or $\field$-$\auxnew$ line and dimension $-2z$ for a $\field$-$\field$ line. Furthermore, each $I_{\auxnew\field}$ contributes one factor of internal momentum due to the vertex. Therefore, the superficial degree of divergence, $D$, of any diagram is
\begin{align}
D &= (z+d)L - z ( I_{\auxnew\field} + 2I_{\field\field} ) + I_{\auxnew\field} \notag \\
&= (z+d)L - 2z I_{\field\field} - (z-1) I_{\auxnew\field} \notag \\
&= (z+d)L - 2zI + (z+1) I_{\auxnew\field}.
\end{align}
Plugging in $z=d=2$ gives
\begin{equation}
D = 4(L-I) + 3I_{\auxnew\field}.
\end{equation}
Indeed, if we reduce the number of internal $\phi$-$\psi$ lines by one, so that $I_{\auxnew\field} \to I_{\auxnew\field} -1$, then the superficial degree of divergence of the diagram is reduced by 3, as previously described.

The above formula gives the superficial degree of divergence of an \emph{individual} diagram. However, this does not preclude the possibility of so-called \emph{renormalons} for which infinite towers of diagrams combine constructively to increase the overall degree of divergence (see e.g., \cite{Beneke:1998ui}). This phenomenon tends to occur at specific discrete values of the system parameters ($\Delta / \kappa$, in this case). There is one such special point at $\Delta / \kappa = 4 \pi$ where the parameter $\kappa$ does, in fact, get renormalized. The beta functions near this point are given in \cite{sezik2026}. In this present work, we instead focus on regions of parameter space away from these specific special points. 


\section{One-Loop Order -- Regularization and Scheme Independence of Log Divergences}
\label{sec:onelooporder}

The simple dimension counting of the previous subsections have very important consequences. For example, for one-loop diagrams, one has
\begin{equation}
D = 4 - 4I + 3 I_{\auxnew\field}.
\end{equation}
The maximum value for $I_{\auxnew\field}$ is $I$, and so
\begin{equation}
D_{\rm max} = 4 - I.
\end{equation}
We consider $I \geq 2$ and $I = 1$ as separate cases. For $I \geq 2$, one has $D_{\rm max} \leq 2$, and so we need only consider the case $I_{\auxnew\field} = I$, as reducing $I_{\auxnew\field}$ would reduce $D$ by at least 3, thereby giving $D \leq -1$ which correspond to UV-finite diagrams that do not require renormalization. The case $I=1$ implies $D_{\rm max} = 3$, where we could either have $I_{\auxnew\field} = 1$ for which $D=3$, or $I_{\auxnew\field} = 0$, for which $D=0$ (i.e., corresponding to a log-divergence). However, if $I_{\auxnew\field} = I$, the process involves the sum of loop diagrams where $\auxnew$-$\field$ and $\field$-$\auxnew$ lines are interchanged, and therefore vanishes according to the non-renormalization theorem in sec. \ref{sec:NRT}. Hence, only one family of diagrams contributes to the renormalization of the parameters of our theory at one loop: $(I, I_{\auxnew\field}) = (1, 0)$, which correspond to log-divergences. 

For the $\auxnew$-$\field$ \emph{inverse} propagator, the relevant one loop diagram is thus given by:
\begin{equation} \label{eq:SGchipsia}
\vcenter{
\hbox{
\begin{tikzpicture}[scale=0.8]
\draw[thick,>=Latex,>-] (-1,0) -- (0,0);
\draw[thick,dashed,>=Latex,-<] (0,0) -- (1,0);
\draw[thick,->] (0,0) to[out=150,in=180] (0,1.8);
\draw[thick] (0,0) to[out=30,in=0] (0,1.8);
\end{tikzpicture}
}
}
= M_{\auxnew\field}^{(1)}.
\end{equation}
Here, we are using the standard convention wherein the external propagators of the diagram are amputated, here denoted, as in Eq.~\eqref{eq:vertices} for the vertices, by the incomplete external lines.\footnote{Although this convention arises from the necessities of LSZ reduction in quantum field theory \cite{1998qft, Peskin:1995ev, Burgess:2020tbq}, it persists in statistical field theory in isolating the interacting parts of the one particle irreducible diagrams given the role of external propagators in simply tracking momentum in and outflow in a perturbative computation \cite{heliasdahmen, tauber2014critical}.} This is why we have denoted this term by $M$, not $\calG$. This evaluates to
\begin{align}
- \frac{i \lambda k_x}{2} \int_{\nu , \bq} \frac{2 \Delta}{( \nu - \lambda q_x - i \diff q^2 ) ( \nu - \lambda q_x + i \diff q^2)} = - i \lambda k_x \scrI,
\end{align}
where
\begin{equation}
\scrI = \Delta \int_{\nu , \bq} \frac{1}{( \nu - \lambda q_x - i \diff q^2 )( \nu - \lambda q_x + i \diff q^2 )} = \Delta \int_{\wt{\nu}, \bq} \frac{1}{\wt{\nu}^2 + ( \diff q^2 )^2},
\end{equation}
and
\begin{equation}
\wt{\nu} = \nu - \lambda q_x.
\end{equation}
The $\wt{\nu}$ integral is straightforward to compute either using the calculus of residues or a trigonometric substitution, and the result is
\begin{equation}
	\label{eq:logdiv}
\scrI = \frac{\Delta}{2 \diff} \int_{\bq} \frac{1}{q^2} \xrightarrow{d=2} \frac{\Delta}{4 \pi \diff} \int_{0}^{\infty} \frac{dq}{q} \xrightarrow{\rm reg} \frac{\Delta}{4 \pi \diff} \ln \frac{\Lambda}{\mu},
\end{equation}
where ``reg'' stands for ``regularized,'' implemented most directly with hard UV and IR cut-offs, $\Lambda$ and $\mu$, respectively. Ideally, we should ensure that this would be the result in any scheme, given that the coefficients of the leading short distance logarithmic divergences dictate the corresponding coefficients of the beta functions, and therefore the renormalization group flow of the associated couplings. 

Fortunately, the coefficients of leading short-distance logarithmic divergences are scheme-independent. That is, for any regularization scheme, whether hard cutoff, zeta-function, Pauli-Villars, point-split, dimensional regularization, etc., the leading log divergences can explicitly be shown to have the same coefficient (see e.g. \cite{Burgess:2020tbq, Negro:2024bbf}). This is an important enough (and often misunderstood) point, that is worth elaborating upon it in the present context.

One can, for example, arrive at the same coefficient for the leading logarithmic divergence in the context of dimensional regularization by first recasting the integral in \eqref{eq:logdiv} in terms of a related integral over $\mathbb R^2$:  
\eq{eq:sf}{\int_0^\infty \frac{dq}{q} = \int_0^\infty \frac{q\,dq}{q^2} = \frac{1}{2\pi}\int_{-\infty}^\infty \frac{d^2q}{q^2},} 
and realizing that for integrals that include an explicit mass scale $m$, that:  
\eq{eq:dr}{I_D(m^2) = \int \frac{d^D k}{(2\pi)^D} \frac{k^{2A}}{(k^2 + m^2)^B} = \frac{\Gamma\left(A+\frac{D}{2}\right) \Gamma\left(B-A-\frac{D}{2}\right)}{(4\pi)^{D/2}\Gamma\left(\frac{D}{2}\right)\Gamma(B)}\left(m^2\right)^{A-B+D/2}.}
Although the integrand \eqref{eq:sf} is scaleless and therefore might appear to nominally vanish when dimensionally regularized, this vanishing can be traced to a cancellation of UV and IR contributions enforced by default in dimensional regularization. In order to see this, we expand the divergent integrand via partial fractions into the sum
\eq{eq:div}{ \frac{1}{2\pi}\int_{-\infty}^\infty \frac{d^2q}{q^2} =  \frac{1}{2\pi}\int_{-\infty}^\infty d^2q\left[\frac{1}{q^2 + m^2} + \frac{m^2}{q^2(q^2 + m^2)}\right],}
where $m^2$ arises as an auxiliary mass scale via the partial fraction expansion that subsequently cancels between the two contributions. Of particular note is the fact that the first term on the right hand side of the above is UV-divergent, whereas the second term is IR-divergent. Evaluating both terms using \eqref{eq:dr} and expanding around $D = 2 - \delta$ dimensions results in equal and opposite contributions of the form
\eq{}{\pm\left[\frac{1}{\delta} - \frac{1}{2}\left(\log \frac{m^2}{4\pi\mu^2} + \gamma_E \right) \right],}
where $\mu$ is some arbitrary mass scale necessitated from dimensional deformation. Were we to label the UV and IR pole contributions separately, one arrives at the result  
\eq{}{\frac{\Delta}{4 \pi \diff} \int_{0}^{\infty} \frac{dq}{q} \xrightarrow{\rm reg} \frac{\Delta}{4 \pi \diff}\left[\frac{1}{\delta_{\rm UV}} - \frac{1}{\delta_{\rm IR}} + \log \frac{\mu_{\rm UV}}{\mu_{\rm IR}} \right],}
which matches with \eqref{eq:logdiv} with the identifications $\mu_{\rm UV} \equiv \Lambda$ and $\mu_{\rm IR} \equiv \mu$. Whatever scheme one works in, provided one is expanding around a background field that is a \textit{stable saddle point} of the MSRDJ action, all IR divergences will cancel in well defined observables (and can be taken as the defining criteria for the latter) \cite{Burgess:2020tbq, Negro:2024bbf}. UV divergences get subtracted by appropriate counterterms, which, remarkably, can be absorbed via the renormalization of the couplings in the action \eqref{eq:Lfree} and \eqref{eq:Lint}, i.e., the theory is itself \textit{renormalizable} in the particle physics sense, and does not need to be treated as an effective theory \cite{Burgess:2020tbq}. Given that it is the coefficient of the leading logarithmic short distance divergences that determine how couplings run with scale, we see that one would have arrived at the same running having implemented dimensional regularization properly, as one would have by na\"ively imposing hard cut-offs. 

An alternate manner in which one could have arrived at this same conclusion would have been through reexamining the statement that \eqref{eq:logdiv} legitimately corresponds to a scaleless integral. We note that implicit in the background field expansion invoked in our calculation is the fact that $\psi_0$ and $\phi_0$ in \eqref{eq:bg} correspond to stable saddle points of the MSRDJ action. Although taking both to be (vanishing) constants certainly meet this criteria, one will in reality have very long wavelength fluctuations of the background fields such that the net effect of expanding around them would be to generate an IR mass scale for the propagator in \eqref{eq:sf} arising from derivatives acting on the background fields\footnote{Note that when a similar integral is encountered, for example, in \cite{Peskin:1995ev} in the context of the non-linear sigma model in 1+1 dimensions, an IR regulator is introduced by hand. Strictly speaking, this should be understood in terms of the argument just presented, with the net result that the one loop beta functions are proportional to the target space curvature felt by slow variations of the background field.}, which we denote as $\mu_{\rm IR} \equiv \mu_{\rm BG}$ so that one instead has to evaluate:
\begin{equation}
    I = \frac{1}{2\pi}\int_{- \infty}^{\infty} \frac{d^2 q}{q^2} \rightarrow   \frac{1}{2\pi}\int_{- \infty}^{\infty} \frac{d^2 q}{q^2 + \mu_{\rm BG}^2}, 
\end{equation}
which now only manifests a UV divergence. Regularizing the above using \eqref{eq:dr} results in
\eq{}{\frac{\Delta}{4 \pi \diff} \frac{1}{2\pi}\int_{- \infty}^{\infty} \frac{d^2 q}{q^2 + \mu_{\rm BG}^2} \xrightarrow{\rm reg} \frac{\Delta}{4 \pi \diff}\left[\frac{1}{\delta_{\rm UV}}  + \log \left(\frac{\mu_{\rm UV}}{\mu_{\rm BG}}\right) \right],}
and will result in the same running as would have been extracted by hard cut-offs, or factorizing and dimensionally regularizing the nominally scaleless integral \eqref{eq:logdiv}.

Thus satisfied by the scheme-independence of the coefficient of the leading UV logarithmic divergence, we find that upon amputation of the external $\auxnew$-$\field$ propagators, the inverse propagator gets the following scale dependent one loop correction:
\begin{equation}
M_{\auxnew\field}^{(1)} = - i \lambda k_x \scrI = - i \lambda k_x \frac{\Delta}{4 \pi \diff} \ln \frac{\Lambda}{\mu}.
\end{equation} 
Given that the bare value is $M_{\auxnew\field}^{(0)} = \wt{O}_{\omega, \bk} = - i ( \omega - \lambda k_x + i \diff k^2 )$, adding these contributions together results in
\begin{equation} \label{eq:auxfield1}
M_{\auxnew\field}^{(0)} + M_{\auxnew\field}^{(1)} = - i \bigl[ \omega - (1 - \scrI ) \lambda k_x + i \diff k^2 \bigr].
\end{equation}

At this point, one might be tempted to conclude that the only quantity that gets renormalized is $\lambda$. As we elaborate upon later, this is incorrect. Instead, $\lambda$, $\Delta$, $\kappa$, $\phi$, and the spatial coordinates $\bx$ all receive anomalous dimensions, as given by \eqref{eq:anomalousdims} and elaborated upon in Appendix \ref{sec:wilsonianRG}.

We furthermore note that the correction to the $\auxnew$-$\field$ inverse propagator is the same as for the $\field$-$\auxnew$  case. In addition, no diagram of this type, that is with $(I , I_{\auxnew \field} ) = (1,0)$, exists with both external fields incident on the vertex being $\auxnew$ fields, since no vertex exists containing two $\auxnew$ fields. Therefore, there is no correction to $M_{\auxnew\auxnew}$, or in other words, $M_{\auxnew\auxnew}^{(1)} = 0$ and
\begin{equation}
M_{\auxnew\auxnew}^{(0)} + M_{\auxnew\auxnew}^{(1)} = M_{\auxnew\auxnew}^{(0)} = 2 \Delta.
\end{equation}
Once more, it might be tempting to conclude that $\Delta$ does not get renormalized because there are no diagrammatic contributions to the above term. This is in fact precisely the argument made by Chat\'e and Solon in \cite{CS1}. However, this argument is not entirely correct. As we discuss in detail in Appendix \ref{sec:wilsonianRG}, $\Delta$ does indeed get an anomalous dimension as per \eqref{eq:anomalousdims}, and this can be traced to the peculiarities of any Lifshitz theory with $z\neq 1$.

Finally, there is no diagram with $(I , I_{\phi \psi} ) = (1,0)$ with both external fields incident on the vertex being $\psi$ fields, since all vertices must contain exactly one $\phi$ field. Therefore, there is no correction to $M_{\psi\psi}$ either, or in other words, $M_{\psi\psi}^{(1)} = 0$, and
\begin{equation}
M_{\field\field}^{(0)} + M_{\field\field}^{(1)} = 0.
\end{equation}

Shifting our attention now to the one-loop correction to the vertices: Each vertex contains the same coupling constant, $\lambda$, and this factor is the same as the one appearing in the propagator. A priori, there is no reason why these should get renormalized uniformly, nor in the same manner as within the propagator. That is, the advection terms $\lambda \bn(\psi)\cdot\nabla \psi$ and  $\lambda \bn(\psi)\cdot\nabla \phi$ could in principle get deformed into an arbitrary derivative expansion that need not resum into an analytic form, let alone the special trigonometric forms appearing in our theory. That it does to all loop orders is a very special property of our model, which is ultimately due to the nonlinear realization of the symmetries and field redundancies identified in Section \ref{sec:symm}. In operational terms, we can see this via the following argument: The only one-loop diagram that contributes to the renormalization of the $(2n+1)$-vertex $V_{2n+1}$ is the $V_{2n+3}$ vertex with two of the internal $\field$ lines contracted to form a loop. The external $\auxnew$ field and $2n$ out of the $2n+1$ external $\field$ fields must be contracted with $2n+1$ of the $2n+3$ internal $\field$ fields. That gives a combinatoric factor of $\frac{1}{2} [ (2n+3)! ]$. The factor of $\frac{1}{2}$ is the symmetry factor of the diagram. The result of the computation of the loop integral proceeds as before and gives the result $\scrI$. Since there is a relative minus sign between $V_{2n+3}$ and $V_{2n+1}$, the one-loop contribution to $V_{2n+1}$ also has a minus sign relative to $V_{2n+1}$ itself. Finally, it is clear that this argument works equally well for $V_{2n+2}$, whose one-loop correction comes from the vertex $V_{2n+4}$ by contracting two of the internal $\field$ fields to form a loop. Thus, letting $V$ with no subscript stand for an arbitrary vertex, we find
\begin{equation} \label{eq:squid}
V^{(0)} + V^{(1)} = \vcenter{
\hbox{
\begin{tikzpicture}
\draw[thick,dashed,>=Latex,>-] (-1,0) -- (0,0);
\draw[thick,>=Latex,-<] (0,0) -- (45:1cm);
\draw[thick,>=Latex,-<] (0,0) -- (22.5:1cm);
\draw[thick,>=Latex,-<] (0,0) -- (-45:1cm);
\draw[thick,dotted]  (-30:0.8) arc(-30:10:0.8);
\end{tikzpicture}
}
}
\ + \
\vcenter{
\hbox{
\begin{tikzpicture}
\draw[thick,dashed,>=Latex,>-] (-1,0) -- (0,0);
\draw[thick,>=Latex,-<] (0,0) -- (45:1cm);
\draw[thick,>=Latex,-<] (0,0) -- (22.5:1cm);
\draw[thick,>=Latex,-<] (0,0) -- (-45:1cm);
\draw[thick,dotted]  (-30:0.8) arc(-30:10:0.8);
\draw[thick,->] (0,0) to[out=170,in=190] (-0.2,0.8);
\draw[thick] (0,0) to[out=55,in=10] (-0.2,0.8);
\end{tikzpicture}
}
}
= (1 - \scrI ) V^{(0)}.
\end{equation}
Since $V^{(0)}$ is proportional to $\lambda$, we find that result of the one-loop computation can be summarized by simply shifting $\lambda$ in precisely the same manner in all of the vertex factors as well as in the propagator:
\begin{equation}
\lambda \rightarrow ( 1- \scrI ) \lambda.
\end{equation}
We contrast this with the argument in \cite{inconvenienttruth}, which argues that the strict form of the vertices in terms of $\lambda$ need not hold as one iterates the dynamical renormalization group treatment. Our computation shows that in fact it does, and is the result of a simple multiplicative renormalization of the coupling $\lambda$, which follows as a direct consequence of the non-linearly realized symmetries and field redundancies and  associated Ward identities elaborated upon in Section \ref{sec:symm}. Were this not the case, the advection interaction potential would have been deformed into an arbitrary power series at any given loop order, subsequently destroying their trigonometric forms and, along with them, the interpretation of $\field$ as a flock angle and $\bn(\psi)$ as a flock director. 

Before we move on to the discussion of higher orders in the loop expansion, we reiterate that, as will be elaborated upon in Appendix \ref{sec:wilsonianRG} and as given in \eqref{eq:anomalousdims}, $\lambda$ is not the only parameter that will end up getting renormalized.


\section{Two- and Higher-Loop Order}
\label{sec:twoloopandhigher}

We now move on to consider the effect of higher order loop corrections. From \eqref{eq:Iprop} and \eqref{eq:Lprop}, one straightforwardly derives the relation
\begin{equation} \label{eq:eulerc}
I = L-1 + V,
\end{equation}
where $V = \sum_{n=1}^{\infty} ( V_{2n+1} + V_{2n+2})$ is the total number of vertices, which is nothing more than a transcription of the fact that the Euler characteristic of the plane is 2, and that the Feynman diagrams in question are planar. The usual expression for the Euler characteristic is given by\footnote{The usage of $\chi$ as the symbol for the Euler characteristic is standard. However, it should not be confused with the roughness exponent, which measures the scaling dimension of $\field$, i.e., $[ \field ] = \chi$, nor with the original $\auxold$ field that we introduced in the MSRDJ procedure before converting to $\auxnew = -i \auxold$.} $\chi = V-E+F$, where $V$ is the number of vertices, $E$ the number of edges, and $F$ the number of faces. We can re-express this in terms of $I$, the number of edges, and where $L+1$ is the number of faces, given that we must count the rest of the plane outside the Feynman diagram as a face. Since $\chi = 2-2g$, where $g$ is the genus, and where $\chi = 2$ for the plane, so that one finally arrives at \eqref{eq:eulerc}.

For $L=1$, we have $I=V$ and so it is possible to have $I_{\auxnew\field} = I$ with each vertex having exactly one $\auxnew$ field, except that the diagram would vanish by the nonrenormalization theorem discussed above. For $L=2$, however, this is not possible because $I = V+1$. Thus, the maximum value of $I_{\auxnew\field}$ is $I-1$. However, with two loops, $I_{\auxnew\field} = I-1$ will force one of the loops to be cyclic in $\auxnew$ and $\field$ and thus vanish by the nonrenormalization theorem. Therefore, the maximum value for $I_{\auxnew\field}$ is in fact $I-2$. Consequently, the maximum superficial degree of divergence is
\begin{equation}
D_{\rm max} = 8 - 4I + 3 (I-2) = 2 - I,
\end{equation}
and so only $(I , I_{\auxnew\field} ) = (2,0)$ contributes. In other words, there is only one contributing diagram:
\begin{equation}
M_{ab}^{(2)} = \!
\vcenter{
\hbox{
\begin{tikzpicture}[scale=0.5]
\draw[white] (0,-1.4) -- (0,1.4);
\draw[thick] (-2,0) -- (2,0);
\draw[thick] (0,0) to[out=170,in=180] (0,1.2);
\draw[thick] (0,0) to[out=10,in=0] (0,1.2);
\draw[thick] (0,0) to[out=-10,in=0] (0,-1.2);
\draw[thick] (0,0) to[out=190,in=180] (0,-1.2);
\node[left] at (-2,0) {$a$};
\node[right] at (2,0) {$b$};
\end{tikzpicture}
}
},
\end{equation}
where $a$ and $b$ can be $\auxnew$ or $\field$. This vanishes for $a = b = \auxnew$ and $a = b = \field$ for the same reason that the one-loop diagrams vanish in those cases. Therefore,
\begin{align}
M_{\field\field}^{(0)} + M_{\field\field}^{(1)} + M_{\field\field}^{(2)} &= 0, &%
M_{\auxnew\auxnew}^{(0)} + M_{\auxnew\auxnew}^{(1)} + M_{\auxnew\auxnew}^{(2)} &= 2 \Delta.
\end{align}
The result for $(a,b) = ( \auxnew, \field )$ is thus 
\begin{equation}
M_{\auxnew\field}^{(2)} = \frac{1}{8} ( 2 \scrI )^2 i k_x = \frac{1}{2} \scrI^2 ik_x.
\end{equation}
Compared to the one-loop diagram, which has a symmetry factor of 2, this diagram has a symmetry factor of $8 = 2^L L!$ for $L=2$. The factor of $L!$ comes from permutations of the loops, and where each loop comes with a factor of $2$ since switching the two fields in each internal loop does not actually change the diagram. Therefore:
\begin{equation}
M_{\auxnew\field}^{(0)} + M_{\auxnew\field}^{(1)} + M_{\auxnew\field}^{(2)} = - i \biggl[ \omega - \biggl( 1 - \scrI + \frac{1}{2} \scrI^2 \biggr) \lambda k_x + i \diff k^2 \biggr].
\end{equation}
In other words, $\lambda$ renormalizes to
\begin{equation}
\lambda \rightarrow \biggl( 1 - \scrI + \frac{1}{2} \scrI^2 \biggr) \lambda.
\end{equation}
By the same argument, the only diagram that contributes to the two-loop vertex correction is the one-loop diagram in \eqref{eq:squid} but with an extra loop added to the same vertex:
\begin{equation} \label{eq:squid2}
V^{(0)} + V^{(1)} + V^{(2)} = \vcenter{
\hbox{
\begin{tikzpicture}
\draw[thick,dashed,>=Latex,>-] (-1,0) -- (0,0);
\draw[thick,>=Latex,-<] (0,0) -- (45:1cm);
\draw[thick,>=Latex,-<] (0,0) -- (22.5:1cm);
\draw[thick,>=Latex,-<] (0,0) -- (-45:1cm);
\draw[thick,dotted]  (-30:0.8) arc(-30:10:0.8);
\end{tikzpicture}
}
}
\ + \
\vcenter{
\hbox{
\begin{tikzpicture}
\draw[thick,dashed,>=Latex,>-] (-1,0) -- (0,0);
\draw[thick,>=Latex,-<] (0,0) -- (45:1cm);
\draw[thick,>=Latex,-<] (0,0) -- (22.5:1cm);
\draw[thick,>=Latex,-<] (0,0) -- (-45:1cm);
\draw[thick,dotted]  (-30:0.8) arc(-30:10:0.8);
\draw[thick,->] (0,0) to[out=170,in=190] (-0.2,0.8);
\draw[thick] (0,0) to[out=55,in=10] (-0.2,0.8);
\end{tikzpicture}
}
}
\ + \
\vcenter{
\hbox{
\begin{tikzpicture}
\draw[thick,dashed,>=Latex,>-] (-1,0) -- (0,0);
\draw[thick,>=Latex,-<] (0,0) -- (45:1cm);
\draw[thick,>=Latex,-<] (0,0) -- (22.5:1cm);
\draw[thick,>=Latex,-<] (0,0) -- (-45:1cm);
\draw[thick,dotted]  (-30:0.8) arc(-30:10:0.8);
\draw[thick,->] (0,0) to[out=170,in=190] (-0.2,0.8);
\draw[thick] (0,0) to[out=55,in=10] (-0.2,0.8);
\draw[thick,->] (0,0) to[out=-170,in=-190] (-0.2,-0.8);
\draw[thick] (0,0) to[out=-55,in=-10] (-0.2,-0.8);
\end{tikzpicture}
}
}
= \biggl( 1 - \scrI + \frac{1}{2} \scrI^2 \biggr) V^{(0)}.
\end{equation}
Thus, up to two-loop order, the result of the loop calculation is to shift $\lambda$ in a way which is consistent across its appearance in the propagator and all of the interaction terms. However, again, this does not imply that $\lambda$ is the \emph{only} parameter that gets renormalized, as we'll flesh out in the next section.

The pattern is clear: In general, $V = I-L + I$ and so $(I_{\auxnew\field} )_{\rm max} = V = I-L+1$ just from topological considerations. However, the nonrenormalization theorem implies that $(I_{\auxnew\field} )_{\rm max}$ is one less than this, given that $(I_{\auxnew\field} )_{\rm max} = V = I-L$. Thus, the maximum superficial degree of divergence is
\begin{equation}
D_{\rm max} = 4(L-I) + 3 (I_{\auxnew\field} )_{\rm max} = L-I.
\end{equation}
Since $I \geq L$, the only possiblity is $I = L$, which is logarithmically divergent and has the structure of $L$ $\field$-$\field$ loops attached to one vertex. Therefore, to all-loop order, the shift of $\lambda$ reads
\begin{equation} \label{eq:fin}
\lambda \rightarrow \sum_{L=0}^{\infty} \frac{(-1)^L}{2^L L!} (2 \scrI)^{L} \lambda = e^{- \scrI} \lambda = \biggl( \frac{\Lambda}{\mu} \biggr)^{- \frac{\Delta}{4 \pi \diff}} \lambda,
\end{equation}
while there are no diagrammatic contributions to the $\Delta$ or $\kappa$ terms.


\subsection{Random Phase Approximation}
\label{sec:RPA}

Instead of amputating the corrections to the propagators, we can instead sum all of the one-particle-irreducible (1PI) diagrams and then multiply each by two copies of the bare propagator. The sum of the amputated 1PI diagrams is the analog of the self energy $\Sigma$. For example,
\begin{equation}
\Sigma_{\auxnew\field} \equiv \sum_{L=1}^{\infty} M_{\auxnew\field}^{(L)} = \sum_{L=1}^{\infty} \frac{(- \scrI )^L}{L!} i \lambda k_x = \bigl( e^{- \scrI} - 1 \bigr) i \lambda k_x = \biggl[ \biggl( \frac{\Lambda}{\mu} \biggr)^{- \frac{\Delta}{4 \pi \diff}} -1 \biggr] i \lambda k_x.
\end{equation}
The 1PI correction to the propagator, as opposed to the inverse propagator, is given by the above flanked by one factor of the bare propagator on either side. The Random Phase Approximation (RPA) simply chains any number of these in series and sums them all up. Thus, the renormalized propagator is given by:
\begin{align}
\calG_{\auxnew\field}^{\rm ren} &= \calG_{\auxnew\field}^{(0)} + \calG_{\auxnew\field}^{(0)} \Sigma_{\auxnew\field} \calG_{\auxnew\field}^{(0)} + \calG_{\auxnew\field}^{(0)} \Sigma_{\auxnew\field} \calG_{\auxnew\field}^{(0)} \Sigma_{\auxnew\field} \calG_{\auxnew\field}^{(0)} + \cdots \notag \\
&= \frac{\calG_{\auxnew\field}^{(0)}}{1 - \calG_{\auxnew\field}^{(0)} \Sigma_{\auxnew\field}} \notag \equiv \frac{1}{\bigl( \calG_{\auxnew\field}^{(0)} \bigr)^{-1} - \Sigma_{\auxnew\field}},
\end{align}
which simplifies to
\begin{equation}
\calG_{\auxnew\field}^{\rm ren} = \frac{1}{i ( \omega - \lambda k_x - i \diff k^2 ) - \bigl[ \bigl( \frac{\Lambda}{\mu} \bigr)^{- \frac{\Delta}{4 \pi \diff}} -1 \bigr] i \lambda k_x} = \frac{-i}{\omega - \bigl( \frac{\Lambda}{\mu} \bigr)^{- \frac{\Delta}{4 \pi \diff}} \lambda k_x - i \diff k^2}.
\end{equation}
Therefore, the renormalized propagator is the same as the original propagator with the $\lambda$ parameter shifted as in \eqref{eq:fin}. There are no diagrams with purely $\field$ internal loops that can contribute to $G_{\auxnew\auxnew}^{(0)} = 0$, and so that persists in vanishing identically.

The corrections to the $\field$-$\field$ propagator take the form of any number of corrections to the $\field$-$\auxnew$ propagator excluding the final bare $G_{\field\auxnew}^{(0)}$, which is instead replaced with a $G_{\field\field}^{(0)}$ insertion, followed by any number of corrections to the $\auxnew$-$\field$ propagator, excluding the initial bare $G_{\auxnew\field}^{(0)}$, schematically depicted below, where the red solid line is the $G_{\psi\psi}^{(0)}$ which replaces the $G_{\psi \phi}^{(0)}$, which would have been on the end of the left-hand portion of the diagram, and the $G_{\phi\psi}^{(0)}$, which would have been at the start of the right-hand portion of the diagram:
\begin{equation}
\vcenter{
\hbox{
\begin{tikzpicture}[scale=0.8]
\draw[thick,->] (-2,0) -- (-1,0);
\draw[thick,dashed] (-1,0) -- (0,0);
\draw[thick,->] (0,0) -- (1,0);
\draw[thick,->] (0,0) to[out=150,in=180] (0,1.8);
\draw[thick] (0,0) to[out=30,in=0] (0,1.8);
\draw[thick,dashed] (1,0) -- (2,0);
\draw[thick,loosely dotted] (2.2,0) -- (2.8,0);
\draw[thick,->] (3,0) -- (4,0);
\draw[thick,dashed] (4,0) -- (5,0);
\draw[thick,->] (5,0) to[out=150,in=180] (5,1.8);
\draw[thick] (5,0) to[out=30,in=0] (5,1.8);
\draw[thick,red,->] (5,0) -- (6,0);
\draw[thick,red] (6,0) -- (7,0);
\draw[thick,->] (7,0) to[out=150,in=180] (7,1.8);
\draw[thick] (7,0) to[out=30,in=0] (7,1.8);
\draw[thick,dashed,->] (7,0) -- (8,0);
\draw[thick] (8,0) -- (9,0);
\draw[thick,loosely dotted] (9.2,0) -- (9.8,0);
\draw[thick,dashed,->] (10,0) -- (11,0);
\draw[thick] (11,0) -- (12,0);
\draw[thick,->] (12,0) to[out=150,in=180] (12,1.8);
\draw[thick] (12,0) to[out=30,in=0] (12,1.8);
\draw[thick,dashed,->] (12,0) -- (13,0);
\draw[thick] (13,0) -- (14,0);
\end{tikzpicture}
}
}
\end{equation}
Again, there can be any number of RPA $G_{\field\auxnew}$ corrections on the left and RPA $G_{\auxnew\field}$ corrections on the right (and need not involve the same number on both sides).
Therefore, this evaluates to
\begin{equation} \label{eq:Gffren}
\calG_{\field\field}^{\rm ren} = \calG_{\field\auxnew}^{\rm ren} \bigl( \calG_{\field\auxnew}^{(0)} \bigr)^{-1} \calG_{\field\field}^{(0)} \bigl( \calG_{\auxnew\field}^{(0)} \bigr)^{-1} \calG_{\auxnew\field}^{\rm ren}.
\end{equation}
Using the fact that
\begin{equation}
\calG_{\field\field}^{(0)} = 2 \Delta \calG_{\field\auxnew}^{(0)} \calG_{\auxnew\field}^{(0)},
\end{equation}
\eqref{eq:Gffren} simplifies to
\begin{equation}
\calG_{\field\field}^{\rm ren} = 2 \Delta \calG_{\field\auxnew}^{\rm ren} \calG_{\auxnew\field}^{\rm ren},
\end{equation}
which is indeed precisely the same as the bare propagator with $\lambda$ shifted as in \eqref{eq:fin}. 

Once again, the complete all-loop calculation seems to na\"ively indicate that the only quantity that gets renormalized is $\lambda$. We finally turn towards this recurring point, along with the remaining details of the complete renormalization of the theory in the next section.


\section{Wilsonian Renormalization}
\label{sec:wilsonianRG}

In what follows, we adopt the Wilsonian perspective on renormalization with hard cutoff regularization\footnote{Given the scheme-independence of the coefficients of the leading logarithmic UV divergences, along with the results of the previous subsections, one can also arrive at the same conclusions by isolating the UV poles via dimensional regularization using the prescription detailed in Section \ref{sec:oneloopcorrections} and integrating the corresponding beta functions.}. We split the modes of the angle and response fields, $\field$ and $\phi$, into ``slow'' modes with momentum $< \frac{\Lambda}{b}$, where $\Lambda$ is some UV scale and $b>1$ is a real number, and ``fast'' modes with momentum between $\frac{\Lambda}{b}$ and $\Lambda$. We integrate out the fast modes, which means that in the result \eqref{eq:fin}, we set $\mu = \frac{\Lambda}{b}$ so that the factor involving the cutoff becomes
\begin{equation} \label{eq:b}
\biggl( \frac{\Lambda}{\mu} \biggr)^{- \frac{\Delta}{4 \pi \diff}} \xrightarrow{\mu \rightarrow \Lambda / b} b^{- \frac{\Delta}{4 \pi \diff}}.
\end{equation}
We then \emph{rescale} the momentum back up so as to restore the domain of the slow modes to be between 0 and $\Lambda$. In so doing, we have to scale all dimensionful quantities according to their \textit{bare} scaling dimension. The bare scaling dimensions have been determined in such a way as to force the action to be dimensionless, of course. 

Though we performed the loop calculation in momentum space, we can, of course, have just as readily worked in real space, especially since the result was so simple: The renormalized action can at first glance be brought into a form that looks exactly the same as the bare action but with $\lambda$ shifted according to \eqref{eq:fin}, or, using \eqref{eq:b}, according to
\begin{equation} \label{eq:r}
\lambda \rightarrow b^{-r} \lambda, \qquad\qquad r \equiv \frac{\Delta}{4 \pi \kappa}.
\end{equation}
Denoting renormalized quantities with an overline, the results of our loop computation to all orders imply
\begin{equation} \label{eq:Sbar}
\ovl{S} = - \int_{\bar{t}, \bx} \biggl\{ \auxnew ( \bar{t}, \bx ) \Bigl[ \partial_{\bar{t}} \ovl{\field} ( \bar{t}, \bx ) + b^{-r} \lambda \nabla \cdot \hbn_{\perp} ( \ovl{\field} ) - \diff \nabla^2 \ovl{\field} ( \bar{t}, \bx ) \Bigr] - \Delta \auxnew ( \bar{t}, \bx )^2 \biggr\},
\end{equation}
where we have anticipated the fact that the field $\field$, being an angle, must remain exactly dimensionless. Since renormalization is premised on the separation of low-\emph{energy} from high-\emph{energy} modes and integrating out the latter, the scaling of time must also be fixed: $\bar{t} = t$. In a Lorentz-invariant theory, that would imply that the scaling of the spatial coordinates is identical to that of time, and, thus, that it cannot change under renormalization either. However, this is no longer true in the absence of boost invariance, such as in a theory with non-trivial Lifshitz scaling. In fact, in such a theory, an anomalous dimension for spatial momentum is implied by the renormalization of the dynamical critical exponent, $z$. Were it not possible for spatial momentum to gain an anomalous dimension, it would likewise be impossible for $z$ to get renormalized. This is an important enough point that additional elaboration is warranted at this juncture.  

To go from the subtraction of ultraviolet divergences and fixing renormalization conditions towards inferring the phase diagram of a theory are distinct parts of the RG analysis. The absence of boost invariance matters because the boost Ward identity relating the temporal and spatial terms of the stochastic action can no longer be taken as a given. Flocks moving through a medium or on a substrate exhibit this lack of boost invariance, and independent kinetic renormalizations must therefore be admitted unless another valid identity relates them. Imposing that the temporal coordinate stay fixed so that a separation of energy scales can be cleanly be defined at every RG step does not furnish such an identity. For example, suppressing a uniform drift, we can rewrite the quadratic sector as
\begin{equation}
	S_2=\int dt\,d^2x\,
	\left[\phi\left(A_t\partial_t-A_x\nabla^2\right)\psi
	-N\phi^2\right].
	\label{eq:revision-quadratic-sector}
\end{equation}
Normalizing the time derivative by redefining the response field gives
\begin{equation}
	\kappa_R=\frac{A_x}{A_t},
	\qquad
	\Delta_R=\frac{N}{A_t^2},
	\label{eq:revision-relative-coefficients}
\end{equation}
before any additional rescaling of $\psi$. Thus the absence of a direct
correction to the noise vertex does not, on its own, establish the
absence of a correction to the normalized noise amplitude. The temporal
normalization must be included. More generally, fixing $t_B=t$ does not
forbid counterterms proportional to $\phi\partial_t\psi$. Such
counterterms can be absorbed by field (aka wave function) and coupling renormalizations without redefining the time coordinate \cite{Tauber2007}.

Time can also be chosen as providing the reference scaling for a pointwise dilation. In this convention, we define
\begin{equation}
	t\mapsto bt,
	\qquad
	\mathbf{x}\mapsto b^{\vartheta}\mathbf{x},
	\qquad
	\vartheta=\frac{1}{z}.
	\label{eq:revision-time-normalized-scaling}
\end{equation}
Where free diffusion corresponds to $\vartheta_0=1/2$. An interacting correction to $\vartheta$ expresses the relative anomalous scaling of space and time, and is not an additional exponent independent of $z$. If
longitudinal and transverse distances scale differently, the same convention gives weights $\zeta/z$ and $1/z$, respectively. This distinction is important when matching counterterms. A spatial coordinate factor and a spatial kinetic factor are not independent physical parameters. If $\mathbf{x}_B=Z_{\bx}\mathbf{x}$ and $\kappa_B=Z_\kappa\kappa$, the relative factor multiplying a second order spatial operator is
\begin{equation}
	Z_{\mathrm{rel}}=Z_\kappa Z_{\bx}^{-2},
	\qquad
	\gamma_{\mathrm{rel}}=\gamma_\kappa-2\gamma_{\bx},
	\label{eq:revision-coordinate-factor}
\end{equation}
where the logarithmic derivatives use the same RG convention. The
coordinate transformation also changes the measure and the
parametrization of the interactions. Redistribution of renormalization
among these factors is a normalization freedom (see also \cite{BenedettiGurauLettera2024}). The matching problem considered in the present context lays bare this freedom. Suppose the only direct logarithmic correction retained is the vertex correction $\lambda\mapsto b^{-r}\lambda$. The matching conditions, with $\overline{t}=t$ and $\overline{\psi}=\psi$, are
\begin{equation}
	\begin{aligned}
		\gamma_\phi&=-2\gamma_{\bx},
		&\gamma_\lambda&=\gamma_{\bx}-r,\\
		\gamma_\kappa&=2\gamma_{\bx},
		&\gamma_\Delta&=2\gamma_{\bx}.
	\end{aligned}
	\label{eq:revision-matching-conditions}
\end{equation}
Here the $\gamma$'s denote the anomalous dimensions separate from the classical scaling dimension. There are four equations for five factors, but the remaining freedom can be parameterized in terms of the spatial normalization. Were $\gamma_{\bx}=0$ for example -- as would be forced in the presence of boost invariance when $\bar t = t$ -- one obtains $\gamma_\lambda=-r$ with the anomalous dimensions also vanishing. On the other hand the combinations
\begin{equation}
	\gamma_\Delta-\gamma_\kappa=0,
	\qquad
	2\gamma_\lambda-\gamma_\kappa=-2r,
	\label{eq:revision-invariant-combinations}
\end{equation}
are independent of that choice. An arbitrary value of $\gamma_{\bx}$
in these matching equations is therefore not evidence for an
undetermined physical exponent.  The KPZ equation gives a useful comparison \cite{KardarParisiZhang1986}. For the usual additive white noise, its Galilean tilt symmetry provides the requisite Ward identity that relates the advective vertex to the temporal response. At a scaling fixed point with nonzero nonlinear coupling, this implies
\begin{equation}
	\alpha_{\mathrm{KPZ}}+z_{\mathrm{KPZ}}=2.
	\label{eq:revision-kpz-relation}
\end{equation}
The symmetry acts jointly on the height and the coordinates and is not a purely internal symmetry. Its Ward identities follow from the
stochastic functional, not from an assumed equality of coordinate
anomalous dimensions \cite{CanetEtAl2011}. The relation supplies
information that a general boost noninvariant theory need not possess.
It does not determine the full phase diagram by itself either. In particular,
the exact one dimensional KPZ exponents also use the stationary
measure, while the tilt relation alone does not determine both
exponents in higher dimensions \cite{CanetEtAl2011}.

For the flock, ordering must ultimately be tested through the
orientation correlation
\begin{equation}
	C_n(\mathbf{R})=
	\left\langle\hat{\mathbf{n}}(\mathbf{R},t)
	\mathbin{\cdot}\hat{\mathbf{n}}(\mathbf{0},t)\right\rangle
	=
	\operatorname{Re}\left\langle
	e^{i[\psi(\mathbf{R},t)-\psi(\mathbf{0},t)]}
	\right\rangle.
	\label{eq:revision-orientation-correlation}
\end{equation}
Long range order requires a nonzero limit at large separation.
Gaplessness alone does not establish this limit. Already in the
Gaussian smooth angle theory with equal time spectrum
$\Delta/(\kappa k^2)$, one obtains
\begin{align}
	\left\langle[\psi(\mathbf{R})-\psi(\mathbf{0})]^2\right\rangle
	&=\frac{\Delta}{\pi\kappa}\ln\frac{R}{a}+O(1),
	\label{eq:revision-gaussian-variance}\\
	C_n(R)&\propto
	\left(\frac{R}{a}\right)^{-\Delta/(2\pi\kappa)}.
	\label{eq:revision-gaussian-ordering}
\end{align}
where $a$ is a microscopic cutoff and $R=|\mathbf{R}|$.
This is a gapless theory without true long range order at positive
noise strength, although it tells us nothing of the interacting flock. Furthermore, the compact theory can additionally admit defects whose effects are not decided by a smooth fluctuation expansion \cite{KosterlitzThouless1973}. 

Therefore in the interacting theory, we are obliged to consider assigning anomalous dimensions  $\gamma_{\lambda}$, $\gamma_{\Delta}$, $\gamma_{\kappa}$, $\gamma_{\phi}$, and $\gamma_{\bx}$. That is, we define the renormalized parameters:
\begin{align}
\ovl{\lambda} &= b^{\gamma_{\lambda}} \lambda, &%
\ovl{\Delta} &= b^{\gamma_{\Delta}} \Delta, &%
\ovl{\kappa} &= b^{\gamma_{\kappa}} \kappa, &%
\ovl{\phi} &= b^{\gamma_{\phi}} \phi, &%
\ovl{\bx} &= b^{\gamma_{\bx}} \bx.
\end{align}
As stressed, the dimensions of time must remain unchanged, and $\psi$ must remain exactly dimensionless in order for the trigonometric interactions to maintain their form, which is a consequence of the shift-rotation subgroup of the symmetry discussed in Section \ref{sec:symm}. That is,
\begin{align}
\ovl{t} &= t, \qquad
\ovl{\psi} = \psi.
\end{align}
Inserting the above into the expression of the bare quantities in \eqref{eq:Sbar} in terms of renormalized quantities results in
\begin{align} \label{eq:Sbarren}
\ovl{S} &= - b^{-2 \gamma_{\bx} - \gamma_{\auxnew}} \int_{\bar{t}, \bar{\bx}} \ovl{\auxnew} ( \bar{t}, \bar{\bx} ) \Bigl[ \partial_{\bar{t}} \ovl{\field} ( \bar{t}, \bar{\bx} ) + b^{\gamma_{\bx} - \gamma_{\lambda} -r} \ovl{\lambda} \, \ovl{\nabla} \cdot \hbn_{\perp} ( \ovl{\field} ) - b^{2 \gamma_{\bx} - \gamma_{\diff}} \ovl{\diff} \ovl{\nabla}^2 \ovl{\field} ( \bar{t}, \bar{\bx} ) \Bigr] \notag \\
&\quad - b^{-2 \gamma_{\bx} - 2 \gamma_{\auxnew} - \gamma_{\Delta}} \ovl{\Delta} \int_{\bar{t}, \bar{\bx}} \ovl{\auxnew} ( \bar{t}, \bar{\bx} )^2 \biggr\},
\end{align}
which immediately implies the following relations
\begin{align} \label{eq:gammas}
\gamma_{\phi} &= - 2 \gamma_{\bx}, &%
\gamma_{\lambda} &= \gamma_{\bx} - r, &%
\gamma_{\kappa} &= 2 \gamma_{\bx}, &%
\gamma_{\Delta} &= 2 \gamma_{\bx}.
\end{align}
Note that consistency requires that $\Delta$ and $\kappa$ have the same anomalous dimension regardless of the anomalous dimension of space. Were this not the case, the ratio $r$ would have failed to remain dimensionless. 

We have only four equations for five unknowns: without a fifth relation, it is not possible to determine the exact values of the anomalous dimensions in the problem. Instead, we can write five of the anomalous dimensions in terms of one of them. The remaining freedom in Eq.~\eqref{eq:gammas} is a freedom in the assignment of anomalous scaling between the spatial coordinates and the diffusion coefficient. To express the resulting anomalous dimensions in terms of the dynamical critical exponent, we adopt the parametrization
\begin{equation}
\gamma_{\bx} = 1 - \frac{2}{z}.
\end{equation}
This relation translates between the chosen spatial-coordinate scaling convention and $z$; it is not an additional renormalization constraint that determines $z$. In particular, the matching conditions fix the combination $\gamma_\kappa-2\gamma_{\bx}$, rather than independently fixing $\gamma_{\kappa}$ and $\gamma_{\bx}$. The physical dynamical exponent characterizes the scaling of frequency with respect to physical spatial momentum and is invariant under a redistribution of renormalization between these two quantities. Since the matching conditions supply no further independent relation, the value of $z$, and hence the numerical values of the anomalous dimensions expressed in terms of it, remain undetermined. The anomalous dimensions expressed in terms of $z$ are
\begin{align} \label{eq:anomalousdims}
\gamma_{\auxnew} &= \frac{4}{z} - 2, &
\gamma_{\lambda} &= 1 - \frac{2}{z} - \frac{\Delta}{4 \pi \diff}, &
\gamma_{\diff} &= \gamma_{\Delta} = 2 - \frac{4}{z}.
\end{align}
Since the bare dimensions of $\auxnew$ and $\lambda$ are $2$ and $1$, respectively, their full dimensions are
\begin{align} \label{eq:lambdadim}
[ \auxnew ] &= \frac{4}{z}, &%
[ \lambda ] &= 2 - \frac{2}{z} - \frac{\Delta}{4 \pi \diff}.
\end{align}
We note that in \cite{CS1, inconvenienttruth}, the authors make the a priori reasonable distinction between the factor $\lambda$ that appears in the $\phi$-$\psi$ and $\psi$-$\phi$ propagators and the factor of $\lambda$ that appears in the interaction vertices. In \cite{inconvenienttruth}, for example, the former is called $\lambda_1$ and the latter is called $- \lambda_2$. The reason for this distinction is that although $\lambda_1 = - \lambda_2 = \lambda$ at the bare level, loop corrections will generically spoil this relation\footnote{The authors go on to argue that there can be no nontrivial fixed point with $\lambda_1 \neq 0$ and $\lambda_2 = 0$ or with $\lambda_1 = 0$ and $\lambda_2 \neq 0$. That is, any fixed point must have $\lambda_1 \neq 0$ and $\lambda_2 \neq 0$, which is consistent with our findings.}. The only manner in which this relation can be preserved is if it is somehow protected by symmetry. As demonstrated in Section \ref{sec:symm}, this protection is in fact provided by a non-linear realization of the one-parameter shift/rotation transformation in \eqref{eq:transfs} that forces relations between the various vertex factors at all loops, such that the recursion relations \eqref{eq:recursion} hold to all orders, and the trigonometric form of the advection interaction is invariant under RG up to a rescaling of $\lambda$. That is, all appearances of $\lambda$, whether in the propagators or in the vertex factors are forced to renormalize in the exact same manner because of this symmetry.

Therefore, indeed, it is true that there cannot be a nontrivial fixed point unless both $\lambda_1$ and $\lambda_2$ are nonzero, and in fact, $\lambda_1 = -\lambda_2 \equiv \lambda$. However, it is not possible to further specify the precise value of $\lambda$ at any fixed point or collection of fixed points unless one specifies some quantity that depends on $\lambda$ at some particular scale, $\bar \mu$ to fix the value of $\lambda$ at that scale as a renormalization condition that subsequently allows us to infer the value of $\lambda$ at any other scale, $\mu$. Doing so, results in the scale-dependent coupling
\begin{equation}
\label{eq:RGE}
\lambda ( \mu ) = \biggl( \frac{\bar\mu}{\mu} \biggr)^{2 - \frac{2}{z} - \frac{\Delta}{4 \pi \diff}} \lambda ( \bar \mu ).
\end{equation}
Which can be viewed as the result of having integrated the corresponding beta function:
\begin{equation}
	\mu \frac{d \lambda}{d\mu} = -\biggl( 2 - \frac{2}{z} - \frac{\Delta}{4 \pi \diff} \biggr) \lambda.
\end{equation}
It is more typical to express RG running in terms of an RG scale $s$, which flows from the UV to the IR (thus encapsulating that it is a monotonic flow resulting from successive coarse graining over short distance modes), and so \textit{increases} as $\mu$ decreases. Hence, 
\begin{equation}
	\frac{d \lambda}{ds} = \biggl( 2 - \frac{2}{z} - \frac{\Delta}{4 \pi \diff} \biggr) \lambda,
\end{equation}
where $s = \log(\bar \mu/\mu)$. We stress in closing this appendix that common to all field-theoretic computations, there is a degree of arbitrariness with respect to field redefinitions (or changes of variable such that the transformed variable also represents a sufficient statistic) that will propagate onto how individual couplings that parameterize individual operators in the action renormalize. All of this arbitrariness will drop out of physical observables, which is the content of the equivalence theorem for on-shell S-matrix elements in quantum field theory \cite{Burgess:2020tbq, Peskin:1995ev}, but is much more simply the invariance of the measure and generating functional for connected correlation functions in statistical field theories up to a suitable redefinition of the source \cite{heliasdahmen, tauber2014critical}. 


\section{Dimension of the Director Field}

The form of the renormalization of this theory is clearly very special. Only the convective derivative term, with its coupling $\lambda$, gets diagrammatic corrections. The coupling $\lambda$ is already present even at the free-theory level. Furthermore, $\lambda$ must be renormalized equally across all of the interaction terms. These points indicate that it should be possible to interpret the result of the all-order loop calculation as a purely classical effect from the perpsective of the composite operator $\hbn_{\perp}$, since the convective derivative is simply $\lambda \nabla \cdot \hbn_{\perp}$. Indeed, the factor $r = \frac{\Delta}{4 \pi \diff}$, which is the result of the all-order loop calculation is precisely the classical scaling dimension of the composite operator $\hbn_{\perp}$. This is a simple corollary of the well-known result that even if $\field$ has vanishing classical scaling dimension, composite operators built out of $\field$, such as $e^{i \alpha \psi}$, famously can have non-vanishing classical scaling dimension. Let us review how this works.

Take the bare $\field$-$\field$ propagator and integrate it over frequency to isolate the momentum-dependence:
\begin{equation} \label{eq:static2pt}
\expect{\field_{- \bk} \field_{\bk}}_0 \equiv \int_{- \infty}^{\infty} \frac{d \omega}{2 \pi} \expect{\field_{- \omega , - \bk} \, \field_{\omega, \bk}}_0 = \int_{- \infty}^{\infty} \frac{d \omega}{2 \pi} \frac{2 \Delta}{( \omega - \lambda k_x )^2 + \diff^2 k^4} = \frac{\Delta}{\diff k^2}.
\end{equation}
At the Gaussian level, Wick's theorem gives 
\begin{equation}
\expect{e^{i \alpha \field ( \bx )} e^{-i \alpha \field (\by )}}_0 = \exp \biggl[ - \frac{\alpha^2}{2} \expect{[ \field ( \bx ) - \field ( \by )]^2}_0 \biggr].
\end{equation}
The expectation value in the above exponential is simply evaluated by the appropriate inverse Fourier transform of Eq.~\eqref{eq:static2pt}:
\begin{equation}
\expect{[ \field ( \bx ) - \field ( \by )]^2}_0 = 2 \int \frac{d^2 k}{( 2 \pi )^2} \, \frac{\Delta}{\diff k^2} \biggl( 1 - e^{i \bk \cdot ( \bx - \by )} \biggr).
\end{equation}
We integrate $k = | \bk |$ up to a UV cut-off $\Lambda$. There is no need for an IR cut-off because one is naturally provided by $| \bx - \by |$. The result is
\begin{equation}
\expect{[ \field ( \bx ) - \field ( \by )]^2}_0 = \frac{\Delta}{\pi \diff} \ln \bigl( \Lambda | \bx - \by | \bigr).
\end{equation}
Therefore,
\begin{equation}
\expect{e^{i \alpha \field ( \bx )} e^{-i \alpha \field (\by )}}_0 = \bigl( \Lambda | \bx - \by | \bigr)^{- \frac{\alpha^2 \Delta}{2 \pi \diff}},
\end{equation}
and we conclude that the scaling dimension of the composite operator $e^{i \alpha \psi}$ is
\begin{equation}
[ e^{i \alpha \psi} ] = \frac{\alpha^2 \Delta}{4 \pi \diff}.
\end{equation}
In particular, the scaling dimension of $\hbn$ and of $\hbn_{\perp}$ is
\begin{equation}
[ \hbn ] = [ \hbn_{\perp} ] = \frac{\Delta}{4 \pi \diff}.
\end{equation}
We make one more important comment regarding this dimension, which depends on the result of Sec.~\ref{sec:RPA} that the diagrams only contribute to the $\lambda$ term in the propagators and simply multiplicatively. Since the $\lambda$ term is simply absorbed by a redefinition of the integration variable $\omega$ in Eq.~\eqref{eq:static2pt} of the above derivation, we find that the loop calculation does not affect the dimension of the director field at all.


\section{\label{sec:comp} Comparisons with Previous Treatments of Flock Scaling}

The recent literature presents the interested reader with differing viewpoints regarding the extent to which symmetry arguments constrain the scaling behavior of two-dimensional polar flocks \cite{CS1, Ikeda2024VicsekNG, inconvenienttruth, comment1, response1, comment2}. Since our treatment deploys an explicit diagrammatic renormalization procedure, it is useful to summarize the assumptions underlying the approaches of \cite{CS1, comment1, comment2} relative to \cite{inconvenienttruth, response1}, and to clarify where and how our conclusions might differ. In addition, while the debate between the two aforementioned sets of authors has dominated the discussion in this area, a notable alternative viewpoint has been proposed by Amoretti et al. \cite{Amoretti:2024obt}, which utilizes boost-agnostic hydrodynamics and thermodynamic constraints to derive a different set of relations among the critical exponents.


\subsection{Chat\'e--Solon: Symmetry-Based Non-Renormalization Arguments}

In \cite{CS1}, Chat\'e and Solon analyze the Goldstone-mode dynamics of the ordered phase and argue that certain nonlinear couplings do not get renormalized. In particular, immediately following their Eq.~(10), they state that the terms originating from the convective derivative are not expected to receive graphical corrections. In their formulation, these terms are proportional to the parameter $\lambda$, and the argument is based on the generalized Galilean invariance introduced in their Appendix B. Our findings lead us to a different conclusion. Our explicit diagrammatic analysis finds nonvanishing renormalization of the couplings associated with the convective derivative. Consequently, the non-renormalization of these couplings does not appear to follow from the symmetries realized in our formulation, which moreover, protects their functional form under loop corrections. It is worth noting that the generalized Galilean invariance invoked in Ref.~\cite{CS1} is derived within a theory truncated at a particular order in the expansions of $\sin\psi$ and $\cos\psi$ and is evidently not a symmetry of the full non-linear theory, as elaborated upon in Section \ref{sec:symm}. Whether an analogous symmetry survives in a more general treatment is a separate question. However, regardless of its interpretation, our diagrammatic calculation indicates that the corresponding couplings acquire nonzero anomalous dimensions.

A second point concerns the variance of the noise $\Delta$. Chat\'e and Solon argue that because all interactions are derivatively coupled, $\Delta$ receives no graphical corrections and therefore acquires no anomalous dimension. We agree with the first statement: in our calculation, $\Delta$ indeed receives no direct diagrammatic corrections. However, we arrive at a different conclusion regarding the second statement. Even in the absence of direct graphical corrections, the scaling dimension of $\Delta$ can be modified through its coupling to other running parameters, including the nonlinear couplings, and this is due to certain subtleties of the renormalization procedure in boost non-invariant theories with non-trivial Lifshitz scaling (i.e. with $z\neq 1$), as discussed in Appendix \ref{sec:wilsonianRG}. Thus, the absence of explicit diagrams contributing to $\Delta$ does not, by itself, imply that it does not acquire an anomalous dimension.

These assumptions lead directly to the exponent relations proposed in \cite{CS1}. For Malthusian flocks, Chat\'e and Solon derive three relations among the scaling exponents $(\chi,\zeta,z)$, from which they obtain exact exponent values. Their first relation,
\begin{equation}
	2\chi+z-\zeta=0,
\end{equation}
follows from the claim that the coupling $\lambda_x$ multiplying the interaction $\lambda_x\partial_x(\psi^3)$ receives no anomalous dimension. Their second relation,
\begin{equation} \label{eq:CSlxrel}
	\chi+z-1=0,
\end{equation}
follows from the analogous claim for the coupling $\lambda_y$ multiplying $\lambda_y\partial_y\psi^2$. In both cases, the justification is the generalized Galilean invariance discussed in Appendix B of Ref.~\cite{CS1}. Their third relation,
\begin{equation} \label{eq:CSDeltarel}
	z-2\chi-1-\zeta=0,
\end{equation}
follows from the assumption that $\Delta$ acquires no anomalous dimension because no diagrams contribute directly to its renormalization. Thus, they derive the critical exponents
\begin{align}
z &= \frac{5}{4}, &%
\zeta &= \frac{3}{4}, &%
\chi &= - \frac{1}{4},
\end{align}
with which we clearly differ.

The generalized Galilean invariance employed in Ref.~\cite{CS1} is expressed through their Eq.~(B1),
\begin{equation}
	\label{css1}
	\psi'(\mathbf r,t)=\psi(\mathbf r-\mathbf r_c,t)+\psi_0,
\end{equation}
with
\begin{equation}
	\label{css2}
	\partial_t x_c = 2\lambda_x\psi_0,
	\qquad
	\partial_t y_c = \lambda_y\psi_0.
\end{equation}
Since the symmetry structure adopted in the present work differs from that used in Ref.~\cite{CS1}, and moreover, is valid for the full non-linear theory, it is perhaps unsurprising that our renormalization analysis does not support the resulting non-renormalization conditions for $\lambda_x$ and $\lambda_y$.\footnote{After the first version of our preprint came out, the study \cite{jentsch2026truequasilongrangeorder} appeared supporting the findings of Chat\'e-Solon. Our conclusions differ for reasons elaborated upon in the main body of our paper -- namely, their treatment implicitly assumes an additional relation that we cannot find in the boost non-invariant context in which we are working.}


\subsection{Chen \emph{et al.}: Limitations of Symmetry Constraints and Dynamical RG}

In \cite{inconvenienttruth}, the authors revisit the derivation of exact exponent relations. Their principal conclusion is that, for Malthusian flocks, only two independent exponent relations can be established. Consequently, they argue that the three critical exponents cannot be determined exactly. An important point is that the disagreement between Chen \emph{et al.} and Chat\'e--Solon is narrower than it may initially appear. Both groups agree on two exponent relations. Indeed, one of the relations obtained by Chen \emph{et al.} is equivalent to the difference between the first two equations in Eq.~(11) of \cite{CS1},
\begin{equation}
	\chi-\zeta+1=0.
\end{equation}
The primary disagreement concerns the existence of a third independent relation. Chen \emph{et al.} also accept that $\Delta$ receives no direct graphical corrections and therefore recover the relation
\begin{equation} \label{eq:ChenDeltarel}
	z-2\chi-1-\zeta=0
\end{equation}
using essentially the same reasoning as Chat\'e and Solon. Their departure from \cite{CS1} instead concerns the nonlinear couplings. Rather than arguing that $\lambda_x$ and $\lambda_y$ have vanishing anomalous dimensions, Chen \emph{et al.} argue only that their anomalous dimensions must be equal because both couplings originate from the same underlying parameter $\lambda$. Equating the anomalous dimensions of $\lambda_x$ and $\lambda_y$ yields a single exponent relation, whereas setting both anomalous dimensions individually to zero (as done in \cite{CS1}) yields two. This difference is ultimately responsible for the differing conclusions regarding the exact determination of the Malthusian-flock exponents.

For immortal flocks, the disagreement becomes more substantial. Chen \emph{et al.} argue that the currently known symmetries of flocking theories are insufficient to derive any exact relations among the scaling exponents. Consequently, they conclude that no exact exponent values can presently be inferred from symmetry considerations alone. Chat\'e and Solon, while not explicitly deriving exact exponent relations for immortal flocks, argue that the renormalization-group treatment of Chen \emph{et al.} does not establish such a no-go theorem and therefore does not exclude the possibility that additional symmetry-based constraints may exist.


\subsection{Amoretti et al.: Leveraging Thermodynamic Relations}

In \cite{Amoretti:2024obt}, the authors make use of thermodynamic relations among the various constants appearing in the hydrodynamic equations (the generalized continuity equation and Euler equation, corresponding to their equations 13a and 13b, respectively) to derive two critical-exponent relations, which are distinct from those of the previous two sets of authors. This argument proceeds by first relating all of these coefficients to parameters and transport coefficients of boost-agnostic hydrodynamics and then claiming that these relations should be preserved under RG. The relevant constants split into two sets of like scaling dimension and setting these dimensions to 0 gives the two relations
\begin{align}
z - 1 + \chi &= 0, &%
z - \zeta + \chi &= 0.
\end{align}
By the same argument as in the other works, namely that $\Delta$ gains no anomous dimenion, these authors again get the relation
\begin{equation}
z - 2 \chi + 1 - d - \zeta = 0,
\end{equation}
where $d$ is the number of spatial dimensions. This relation reduces to Eq.~\eqref{eq:CSDeltarel} and \eqref{eq:ChenDeltarel} of the previous two works when $d=2$. Thus, their critical exponents are
\begin{align}
z &= \frac{4}{3}, &%
\zeta &= 1, &%
\chi = - \frac{1}{3}.
\end{align}
We agree with $\zeta = 1$, but not with the other two critical exponents.

It should be noted that this work utilizes the generalized continuity equation. That is the evolution equation for density perturbations, which should be ignored for Malthusian flocks. In fact, five out of the eight coefficients considered in this work involve density fluctuations and should be removed in the Malthusian theory. Of the eight coefficients that they consider, only two coefficients of dimension $z-1 + \chi$ (their $g_4$ and $g_5$) and one coefficient of dimension $z- \zeta + \chi$ (their $w_3$) do not multiply terms involving the density fluctuation. However, $w_3$ is a coefficient appearing in the evolution equation of the density fluctuation. Therefore, if that is to be dropped altogether for Malthusian flocks, then only their relation $z -1 + \chi = 0$ would remain.  Note that this would be consistent with Eq.~\eqref{eq:CSlxrel} of Chat\'e and Solon, though it would lack a third equation to completely determine the values of the critical exponents.


\subsection{Relation to this Work}

The present analysis differs from these three approaches in that it is based on a different set of symmetries and field redundancies as those considered in \cite{CS1} and a different set of renormalization conditions, particularly as pertains to the $\Delta$ parameter, as in all three approaches. We back up the RG predictions derived from the associated Ward identities with an explicit diagrammatic renormalization procedure. Of course, we cannot claim to arrive at any firm conclusions beyond the confines of the very symmetric limit in which we have worked, and much more work remains to be done before we can address any of the relevant questions away from this limit. 



\bibliography{flocking_RG_bib}{}

@ARTICLE{CS1,
       author = {{Chat{\'e}}, Hugues and {Solon}, Alexandre},
        title = "{Dynamic Scaling of Two-Dimensional Polar Flocks}",
      journal = {\prl},
         year = 2024,
        month = jun,
       volume = {132},
       number = {26},
          eid = {268302},
        pages = {268302},
          doi = {10.1103/PhysRevLett.132.268302},
archivePrefix = {arXiv},
       eprint = {2403.03804},
 primaryClass = {cond-mat.stat-mech},
       adsurl = {https://ui.adsabs.harvard.edu/abs/2024PhRvL.132z8302C}
}

@article{Ikeda2024VicsekNG,
	author  = {Harukuni Ikeda},
	title   = {Minimum Scaling Model and Exact Exponents for the {N}ambu-{G}oldstone Modes in the {V}icsek Model},
	journal = {Physical Review Letters},
	volume   = {133},
	number   = {25},
	pages    = {258301},
	year     = {2024},
	doi      = {10.1103/PhysRevLett.133.258301},
	publisher = {American Physical Society}
}

@article{TonerTu1998,
	author  = {Toner, John and Tu, Yuhai},
	title   = {Flocks, Herds, and Schools: A Quantitative Theory of Flocking},
	journal = {Physical Review E},
	volume  = {58},
	number  = {4},
	pages   = {4828--4858},
	year    = {1998},
	doi     = {10.1103/PhysRevE.58.4828}
}

@article{TonerTuRamaswamy2005,
	author  = {Toner, John and Tu, Yuhai and Ramaswamy, Sriram},
	title   = {Hydrodynamics and Phases of Flocks},
	journal = {Annals of Physics},
	volume  = {318},
	number  = {1},
	pages   = {170--244},
	year    = {2005},
	doi     = {10.1016/j.aop.2005.04.011}
}

@article{Vicsek1995,
	author  = {Vicsek, Tam{\'a}s and Czir{\'o}k, Andras and Ben-Jacob, Eshel
	and Cohen, Inon and Shochet, Ofer},
	title   = {Novel Type of Phase Transition in a System of Self-Driven Particles},
	journal = {Physical Review Letters},
	volume  = {75},
	number  = {6},
	pages   = {1226--1229},
	year    = {1995},
	doi     = {10.1103/PhysRevLett.75.1226}
}

@article{Marchetti2013,
	author  = {Marchetti, M. Cristina and Joanny, Jean-Fran{\c c}ois and
	Ramaswamy, Sriram and Liverpool, Tanniemola B. and
	Prost, Jacques and Rao, Madan and Simha, R. Aditi},
	title   = {Hydrodynamics of Soft Active Matter},
	journal = {Reviews of Modern Physics},
	volume  = {85},
	number  = {3},
	pages   = {1143--1189},
	year    = {2013},
	doi     = {10.1103/RevModPhys.85.1143}
}

@article{VicsekZafeiris2012,
	author  = {Vicsek, Tam{\'a}s and Zafeiris, Anna},
	title   = {Collective Motion},
	journal = {Physics Reports},
	volume  = {517},
	number  = {3--4},
	pages   = {71--140},
	year    = {2012},
	doi     = {10.1016/j.physrep.2012.03.004}
}

@article{GinelliChate2010,
	author={Ginelli, Francesco and Chate, Hugues},
	title={Relevance of Metric-Free Interactions in Flocking Phenomena},
	journal={Phys. Rev. Lett.},
	volume={105},
	pages={168103},
	year={2010}
}

@article{SolonTailleur2013,
	author={Solon, A. P. and Tailleur, J.},
	title={Revisiting the Flocking Transition Using Active Spins},
	journal={Phys. Rev. Lett.},
	volume={111},
	pages={078101},
	year={2013}
}

@article{SolonTailleur2015,
	author={Solon, A. P. and Tailleur, J.},
	title={The Two-Dimensional Active Ising Model},
	journal={Phys. Rev. E},
	volume={92},
	pages={042119},
	year={2015}
}

@article{Solon2022,
       author = {{Solon}, Alexandre and {Chat{\'e}}, Hugues and {Toner}, John and {Tailleur}, Julien},
        title = "{Susceptibility of Polar Flocks to Spatial Anisotropy}",
      journal = {\prl},
         year = 2022,
        month = may,
       volume = {128},
       number = {20},
          eid = {208004},
        pages = {208004},
          doi = {10.1103/PhysRevLett.128.208004},
archivePrefix = {arXiv},
       eprint = {2201.00704},
 primaryClass = {cond-mat.stat-mech},
       adsurl = {https://ui.adsabs.harvard.edu/abs/2022PhRvL.128t8004S}
}

@article{Karmakar2024,
	author={{Mintu Karmakar} and {Swarnajit Chatterjee} and {Raja Paul} and {Heiko Rieger}},
	title={Consequence of anisotropy on flocking: the discretized {V}icsek model},
	journal={New J. Phys.},
    volume={26},
    pages={043023},
	year={2024}
}

@BOOK{1998qft,
	author = {{Huang}, Kerson},
	title = "{Quantum Field Theory: From Operators to Path Integrals}",
	year = 1998,
	adsurl = {https://ui.adsabs.harvard.edu/abs/1998qftf.book.....H}
}

@book{Peskin:1995ev,
	author = "Peskin, Michael E. and Schroeder, Daniel V.",
	title = "{An Introduction to quantum field theory}",
	doi = "10.1201/9780429503559",
	isbn = "978-0-201-50397-5, 978-0-429-50355-9, 978-0-429-49417-8",
	publisher = "Addison-Wesley",
	address = "Reading, USA",
	year = "1995"
}

@article{Negro:2024bbf,
	author = "Negro, Anna and Patil, Subodh P.",
	title = "{An {\'E}tude on the regularization and renormalization of divergences in primordial observables}",
	eprint = "2402.10008",
	archivePrefix = "arXiv",
	primaryClass = "hep-th",
	doi = "10.1007/s40766-024-00053-0",
	journal = "Riv. Nuovo Cim.",
	volume = "47",
	number = "3",
	pages = "179--228",
	year = "2024"
}

@article{deBoer:2017ing,
	author = "de Boer, Jan and Hartong, Jelle and Obers, Niels A. and Sybesma, Watse and Vandoren, Stefan",
	title = "{Perfect Fluids}",
	eprint = "1710.04708",
	archivePrefix = "arXiv",
	primaryClass = "hep-th",
	doi = "10.21468/SciPostPhys.5.1.003",
	journal = "SciPost Phys.",
	volume = "5",
	number = "1",
	pages = "003",
	year = "2018"
}

@article{Grosvenor:2024vcn,
	author = "Grosvenor, Kevin T. and Obers, Niels A. and Patil, Subodh P.",
	title = "{Hydrodynamics without boost-invariance from kinetic theory: From perfect fluids to active flocks}",
	eprint = "2501.00025",
	archivePrefix = "arXiv",
	primaryClass = "hep-th",
	reportNumber = "NORDITA 2024-051",
	doi = "10.21468/SciPostPhys.19.3.071",
	journal = "SciPost Phys.",
	volume = "19",
	number = "3",
	pages = "071",
	year = "2025"
}

@article{Nicolis:2008in,
	author = "Nicolis, Alberto and Rattazzi, Riccardo and Trincherini, Enrico",
	title = "{The Galileon as a local modification of gravity}",
	eprint = "0811.2197",
	archivePrefix = "arXiv",
	primaryClass = "hep-th",
	doi = "10.1103/PhysRevD.79.064036",
	journal = "Phys. Rev. D",
	volume = "79",
	pages = "064036",
	year = "2009"
}

@article{Nicolis:2015sra,
	author = "Nicolis, Alberto and Penco, Riccardo and Piazza, Federico and Rattazzi, Riccardo",
	title = "{Zoology of condensed matter: Framids, ordinary stuff, extra-ordinary stuff}",
	eprint = "1501.03845",
	archivePrefix = "arXiv",
	primaryClass = "hep-th",
	doi = "10.1007/JHEP06(2015)155",
	journal = "JHEP",
	volume = "06",
	pages = "155",
	year = "2015"
}

@article{kardar1986dynamic,
	title={Dynamic scaling of growing interfaces},
	author={Kardar, Mehran and Parisi, Giorgio and Zhang, Yi-Cheng},
	journal={Physical Review Letters},
	volume={56},
	number={9},
	pages={889},
	year={1986},
	publisher={APS}
}

@book{tauber2014critical,
	title={Critical dynamics: a field theory approach to equilibrium and non-equilibrium scaling behavior},
	author={T{\"a}uber, Uwe C},
	year={2014},
	publisher={Cambridge University Press}
}

@article{Tauber2007,
	author = {T{\"a}uber, Uwe C.},
	title = {Field Theory Approaches to Nonequilibrium Dynamics},
	journal = {Lecture Notes in Physics},
	volume = {716},
	pages = {295--348},
	year = {2007},
	doi = {10.1007/3-540-69684-9_7},
	eprint = {cond-mat/0511743},
	archivePrefix = {arXiv}
}

@article{CanetEtAl2011,
	author = {Canet, L{\'e}onie and Chat{\'e}, Hugues
	and Delamotte, Bertrand and Wschebor, Nicol{\'a}s},
	title = {Nonperturbative renormalization group for the
	{Kardar-Parisi-Zhang} equation: General framework and
	first applications},
	journal = {Physical Review E},
	volume = {84},
	pages = {061128},
	year = {2011},
	doi = {10.1103/PhysRevE.84.061128},
	eprint = {1107.2289},
	archivePrefix = {arXiv},
	note = {Erratum: Physical Review E 86, 019904 (2012)}
}

@article{BenedettiGurauLettera2024,
	author = {Benedetti, Dario and Gurau, Razvan and Lettera, Davide},
	title = {Dynamic critical exponent in quantum long-range models},
	journal = {Physical Review B},
	volume = {110},
	pages = {104102},
	year = {2024},
	doi = {10.1103/PhysRevB.110.104102},
	eprint = {2404.13963},
	archivePrefix = {arXiv}
}

@article{Armas:2024iuy,
	author = "Armas, Jay and Jain, Akash and Lier, Ruben",
	title = "{Hydrodynamics of thermal active matter}",
	eprint = "2405.11023",
	archivePrefix = "arXiv",
	primaryClass = "cond-mat.soft",
	doi = "10.1103/yclc-jmcj",
	journal = "Phys. Rev. E",
	volume = "112",
	number = "5",
	pages = "055401",
	year = "2025"
}

@article{KardarParisiZhang1986,
	author = {Kardar, Mehran and Parisi, Giorgio and Zhang, Yi-Cheng},
	title = {Dynamic Scaling of Growing Interfaces},
	journal = {Physical Review Letters},
	volume = {56},
	pages = {889--892},
	year = {1986},
	doi = {10.1103/PhysRevLett.56.889}
}

@book{Burgess:2020tbq,
	author = "Burgess, C. P.",
	title = "{Introduction to Effective Field Theory}",
	doi = "10.1017/9781139048040",
	isbn = "978-1-139-04804-0, 978-0-521-19547-8",
	publisher = "Cambridge University Press",
	month = "12",
	year = "2020"
}

@ARTICLE{inconvenienttruth,
       author = {{Chen}, Leiming and {Jentsch}, Patrick and {Lee}, Chiu Fan and {Maitra}, Ananyo and {Ramaswamy}, Sriram and {Toner}, John},
        title = "{The inconvenient truth about flocks}",
      journal = {arXiv e-prints},
         year = 2025,
        month = mar,
          doi = {10.48550/arXiv.2503.17064},
archivePrefix = {arXiv},
       eprint = {2503.17064},
 primaryClass = {cond-mat.soft},
       adsurl = {https://ui.adsabs.harvard.edu/abs/2025arXiv250317064C}
}

@ARTICLE{comment1,
       author = {{Chat{\'e}}, Hugues and {Solon}, Alexandre},
        title = "{Comment on ``The inconvenient truth about flocks'' by Chen et al}",
      journal = {arXiv e-prints},
         year = 2025,
        month = apr,
          eid = {arXiv:2504.13683},
        pages = {arXiv:2504.13683},
          doi = {10.48550/arXiv.2504.13683},
archivePrefix = {arXiv},
       eprint = {2504.13683},
 primaryClass = {cond-mat.soft},
       adsurl = {https://ui.adsabs.harvard.edu/abs/2025arXiv250413683C}
}

@ARTICLE{response1,
       author = {{Chen}, Leiming and {Jentsch}, Patrick and {Lee}, Chiu Fan and {Maitra}, Ananyo and {Ramaswamy}, Sriram and {Toner}, John},
        title = "{Response to the comment on The inconvenient truth about flocks by Chat{\'e} and Solon}",
      journal = {arXiv e-prints},
         year = 2025,
        month = may,
          eid = {arXiv:2505.21602},
        pages = {arXiv:2505.21602},
          doi = {10.48550/arXiv.2505.21602},
archivePrefix = {arXiv},
       eprint = {2505.21602},
 primaryClass = {cond-mat.soft},
       adsurl = {https://ui.adsabs.harvard.edu/abs/2025arXiv250521602C}
}

@ARTICLE{comment2,
       author = {{Chat{\'e}}, Hugues and {Solon}, Alexandre},
        title = "{Continuing the discussion on ``The inconvenient truth about flocks'' by Chen et al}",
      journal = {arXiv e-prints},
         year = 2025,
        month = jun,
          eid = {arXiv:2506.13437},
        pages = {arXiv:2506.13437},
          doi = {10.48550/arXiv.2506.13437},
archivePrefix = {arXiv},
       eprint = {2506.13437},
 primaryClass = {cond-mat.stat-mech},
       adsurl = {https://ui.adsabs.harvard.edu/abs/2025arXiv250613437C}
}

@article{tonertu1,
  title = {Long-Range Order in a Two-Dimensional Dynamical $\mathrm{XY}$ Model: How Birds Fly Together},
  author = {Toner, John and Tu, Yuhai},
  journal = {Phys. Rev. Lett.},
  volume = {75},
  issue = {23},
  pages = {4326--4329},
  numpages = {0},
  year = {1995},
  month = {Dec},
  publisher = {American Physical Society},
  doi = {10.1103/PhysRevLett.75.4326},
  url = {https://link.aps.org/doi/10.1103/PhysRevLett.75.4326},
  eprint = {arXiv:adap-org/9506001}
}

@article{QAssess,
  title = {Quantitative Assessment of the {T}oner and {T}u Theory of Polar Flocks},
  author = {Mahault, Beno\^{\i}t and Ginelli, Francesco and Chat\'e, Hugues},
  journal = {Phys. Rev. Lett.},
  volume = {123},
  issue = {21},
  pages = {218001},
  numpages = {6},
  year = {2019},
  month = {Nov},
  publisher = {American Physical Society},
  doi = {10.1103/PhysRevLett.123.218001},
  url = {https://link.aps.org/doi/10.1103/PhysRevLett.123.218001}
}

@article{PhysRevA.8.423,
  title = {Statistical Dynamics of Classical Systems},
  author = {Martin, P. C. and Siggia, E. D. and Rose, H. A.},
  journal = {Phys. Rev. A},
  volume = {8},
  issue = {1},
  pages = {423--437},
  numpages = {0},
  year = {1973},
  month = {Jul},
  publisher = {American Physical Society},
  doi = {10.1103/PhysRevA.8.423},
  url = {https://link.aps.org/doi/10.1103/PhysRevA.8.423}
}

@article{Janssen:1976qag,
    author = "Janssen, Hans-Karl",
    title = "{On a Lagrangean for classical field dynamics and renormalization group calculations of dynamical critical properties}",
    doi = "10.1007/BF01316547",
    journal = "Z. Phys. B",
    volume = "23",
    number = "4",
    pages = "377--380",
    year = "1976"
}

@article{PhysRevB.18.4913,
  title = {Dynamics as a substitute for replicas in systems with quenched random impurities},
  author = {De Dominicis, C.},
  journal = {Phys. Rev. B},
  volume = {18},
  issue = {9},
  pages = {4913--4919},
  numpages = {0},
  year = {1978},
  month = {Nov},
  publisher = {American Physical Society},
  doi = {10.1103/PhysRevB.18.4913},
  url = {https://link.aps.org/doi/10.1103/PhysRevB.18.4913}
}

@BOOK{heliasdahmen,
       author = {{Helias}, Moritz and {Dahmen}, David},
        title = "{Statistical Field Theory for Neural Networks}",
         year = 2020,
       volume = {970},
          doi = {10.1007/978-3-030-46444-8},
       adsurl = {https://ui.adsabs.harvard.edu/abs/2020LNP...970.....H}
}

@article{Grosvenor:2021eol,
    author = "Grosvenor, Kevin T. and Jefferson, Ro",
    title = "{The edge of chaos: quantum field theory and deep neural networks}",
    eprint = "2109.13247",
    archivePrefix = "arXiv",
    primaryClass = "hep-th",
    doi = "10.21468/SciPostPhys.12.3.081",
    journal = "SciPost Phys.",
    volume = "12",
    number = "3",
    pages = "081",
    year = "2022"
}

@article{Isakov:2010he,
    author = "Isakov, S. V. and Fendley, P. and Ludwig, A. W. W. and Trebst, S. and Troyer, M.",
    title = "{Dynamics at and near conformal quantum critical points}",
    eprint = "1012.3806",
    archivePrefix = "arXiv",
    primaryClass = "cond-mat.str-el",
    doi = "10.1103/PhysRevB.83.125114",
    journal = "Phys. Rev. B",
    volume = "83",
    pages = "125114",
    year = "2011"
}

@article{Hohenberg:1977ym,
    author = "Hohenberg, P. C. and Halperin, B. I.",
    title = "{Theory of Dynamic Critical Phenomena}",
    doi = "10.1103/RevModPhys.49.435",
    journal = "Rev. Mod. Phys.",
    volume = "49",
    pages = "435--479",
    year = "1977"
}

@article{Chapman:2020vtn,
    author = "Chapman, Shira and Di Pietro, Lorenzo and Grosvenor, Kevin T. and Yan, Ziqi",
    title = "{Renormalization of Galilean Electrodynamics}",
    eprint = "2007.03033",
    archivePrefix = "arXiv",
    primaryClass = "hep-th",
    doi = "10.1007/JHEP10(2020)195",
    journal = "JHEP",
    volume = "10",
    pages = "195",
    year = "2020"
}

@article{Horava:2009uw,
	author = "Horava, Petr",
	title = "{Quantum Gravity at a Lifshitz Point}",
	eprint = "0901.3775",
	archivePrefix = "arXiv",
	primaryClass = "hep-th",
	doi = "10.1103/PhysRevD.79.084008",
	journal = "Phys. Rev. D",
	volume = "79",
	pages = "084008",
	year = "2009"
}

@article{Visser:2009fg,
	author = "Visser, Matt",
	title = "{Lorentz symmetry breaking as a quantum field theory regulator}",
	eprint = "0902.0590",
	archivePrefix = "arXiv",
	primaryClass = "hep-th",
	doi = "10.1103/PhysRevD.80.025011",
	journal = "Phys. Rev. D",
	volume = "80",
	pages = "025011",
	year = "2009"
}

@article{Fujimori:2015wda,
	author = "Fujimori, Toshiaki and Inami, Takeo and Izumi, Keisuke and Kitamura, Tomotaka",
	title = "{Power-counting and Renormalizability in Lifshitz Scalar Theory}",
	eprint = "1502.01820",
	archivePrefix = "arXiv",
	primaryClass = "hep-th",
	doi = "10.1103/PhysRevD.91.125007",
	journal = "Phys. Rev. D",
	volume = "91",
	number = "12",
	pages = "125007",
	year = "2015"
}

@article{Fujimori:2015mea,
	author = "Fujimori, Toshiaki and Inami, Takeo and Izumi, Keisuke and Kitamura, Tomotaka",
	title = "{Tree-Level Unitarity and Renormalizability in Lifshitz Scalar Theory}",
	eprint = "1510.07237",
	archivePrefix = "arXiv",
	primaryClass = "hep-th",
	doi = "10.1093/ptep/ptv185",
	journal = "PTEP",
	volume = "2016",
	number = "1",
	pages = "013B08",
	year = "2016"
}

@article{Berezinskii1971,
	author = {Berezinskii, V. L.},
	title = {Destruction of Long-range Order in One-dimensional and Two-dimensional Systems Possessing a Continuous Symmetry Group. I. Classical Systems},
	journal = {Soviet Journal of Experimental and Theoretical Physics},
	volume = {32},
	pages = {493--500},
	year = {1971}
}

@article{Berezinskii1972,
	author = {Berezinskii, V. L.},
	title = {Destruction of Long-range Order in One-dimensional and Two-dimensional Systems Possessing a Continuous Symmetry Group. II. Quantum Systems},
	journal = {Soviet Journal of Experimental and Theoretical Physics},
	volume = {34},
	pages = {610--616},
	year = {1972}
}

@article{KosterlitzThouless1973,
	author = {Kosterlitz, J. M. and Thouless, D. J.},
	title = {Ordering, Metastability and Phase Transitions in Two-Dimensional Systems},
	journal = {Journal of Physics C: Solid State Physics},
	volume = {6},
	number = {7},
	pages = {1181--1203},
	year = {1973},
	doi = {10.1088/0022-3719/6/7/010}
}

@article{Kosterlitz1974,
	author = {Kosterlitz, J. M.},
	title = {The Critical Properties of the Two-Dimensional {XY} Model},
	journal = {Journal of Physics C: Solid State Physics},
	volume = {7},
	number = {6},
	pages = {1046--1060},
	year = {1974},
	doi = {10.1088/0022-3719/7/6/005}
}

@article{Kosterlitz1977,
	author = {Kosterlitz, J. M.},
	title = {The Critical Properties of the Two-Dimensional {XY} Model},
	journal = {Journal of Physics C: Solid State Physics},
	volume = {10},
	number = {19},
	pages = {3753--3760},
	year = {1977},
	doi = {10.1088/0022-3719/10/19/010}
}

@article{JoseKadanoffKirkpatrickNelson1977,
	author = {Jos\'e, Jorge V. and Kadanoff, Leo P. and Kirkpatrick, Scott and Nelson, David R.},
	title = {Renormalization, Vortices, and Symmetry-Breaking Perturbations in the Two-Dimensional Planar Model},
	journal = {Physical Review B},
	volume = {16},
	number = {3},
	pages = {1217--1241},
	year = {1977},
	doi = {10.1103/PhysRevB.16.1217}
}

@article{AmitGoldschmidtGrinstein1980,
	author = {Amit, Daniel J. and Goldschmidt, Yoseph Y. and Grinstein, G.},
	title = {Renormalisation Group Analysis of the Phase Transition in the 2D {C}oulomb Gas, {S}ine-{G}ordon Theory and {XY}-Model},
	journal = {Journal of Physics A: Mathematical and General},
	volume = {13},
	number = {2},
	pages = {585--620},
	year = {1980},
	doi = {10.1088/0305-4470/13/2/024}
}

@article{Amoretti:2024obt,
    author = "Amoretti, Andrea and Brattan, Daniel K. and Martinoia, Luca",
    title = "{Thermodynamic constraints and exact scaling exponents of flocking matter}",
    eprint = "2405.02283",
    archivePrefix = "arXiv",
    primaryClass = "cond-mat.stat-mech",
    doi = "10.1103/PhysRevE.110.054108",
    journal = "Phys. Rev. E",
    volume = "110",
    number = "5",
    pages = "054108",
    year = "2024"
}

@article{Cheung:2014dqa,
	author = "Cheung, Clifford and Shen, Chia-Hsien and Trnka, Jaroslav",
	title = "{Exact Symmetries of Scattering Amplitudes}",
	eprint = "1410.6395",
	archivePrefix = "arXiv",
	primaryClass = "hep-th",
	reportNumber = "CALT-TH-2014-162",
	doi = "10.1007/JHEP06(2015)118",
	journal = "JHEP",
	volume = "06",
	pages = "118",
	year = "2015"
}

@article{Cheung:2016drk,
	author = "Cheung, Clifford and Kampf, Karol and Novotny, Jiri and Shen, Chia-Hsien and Trnka, Jiri",
	title = "{A Comprehensive Behavior of Effective Field Theories from Soft Limits}",
	eprint = "1610.04381",
	archivePrefix = "arXiv",
	primaryClass = "hep-th",
	reportNumber = "CALT-TH-2016-030",
	doi = "10.1103/PhysRevLett.116.041601",
	journal = "Phys. Rev. Lett.",
	volume = "116",
	number = "4",
	pages = "041601",
	year = "2016"
}

@misc{sezik2026,
      title={Large Spin-Wave Fluctuations Suppress Activity in Malthusian Flocks}, 
      author={Emir Sezik and Gunnar Pruessner},
      year={2026},
      eprint={2608.05805},
      archivePrefix={arXiv},
      primaryClass={cond-mat.stat-mech},
      url={https://arxiv.org/abs/2608.05805}, 
}

@article{Beneke:1998ui,
    author = "Beneke, M.",
    title = "{Renormalons}",
    eprint = "hep-ph/9807443",
    archivePrefix = "arXiv",
    reportNumber = "CERN-TH-98-233",
    doi = "10.1016/S0370-1573(98)00130-6",
    journal = "Phys. Rept.",
    volume = "317",
    pages = "1--142",
    year = "1999"
}

@misc{jentsch2026truequasilongrangeorder,
      title={True and Quasi Long-Range Order in Malthusian Flocks}, 
      author={Patrick Jentsch and Anna Erzberger},
      year={2026},
      eprint={2608.05851},
      archivePrefix={arXiv},
      primaryClass={cond-mat.stat-mech},
      url={https://arxiv.org/abs/2608.05851},
}

	
\end{document}